\documentclass[aps,pre,twocolumn]{revtex4-2}

\usepackage{ulem}
\usepackage{dcolumn}
\usepackage{bm}
\usepackage{amsmath}
\usepackage{amssymb}
\usepackage{amsfonts} 
\usepackage{wasysym}  
\usepackage{graphics}  
\usepackage{psfrag} 
\usepackage{graphicx} 
\usepackage{epsfig}    
\usepackage{rotating}
\usepackage[latin1]{inputenc}
\usepackage{color}
\usepackage{rotating}
\usepackage{csquotes}
\usepackage{multirow}
\usepackage{nicefrac}
\usepackage{cases}
\usepackage[nooneline, tight,raggedright,FIGTOPCAP]{subfigure}
\usepackage[linkcolor=red,citecolor=blue,urlcolor=blue,colorlinks=true]{hyperref}
\usepackage{color}
\usepackage{epsfig}
\usepackage{comment}

\begin{document}

\title{Snowdrift game induces pattern formation in systems of self-propelled particles}

\author{Johanna Mayer$^1$}
\thanks{J.M. and M.O. contributed equally to this work.}
\author{Michael Oberm\"uller$^1$}
\thanks{J.M. and M.O. contributed equally to this work.}
\author{Jonas Denk$^{1,2,3}$}
\author{Erwin Frey$^1$}
\email{frey@lmu.de}
\affiliation{$^1$Arnold Sommerfeld Center for Theoretical Physics (ASC) and Center for NanoScience (CeNS), Department of Physics, Ludwig-Maximilians-Universit\"at M\"unchen, Theresienstrasse 37, D-80333 M\"unchen, Germany}
\affiliation{$^2$Department of Physics, University of California, Berkeley, CA 94720, USA}
\affiliation{$^3$Department of Integrative Biology, University of California, Berkeley, CA 94720, USA}

\date{\today}

\begin{abstract}
Evolutionary games between species are known to lead to intriguing spatio-temporal patterns in systems of diffusing agents. However, the role of inter-species interactions is hardly studied when agents are (self-)propelled, as is the case in many biological systems.
Here, we combine aspects from active matter and evolutionary game theory and study a system of two species whose individuals are (self-)propelled and interact through a snowdrift game.
We derive hydrodynamic equations for the density and velocity fields of both species from which we identify parameter regimes in which one or both species form macroscopic orientational order as well as regimes of propagating wave patterns.
Interestingly, we find simultaneous wave patterns in both species that result from the interplay between alignment and snowdrift interactions - a feedback mechanism that we call game-induced pattern formation.
We test these results in agent-based simulations and confirm the different regimes of order and spatio-temporal patterns as well as game-induced pattern formation. 
\end{abstract}

\maketitle

\section{\label{sec:Introduction}Introduction}
The composition and temporal evolution of microbial as well as other populations depends on a variety of factors including environmental conditions, the population structure, and the degree of mobility of each species~\cite{CremerFrey19, Dobramysl:2018, friedman_ecological_2017}.
Evolutionary game theory (EGT)~\cite{Maynard, Book:HofbauerSigmund98} has been used as a mathematical framework to conceptually describe the evolutionary dynamics of such  populations employing methods from nonlinear dynamics~\cite{Frey.2010}. 
Two-player games, including the prisoner's dilemma and the snowdrift game, have played an important role in the development of the research field. 
In EGT, games are regarded as a population dynamics problem, where 
individuals who follow different strategies (``\textit{species}'') interact over many generations according to rules set by the respective game.
The ensuing (nonlinear) dynamic process then determines which species survive, i.e.\/ `win the game'.

The mobility of individuals in particular has been shown to play a crucial role in the evolutionary success of a species.
Previous theoretical studies have modeled mobility either as a diffusive process, in which individuals perform random walks (see e.g.\ Refs.~\cite{Reichenbach.Frey.2007, Gelimson.Frey.2013, Mobilia:2006}) or alternatively as a migration process, in which individuals move by winning a competition with other species that are spatially adjacent to them (see e.g.\ Refs.~\cite{Nowak.May.1992, Nowak.May.1994, Nakamuru:1997, Helbing:2008, Szabo:2002}). 
To the best of our knowledge, however, there are no studies that deal with the effect of self-propelled movement --- a characteristic feature of many living beings --- on the outcome of spatially extended evolutionary games. 
Since collectives of self-propelled particles (SPP) are known to generically exhibit intriguing self-organisation phenomena like flocking, clustering, and motility-induced phase separation(for reviews see e.g.\ Refs.~\cite{RamaswamyReview10, VicsekReview12, MarchettiReview12,  Cates_Tailleur:2015, ChateReview20}), this will affect the spatial proximity between individuals and thereby their competitive interaction (described as a game). Changes in the composition of the population could in turn affect the ordering processes of the SPPs.

Here, we investigate this interplay of collective effects due to self-propelled motion and nonlinear dynamics based on competitive (game theory) interactions. For specificity, we will address this question for the snowdrift game, one of the classical two-player games.
Using both agent-based simulations and a hydrodynamic approach, we discover a phenomenon that we term \textit{game-induced pattern formation}: When one of the two species starts to exhibit collectively moving patterns due to self-propelled motion and alignment of its agents, a pattern is induced in the second species.
We attribute this to a non-linear feedback mechanism between the local snowdrift game interaction and the alignment between particles.

Upon exploring the model parameters that determine  the alignment and the interaction probability, we find that the system exhibits five distinct phases, each with qualitatively different collective behaviour.
If the game interaction is sufficiently weak, we find that reducing the noise level leads to  a phase transition from a disordered phase to an ordered phase in which one species induces a polar wave pattern in the second species.
These patterns vanish when the noise strength is reduced, leading to a partially ordered phase in which one species shows uniform polar order while the second species is disordered.
A further reduction of the noise strength then also leads to an ordering transition in the second species, which first induces a wave pattern in the previously uniformly ordered species and finally also shows uniform polar order.
All of these phases are generalizations of the phases observed in single-species systems of SPPs, and we show that their existence depends on the interplay between alignment and inter-species interactions.

This paper is organized as follows: In the remainder of the introduction we give a concise overview over previous work on two-player games in a well-mixed environment and in spatially-extended systems. This is mainly intended to inform readers in the active matter field not familiar with this topic about those results relevant for the present study. 
In Section~\ref{sec:Model} we first summarize the Boltzmann approach for SPPs and the dynamics of the snowdrift game to then combine both models to find equations governing the dynamics of two SPP species with inter-species snowdrift game interaction. 
Section~\ref{sec:Hydrodynamic_Theory} takes these equations as a starting point to derive simpler hydrodynamic equations and analyses them using both analytical and numerical methods.  
In this section we obtain the phase diagram of the system in the space spanned by parameters of self-propelled motion and game interaction and discuss their properties.
The results of the hydrodynamic approach and the overall phenomenology of the system are qualitatively validated in Section~\ref{sec:Agent-based} by agent-based simulations.
We conclude with a summary and discussion of our results in Section~\ref{sec:DiscussionAndConclusions}.

\subsection{Two-player public good games}

Originally, the `snowdrift' game (also known as `hawk-dove' game) was formulated as a strategic `cooperator-defector' game \cite{Osborne_game_theory}. 
It owes its name to its prominent explanatory framework:
Two drivers are caught in a winter storm and a snowdrift is blocking the road. 
Both drivers now have the option of either remaining in the car or getting out and removing the snowdrift. 
Since both drivers want to continue their journey, but getting out is not really desirable, the best strategy for each driver is to do exactly the opposite of what the other driver does. 

This strategic game can be transferred to the field of EGT and microbial systems such as yeast cells competing for resources and using different survival strategies~\cite{Gore09}:
In a metabolically costly process, part of the yeast cells (producer cells) convert sucrose to glucose, which it can use as an energy source.
This process yields more glucose than can be metabolized, and the extra glucose diffuses into the surrounding medium. 
There it can be used by all yeast cells, regardless of whether they produce glucose or not.
As long as there is an excess of glucose in the system, it is therefore advantageous to save the metabolic costs for its production and only consume it (non-producer cells). 
In contrast, when there is a lack of glucose, it becomes advantageous to produce glucose as this provides an energy source.
This leads to the coexistence of producer (cooperator) and non-producer (defector) cells in the yeast population.

The snowdrift game is one out of four possible two-player (`cooperator-defector') games. These include the prisoner's dilemma, the coordination, and the harmony game. Their dynamics in a well-mixed environment (where diffusion is much faster than the reaction kinetics) can be formulated in terms of concepts from nonlinear dynamics~\cite{Frey.2010}. 
The main difference between these four games is their evolutionary outcome, i.e.\/ the composition of the population in the long run. 
While, for example, in the snowdrift game, cooperators and defectors coexist, defectors always dominate in the prisoner dilemma and cooperators go extinct. 
This is related to the rules of the prisoner's dilemma, which can be formulated as a public goods game where an individual can either cooperate by providing some kind of public good or defect by exploiting this public good. 
Under these conditions, defectors will prevail as they can benefit from the interaction with cooperators without paying any costs.

\subsection{Evolutionary games in spatially extended systems}

The spatial extension of a population and the ensuing spatio-temporal arrangement of individuals can strongly influence the evolutionary dynamics~\cite{Korolev:2010, Drescher:2014, KayserReview18, Excoffier.Petit.2009}. 
It is well known that this plays an important role for microbial life, as many microbes on our planet are found in the form of dense microbial communities such as colonies and biofilms~\cite{Drescher:2014, Yanni:2019ip}. 

A key factor for the dynamics is the spatial clustering of individuals, which has already been argued by Hamilton to support cooperation in populations~\cite{Hamilton:1964}, because cooperating individuals are more likely to interact with each other and therefore are less likely to be exploited by defectors. 
To what degree this determines the composition of a population has been studied extensively on different kinds of lattices as well as for different kinds of interaction rules~\cite{Nowak.May.1992, Killingback99, Nowak.May.1994, Fu.Hauert.2010, Helbing:2009, Langer:2008, Nakamuru:1997, Gelimson.Frey.2013, Bauer.Frey.2018, Bauer.Frey.2018y3b, Bauer.Frey.2018qp}.
Specifically, for the snowdrift game, it was shown that accounting for spatial extension can change population ratios compared to well-mixed systems~\cite{Hauert:2004,  Fu.Hauert.2010}. 

While all the studies mentioned above consider diffusive dynamics, self-propulsion is an essential part of biological reality. In fact, there are a large variety of microbes that show directed mobility, often along chemotactic gradients~\cite{Cremer:2019, Jeckel:2019, Liu:51NDZHe4, Spormann.1999} or due to chemical warefare~\cite{weber_chemical_2014}.
Here, motility is an important factor of their strategy, e.g.~for directed motion away from hostile environments, chemotaxis or self-organisation into flocks.

\section{\label{sec:Model}Model description }

\subsection{\label{subsec:AMModel}Active matter model}

\begin{figure}[ht]
\centering
\includegraphics[width = \linewidth]{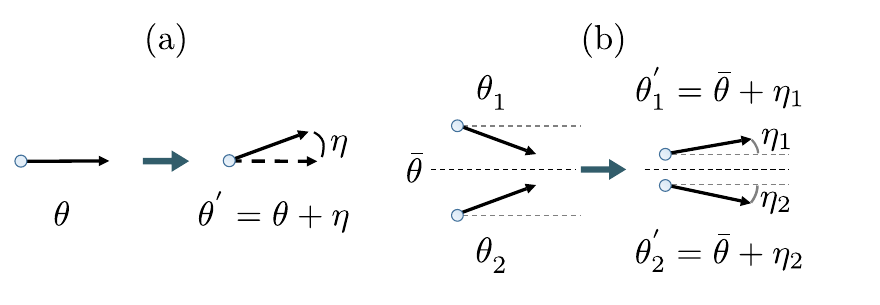}
\caption{Illustration of  self-propelled dynamics: (a) Particles change their orientation $\theta$ by a stochastic angle $\eta$ to $\theta' = \theta + \eta$ during self-diffusion events. (b) Binary collisions between particles of the same species lead to polar half-angle alignment of their orientations. We allow for additional stochastic noise terms $\eta_1$ and $\eta_2$ during the collision event.}
\label{fig:AM_interaction_model} 
\end{figure}

In order to implement propulsion of particles of both species, $A$ and $B$, we employ a kinetic Boltzmann approach~\cite{Bertin06,Bertin09}.
Here, particles with a diameter $d_0$ move ballistically on a two-dimensional substrate with constant speed $v_0$ and orientation $\theta$.
Ballistic motion is interrupted by stochastic \textit{self-diffusion} events: 
Particles change their orientation
by an angle $\eta$ at a rate $\omega$ to
\begin{equation}
\theta' = \theta + \eta,
\end{equation}
where $\eta$ is a stochastic variable (noise); see Fig.~\ref{fig:AM_interaction_model}a. 
Hence, the particles move in a run-and-tumble-like fashion with a mean time $\omega^{-1}$ between tumbling events. 

Further, we assume that \textit{intra-species} interactions are due to binary collisions between particles, that lead to alignment of their orientations, while \textit{inter-species} interactions follow a snowdrift game (see the following section \ref{subsec:GTModel}).
We implement intra-species aligning collisions in terms of half-angle alignment rules; see Fig.~\ref{fig:AM_interaction_model}b. 
The particle orientations after a collision event are given by
\begin{equation}
\theta_1'= \frac{\theta_1+\theta_2}{2} + \eta_1 , \qquad \theta_2'= \frac{\theta_1+\theta_2}{2} + \eta_2,
\end{equation}
where we allowed for additional noise terms $\eta_1$ and $\eta_2$. We have chosen these rules to keep our analysis conceptually simple. In actual systems one would need to account for the problem-specific interaction between the agents of the active system. This has, for example, been achieved for the actin-myosin motility assay \cite{Huber18, Suzuki:2015, Denk:2020}.
For simplicity, diffusion and collision noise variables are drawn from a Gaussian distribution with the same standard deviation $\sigma$.

For a dilute system of only one species of propelled particles, Bertin et al.~\cite{Bertin06, Bertin09} proposed a kinetic Boltzmann equation. 
It describes the dynamics of the one-particle probability density function, $f\left(\mathbf{r} ,\theta,t\right)$, which indicates the probability that a particle is located at position $\mathbf{r}$ with the orientation $\theta$ at time $t$.
The kinetic Boltzmann equation for self-propelled particles for one species reads~\cite{Bertin06,Bertin09}
\begin{equation}
\label{eq:BertinBoltzmann}
    \partial_t f\left(\mathbf{r} ,\theta,t\right)  + v_0 \, \mathbf{e}_{\theta} \cdot \nabla f  = I_{\text{diff}} \left[f\right]  + I_{\text{coll}} \left[f,f\right],
\end{equation} 
where $I_{\text{diff}}$ and $I_{\text{coll}}$ denote contributions resulting from diffusion and collision events, respectively, while $\mathbf{e}_{\theta}$ is the unit vector pointing in $\theta$-direction. 
Their explicit expressions can be found in Appendix~\ref{app:Boltzmann}.
In the present context, the most important feature of $I_{\text{diff}}$ and $I_{\text{coll}}$ is that they depend on the parameter $\sigma$, which characterizes the noise strength.

\subsection{\label{subsec:GTModel}Snowdrift game}
\nobreak 

\begin{figure}[t]
\centering
\includegraphics[width = \linewidth]{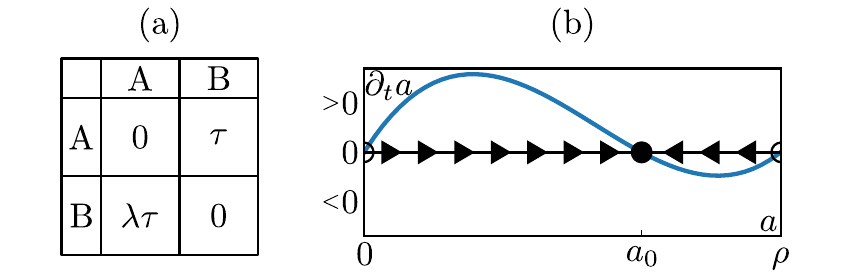}
\caption{\label{fig:GT_interaction_model} Illustration of the snowdrift game. 
(a) Payoff matrix: 
There is a non-zero payoff only if individuals belonging to different species meet. Then, $A$ receives a payoff $\tau$, while $B$ receives a payoff $\lambda \tau$.
The \textit{relative fitness} $\lambda$ and \textit{game strength} $\tau$ serve as control parameters.
(b) Flow diagram of the replicator equation, Eq.~\eqref{eq:ReplicatorA}, for species $A$ with population density $a$: Unstable fixed points are marked with an open circle, while the stable fixed point at $a_0$ is marked with a filled circle. The arrows indicate the flow towards the stable fixed point.}
\end{figure}

In addition to the alignment interactions, we assume that particles during collisions also play a snowdrift game.
The outcome of this interaction is specified by the payoff matrix (Fig.~\ref{fig:GT_interaction_model}a), which is non-zero only if two different species meet.
The fitness of a particular strategy, $A$ or $B$, is defined as a constant background fitness (set to $1$) plus the average payoff
obtained from playing the game~\cite{BacterialGames, Frey.2010}:
\begin{subequations}
\label{eq:Fitness}
\begin{alignat}{2}
    f_{A} &=  1 + \tau \, b , \\	    
	f_{B} &=  1 +  \tau \, \lambda \, a 
	\, .
\end{alignat}
\end{subequations}
Here, $a$ and $b$ denote the densities of species $A$ and $B$, respectively. 
In the following, we will refer to $\lambda$ as the \textit{relative fitness} parameter of particles of species $B$ relative to $A$, and $\tau$ as the \textit{game strength}.

The time evolution now follows `survival of the fittest':
The species with higher fitness value increases in population size, while the other species decreases in number.
Specifically, the dynamics of the densities $a$ and $b$ are given by the replicator equations:
\begin{subequations}
\label{eq:ReplicatorEqs}
\begin{align}
	\partial_t a 
	&= \left(f_A-f_B\right) b\, a,\\
	\partial_t b 
	&= \left(f_B-f_A\right) a\, b .
\end{align}
\end{subequations}
These \textit{replicator equations} govern the dynamics of the game.
The fitness difference $(f_A-f_B)$ indicates the relative strength of the interaction partner, while the factor $a \, b$ encodes the probability that particles of species $A$ and $B$ meet and interact with each other. 
Note that the overall density $\rho=a+b$ is constant. 
Inserting the expressions for the fitness values $f_A$ and $f_B$, Eqs.~\eqref{eq:Fitness}, into Eqs.~\eqref{eq:ReplicatorEqs} one can rewrite the replicator equations as
\begin{subequations}
\label{eq:ReplicatorEquations}
\begin{align}
	\partial_t a 
	&= \tau\left( b - \lambda a\right) a \, b, \\
	\partial_t b 
	&= \tau\left(\lambda a - b\right) a \, b .
\end{align}
\end{subequations}
The meaning of the terms \textit{relative fitness} and \textit{game strength} for $\lambda$ and $\tau$, respectively, now become evident:
If $b \,{>}\, \lambda a$, there is a fitness advantage for species $A$ and its density grows relative to species $B$.
In contrast, if $b \,{<}\, \lambda a$, the growth of  species $B$ is favored relative to species $A$. 
Since the equations are anti-symmetric under the exchange of species $A$ and $B$, we can, without loss of generality, impose that species $A$ is always fitter than species $B$, i.e.~constraining $\lambda \,{\in}\, [0,1]$. 
By changing $\lambda$ we can tune the relative fitness from the two species being equally competitive ($\lambda \,{=}\, 1$) to species $A$ being strongly dominant over $B$ ($\lambda \,{\ll}\, 1$). 
The second game parameter, $\tau \,{\in}\, [0,\infty)$, can be understood as the \textit{game strength}.

Since the total particle number is conserved ($b \,{=} \rho \,{-}\, a$), Eqs.~\eqref{eq:ReplicatorEquations} can be reduced to the temporal evolution of a single species (say $A$):
\begin{equation}
\label{eq:ReplicatorA}
	\partial_t a = \tau\left( b - \lambda a\right)ab = \tau\left[ \left(\rho-a\right) - \lambda a\right]a\left(\rho-a\right).
\end{equation}
Figure~\ref{fig:GT_interaction_model}b shows the flow diagram of Eq.~\eqref{eq:ReplicatorA} with arrows pointing to the right for increasing $a$ ($\partial_t a \,{>}\, 0$) and to the left for $\partial_t a \,{<}\, 0$. 
There are three fixed points (stationary points with $\partial_t a =0$): 
\begin{equation}
	a_B = 0,~ a_A = \rho, \textrm{ and } a_{0}=\frac{\rho}{1+ \lambda}.
\end{equation}
The fixed points $a_B$ and $a_A$ are unstable in the snowdrift game, while $a_{0}$ is stable.
For $a_B \,{=}\, 0$ and $a_A \,{=}\, 0$ species $A$ and $B$ go extinct, respectively, and therefore the game ends (absorbing fixed points).
The fixed point $a_{0}$ refers to a stable coexistence of both species with population ratio $b/a \,{=}\, \lambda$. 

\subsection{\label{subsec:ComboModel}Active matter snowdrift game}

To describe our system of two interacting self-propelled particle species, we introduce the one-particle probability density functions $\alpha\left(\mathbf{r} ,\theta,t\right) $ and $\beta\left(\mathbf{r},\theta,t\right)$, governing the temporal evolution of species $A$ and $B$, respectively.
We use the kinetic Boltzmann equation, Eq.~\eqref{eq:BertinBoltzmann}, and extend it for each species separately by \textit{snowdrift interaction terms}, $ I_{\text{game}}^{A} \left[\alpha,\beta\right]$ and $ I_{\text{game}}^{B} \left[\beta,\alpha \right]$, chosen in analogy to the replicator equations, Eqs.~\eqref{eq:ReplicatorEquations}. 
This leads to the general form of the two-species Boltzmann equations
\begin{subequations}
\label{eq:Boltzmann}
\begin{align}
\partial_t \alpha {+}  v_0 \,\mathbf{e}_\theta \nabla  \alpha
 &=  I_{\text{diff}} \left[\alpha\right] {+} I_{\text{coll}} \left[\alpha,\alpha\right]  {+} I_{\text{game}}^{A} \left[\alpha,\beta\right], 
\\
\partial_t \beta  {+}  v_0 \,\mathbf{e}_\theta \nabla  \beta 
 &=  I_{\text{diff}} \left[\beta\right]  {+} I_{\text{coll}} \left[\beta,\beta\right]  {+} I_{\text{game}}^{B} \left[\beta,\alpha\right]. 
\end{align}
\end{subequations}
We assume snowdrift interactions to be local and independent of particle orientation as in Eqs.~\eqref{eq:ReplicatorEquations}.
With the local densities $a\left( \mathbf{r},t\right)$ and $b\left( \mathbf{r},t\right)$ of species $A$ and $B$, related to $\alpha$ and $\beta$ by
\begin{subequations}
\label{eq:density_a_b}
\begin{align}
a \left( \mathbf{r},t\right)& = \int_{-\pi}^{\pi} d\theta \,\alpha\left( \mathbf{r},\theta,t\right) ,
\\
b \left( \mathbf{r},t\right)& = \int_{-\pi}^{\pi} d\theta \,\beta\left( \mathbf{r},\theta,t\right),
\end{align}
\end{subequations}
the snowdrift interaction terms read
\begin{subequations}
\label{eq:GameInteractionTerm}  
\begin{align}
I_{\text{game}}^{A}  \left[\alpha,\beta\right] &= \tau \left(  b - \lambda a \right) \, b  \,\alpha , \\
I_{\text{game}}^{B}  \left[\beta , \alpha \right] &= \tau \left( \lambda a - b \right) \,a \,\beta.
\end{align}
\end{subequations}
The dependency on $\alpha$ and $\beta$ in Eq.~(\ref{eq:GameInteractionTerm})a and Eq.~(\ref{eq:GameInteractionTerm})b, respectively, is chosen to match the form of the Boltzmann equation with its dependency on the one-particle densities $\alpha$ and $\beta$.
The game strength $\tau$ tunes the relative weight of the snowdrift interaction compared to the alignment interaction (collision terms).
If one sets $\tau \,{=}\, 0$, the snowdrift interaction terms vanish and Eqs.~\eqref{eq:Boltzmann} reduce to two independent equations for species $A$ and $B$. 
On the other hand, in the absence of convection ($v_0 \,{=}\, 0$), diffusion ($I_{\text{diff}} \,{=}\, 0$), and collision ($I_{\text{coll}} \,{=}\, 0 $), one recovers the replicator equations~\eqref{eq:ReplicatorEquations} by integrating Eqs.~\eqref{eq:Boltzmann} over the angle $\theta$.

In the following, time, space, and densities are rescaled such that one measures time in units of $1/\omega$, position in units of $v_0/\omega$ and spatial densities in units of $\omega/(v_0 d_0)$.
This amounts to setting $v_0 \,{=}\, d_0 \,{=}\, \omega \,{=}\, 1$.

\section{\label{sec:Hydrodynamic_Theory}Hydrodynamic Theory}

In the previous section, we have formulated an extended version of two coupled kinetic Boltzmann equations for a dilute system of two species of self-propelled particles with a snowdrift interaction. 
It includes convection, self-diffusion, intra-species binary collision (with polar alignment), as well as a snowdrift interaction between particles.
As solving the kinetic Boltzmann equations, Eqs.~\eqref{eq:Boltzmann}, analytically is generally impractical, one has to resort to numerical implementations~\cite{Thueroff14, DenkHuber16, Denk:2020}.
However, as exemplified by studies of the kinetic Boltzmann equation for a single species~\cite{Bertin06, Peshkov14, Bertin09, DenkHuber16, PeshkovAranson12, PeshkovNgo12}, one can also use Eqs.~\eqref{eq:Boltzmann} to derive hydrodynamic equations for those collective variables that vary slowly in space and time.

In the next section, we first derive these hydrodynamic equations for the system of two species, Eqs.~\eqref{eq:Boltzmann}, and then study them analytically for spatially uniform systems. 
Then we investigate spatially non-uniform solutions using linear stability analysis of the uniform states and numerical solutions of the full hydrodynamic equations.

\subsection{Derivation of hydrodynamic equations}

As the probability densities $\alpha(\mathbf{r},\theta,t)$ and $\beta(\mathbf{r},\theta,t)$ are periodic functions in $\theta$, it is convenient to work with their Fourier modes, defined as $\alpha_k(\mathbf{r},t) := \int_{-\pi}^\pi d\theta \, e^{i \theta k} \alpha(\mathbf{r},\theta,t)$ and  $\beta_k(\mathbf{r},t) := \int_{-\pi}^\pi d\theta \, e^{i \theta k} \beta(\mathbf{r},\theta,t)$. 
Note that the zeroth Fourier modes are the densities of the respective species, $\alpha_0 \,{=}\, a, \beta_0 \,{=}\, b$, defined by Eqs.~\eqref{eq:density_a_b}.
Besides the densities of species $A$ and B, the second set of slow, hydrodynamic variables are the polar order fields \(\mathbf{P}_A (\mathbf{r},t)\) and  \(\mathbf{P}_B (\mathbf{r},t) \), which are related to the first Fourier modes
\begin{align}
a\,\mathbf{P}_A   = \left( \begin{array}{c} \text{Re}(\alpha_1 ) \\ \text{Im}(\alpha_1) \end{array}\right),
\quad
b\, \mathbf{P}_B   = \left( \begin{array}{c} \text{Re}(\beta_1) \\ \text{Im}(\beta_1) \end{array}\right).
\end{align}
Hence, the modes $\alpha_1$ and $\beta_1$ are a measure for the polar order multiplied by the density at a given point in space and time of the system. 
If $\alpha_1 \,{=}\, \beta_1 \,{=}\, 0$, the system is disordered. 
If $\alpha_1$ or $\beta_1 $ is nonzero, there is a preferred direction of motion, meaning the respective species moves collectively. 
We will refer to $\alpha_1$ and $\beta_1$ in the following as \textit{velocity fields}.

The expansion of the kinetic Boltzmann equations, Eqs.~\eqref{eq:Boltzmann}, in Fourier modes is given by  (see Appendix~\ref{app:A_DerivationHydro} for details)
\begin{subequations}
\label{eq:FourierBoltzmann}
\begin{align}
	\partial_t \alpha_k 
	+ \frac{1}{2} &\bigl( \nabla \alpha_{k-1} + \nabla^{*} \alpha_{k+1} \bigr) 
	= - \bigl( 1-e^{-(k\sigma)^{2}/2} \bigr) 
	\, \alpha_k \nonumber \\
    &+ \sum\limits_p \alpha_{k-p} \alpha_p\, \mathcal{I}_{p,k} +\tau (b - \lambda a) \,b\,\alpha_k, 
\\
	\partial_t \beta_k 
	+ \frac{1}{2}& \bigl( \nabla \beta_{k-1} + \nabla^{*} \beta_{k+1} \bigr) 
	= - \bigl( 1- e^{-(k\sigma)^{2}/2} \bigr) 
	\, \beta_k \nonumber \\
    &+ \sum\limits_p \beta_{k-p} \beta_p\, \mathcal{I}_{p,k} - \tau( b - \lambda a ) 
    \, a \, \beta_k ,
\end{align}
\end{subequations}
where we introduced the abbreviations $\nabla \,{=}\, \partial_x + i\partial_y$ and $\nabla^{*} \,{=}\,  \partial_x - i\partial_y$. The explicit expressions for the collision kernels $\mathcal{I}_{p,k}(\sigma)$  are given in Appendix~\ref{app:A_DerivationHydro}.
Equation~\eqref{eq:FourierBoltzmann} constitutes an infinite set of differential equations that couple lower- with higher-order Fourier modes.
For $k \,{=}\, 0$, Eqs.~\eqref{eq:FourierBoltzmann} yield 
\begin{subequations}
\label{eq:Hydro_densities}
\begin{align}
\label{subeq:Hydro_a}
\partial_t a &=
-\frac{1}{2} \left( \nabla  \alpha_1^* + \nabla^*  \alpha_1 \right) + \tau \left( b - \lambda a \right) \, b \, a ,
\\
 \partial_t b &=
-\frac{1}{2} \left( \nabla  \beta_1^* + \nabla^*  \beta_1 \right) + \tau\left( \lambda a -  b\right) \, a \, b.
\end{align}
\end{subequations}
The first terms in Eqs.~\eqref{eq:Hydro_densities} can be understood as the temporal change of the densities due to spatial variations in the velocity fields, while the last terms can be interpreted as source or sink terms given by the replicator equations, Eqs.~\eqref{eq:ReplicatorEquations}. 
These source terms couple the densities of species $A$ and $B$ and we will refer to them in the following as \textit{snowdrift terms}.
Note, that the sum of both snowdrift terms is zero at every point in space and time. Hence, the overall density $\rho \,{=}\, a+b$ follows a continuity equation
\begin{equation}
\partial_t \rho =
-\frac{1}{2} \Bigl[ \nabla  (\alpha_1^*+\beta_1^*) + \nabla^*  (\alpha_1 +\beta_1) \Bigr].
\end{equation}
The spatial average of the density, $\bar{\rho}$, is therefore a conserved quantity and a control parameter of the system.

For a spatially uniform system, the spatial derivatives in the equations for the particle densities, Eqs.~\eqref{eq:Hydro_densities}, vanish and Eqs.~\eqref{eq:Hydro_densities} reduce to the  replicator equations.
The corresponding steady states are given by the stable snowdrift fixed point with $a^{(0)} \,{=}\, \bar{\rho}/(1+\lambda)$ and $b^{(0)} \,{=}\, \bar{\rho} \lambda/(1+\lambda)$. 
Hence, a state with uniform densities $a \,{=}\, a^{(0)}$ and $b \,{=}\, b^{(0)}$ and all higher Fourier modes vanishing is a solution to Eqs.~\eqref{eq:FourierBoltzmann}. 
This solution describes a spatially uniform, disordered system with densities set to their snowdrift fixed points.
Small spatially uniform perturbations $\delta \alpha_k, \delta \beta_k$ to this solution evolve to linear order as
\begin{subequations}
    \begin{align}
    \partial_t \delta \alpha_k &= \mu_{k}^A(a^{(0)},b^{(0)})\, \delta \alpha_k, \\
    \partial_t \delta \beta_k &= \mu^B_k(a^{(0)},b^{(0)}) \, \delta \beta_k,
    \end{align}
\end{subequations}
with
 \begin{subequations}
 \label{eq:MuaMub}
 \begin{align}
  	\mu^A_k(a,b)  =&
     \bigl(e^{{-}\frac{\sigma^{2}}{2}} {-} 1 \bigr) 
     + (\mathcal{I}_{0,k}{+}\mathcal{I}_{k,k})a  
     + \tau(b {-} \lambda a)b,  
\\
  	\mu^B_k(a,b)  =&
     \bigl( e^{{-}\frac{\sigma^{2}}{2}} {-} 1 \bigr) 
     + (\mathcal{I}_{0,k}{+}\mathcal{I}_{k,k})  b 
     + \tau(\lambda a {-} b)a . 
 \end{align}
 \end{subequations}
The respectively first two terms in $\mu^A_k$ and $\mu^B_k$ are due to the self-propelled motion of both species and are familiar expressions from one-species SPP systems~\cite{Bertin06,Bertin09}.
Here, we encounter an additional term due to the snowdrift game, which implies that the growth rate of every Fourier mode depends on the densities of both species.
When we insert the stable solution for the densities, $a^{(0)} \,{=}\, \bar{\rho}/(1+\lambda)$ and $b^{(0)} \,{=}\, \bar{\rho}\lambda/(1+\lambda)$, into the expressions for $\mu^A_k$ and $\mu^B_k$, the snowdrift terms vanish and we are left with the first two terms, which are then functions of $\sigma$, $\bar{\rho}$ and $\lambda$.
Varying these parameters, one finds that all Fourier modes $\mu_k^A$ and $\mu_k^B$ with $k\ge 2$ are negative and only the growth rates for the first Fourier mode, $\mu^A_1$ and $\mu^B_1$, can become positive, meaning that small perturbations in the velocity fields $\alpha_1$ or $\beta_1$, respectively, will grow exponentially. 
For fixed overall density, $\bar{\rho}$, and relative game strength, $\lambda$, the growth rates $\mu^A_1$ and $\mu^B_1$ are negative for high values of the noise strength $\sigma$ and become positive when the noise strength is low enough. 
This defines threshold values for the noise strength, $\sigma^A_{t}(\lambda,\bar{\rho})$ and $\sigma^B_{t}(\lambda,\bar{\rho})$, at $\mu_1^A |_{\sigma^A_t,\bar{\rho},\lambda} \,{=}\, 0$ and $\mu_1^B |_{\sigma^B_t,\bar{\rho},\lambda} \,{=}\, 0$, respectively (see Fig.~\ref{fig:uniformSol}b). 
Below these thresholds ($\sigma \,{<}\, \sigma^A_t$ or $\sigma \,{<}\, \sigma^B_t$), the spatially uniform solution is unstable and small perturbations in the velocity fields, $ \alpha_1$ or $\beta_1$, have positive growth rates.

Close to the threshold values $\sigma_t^A$ and $\sigma_t^B$, a weakly nonlinear analysis yields further insights into the dynamics of the system and the resulting steady states. 
Here we follow Bertin et al.~\cite{Bertin06} and use the following truncation scheme: 
We assume that close to the onset of polar order the velocity fields are small, $\alpha_1, \beta_1 \,{\ll}\, 1$, and obey the scaling relations $(a- a^{(0)}) \,{\sim}\,  \alpha_1$, $\alpha_k \,{\sim}\,  \alpha_1^{|k|}$ (and analogously for $\beta_k$). 
Furthermore, we assume that spatial and temporal variations $\nabla, \, \partial_t \,{\sim}\,  \alpha_1$ are small.
With these assumptions, one can truncate and close the Boltzmann equation in Fourier space, Eq.~\eqref{eq:FourierBoltzmann}, at the third order term for the velocity field, $\alpha_1^3$.
We refer the interested reader to Appendix~\ref{app:A_DerivationHydro} and Refs.~\cite{Bertin06,Bertin09,Peshkov14} for a more detailed discussion on the truncation scheme.
Following the described procedure, we obtain hydrodynamic equations for the velocity fields of species $A$ and $B$:
\begin{subequations}
\label{eq:Hydro_velocities}
\begin{align}
\begin{split}
        \partial_t \alpha_1 &=
        \left( \mu^A_1(a,b) - \xi(a) |\alpha_1 |^2 \right) \alpha_1+ \nu(a) \, \nabla^* \nabla \alpha_1 
        \\
        & \qquad - \gamma(a) \, \alpha_1 \nabla^* \alpha_1 - \kappa(a)\, \alpha_1^* \nabla \alpha_1 - \frac{1}{2} \nabla a,
\end{split}
\\
\begin{split}
          \partial_t \beta_1 &=
          \left( \mu^B_1(a,b) - \xi(b) |\beta_1|^2 \right) \beta_1+ \nu(b) \, \nabla^* \nabla \beta_1
          \\
          &\qquad  - \gamma(b) \, \beta_1 \nabla^* \beta_1 - \kappa(b)\, \beta_1^* \nabla \beta_1 - \frac{1}{2} \nabla b.
\end{split}
\end{align}
\end{subequations}
For explicit expressions of the coefficients please refer to Appendix~\ref{app:A_DerivationHydro}. 
The essential difference between Eqs.~\eqref{eq:Hydro_velocities} and the hydrodynamic equation for the velocity field of one-species SPP~\cite{Bertin06} lies in the coefficients $\mu^A_1$ and $\mu^B_1$, which here couple the velocity fields to the density of the respective other species (see Eqs.~\eqref{eq:MuaMub}). 
Note that the snowdrift interaction term only appears in the term linear in the velocity fields in Eqs.~\eqref{eq:Hydro_velocities}. This is a consequence of the truncation scheme and is explained in more detail in Appendix~\ref{app:A_DerivationHydro}. 

Taken together, Eqs.~\eqref{eq:Hydro_densities} and Eqs.~\eqref{eq:Hydro_velocities} constitute a minimal set of hydrodynamic equations that describe the dynamics of the slow variables of our system, the densities and velocity fields of both species. 
Overall, we identify four control parameters: the noise strength $\sigma$, the overall density $\bar{\rho}$, the game strength $\tau$, and the relative fitness $\lambda$.
For specificity, we use a fixed value for the average density, $\bar{\rho} \,{=}\, 1$ (in units of $\omega/(v_0 d_0)$) and vary the remaining parameters. 
We observe the same phenomenology if we instead fix $\sigma$ and use $\bar{\rho}$ as control parameter \cite{JohannaMasterthesis} (this is due to the opposing roles of density and noise changes in the self-organization of active matter systems \cite{ChateReview20, VicsekReview12, MarchettiReview12}).
 
In the next sections we will first determine solutions for the spatially uniform case of Eqs.~\eqref{eq:Hydro_densities} and Eqs.~\eqref{eq:Hydro_velocities} and subsequently analyse their linear stability against spatially non-uniform perturbations.

\subsection{\label{sec:uniform_Sol}Spatially uniform solutions}

The stationary uniform solutions of the hydrodynamic equations, Eqs.~\eqref{eq:Hydro_densities} and~\eqref{eq:Hydro_velocities}, can be found by setting all temporal and spatial derivatives to zero. 
Then, the equations for the densities and velocity fields decouple.
For the densities, Eqs.~\eqref{eq:Hydro_densities}, one obtains
\begin{subequations}
\label{eq:Uniform_densities}
\begin{align}
\label{eq:Uniform_density_a}
    \partial_t a  
    &= \tau (b - \lambda a)\, a \, b,
    \\
\label{eq:Uniform_density_b}
    \partial_t b 
    &= \tau (\lambda a-b)\, a \, b.
\end{align}
\end{subequations}
These are precisely the replicator equations, Eqs.~\eqref{eq:ReplicatorEquations}, with the stable solution (note we have set $\bar \rho =1$)
\begin{equation}
\label{eq:UnifSolDensities}
    a^{(0)} 
    = \frac{1}{1+\lambda},\, 
    \quad b^{(0)} 
    = \frac{\lambda}{1+\lambda}.
\end{equation}
Hence, the spatially uniform densities of each species are determined by the relative fitness $\lambda$, and are independent of the game strength $\tau$.

\begin{figure}[t]
    \centering
    \includegraphics[width=\linewidth]{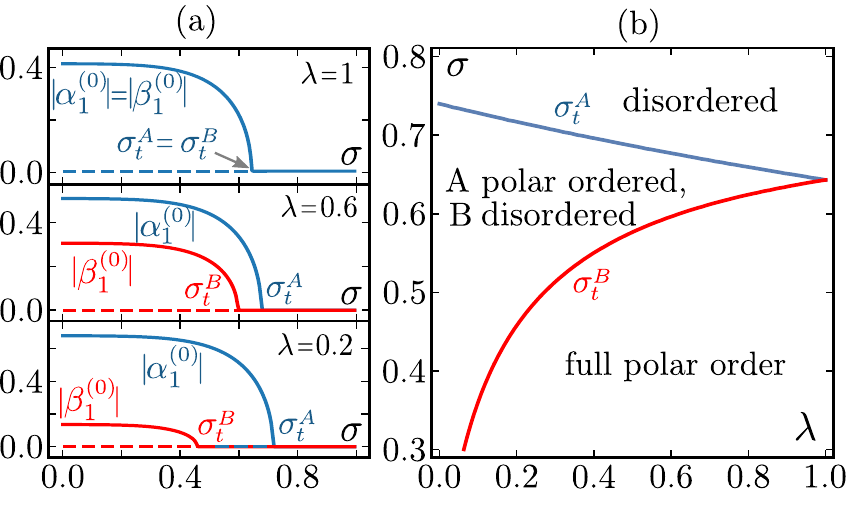}
    \caption{Solutions of the spatially uniform equations for the velocity fields $\alpha_1$ and $\beta_1$ (see Eqs.~\eqref{eq:uniform_order_fields}).
    (a) Amplitudes of the solutions as functions of the noise strength $\sigma$ with the relative fitness set to $\lambda \,{=}\, 1$, $\lambda \,{=}\, 0.6$, and $\lambda \,{=}\, 0.2$, as indicated in the graph. Solid lines are stable solutions, dashed lines unstable solutions.
    If the noise strength is smaller than a threshold value, $\sigma \,{\leq}\,  \sigma_t^A \, (\sigma \,{\leq}\, \sigma_t^B)$, the disordered solution $|\alpha_1^{(0)}| \,{=}\, 0$ $(|\beta_1^{(0)}| \,{=}\, 0)$ becomes unstable and a new stable solution with  $|\alpha_1^{(0)}| \,{>}\,0 $ 
    $( |\beta_1^{(0)}| \,{>}\, 0)$ 
    appears.
    (b) Bifurcation (phase) diagram for the spatially uniform solution in control parameter space ($\lambda,\sigma$).
   At  $\sigma_t^A(\lambda)$, there is a transitions from a disordered state (fixed point $|\alpha_1^{(0)}| \,{=}\,| \beta_1^{(0)}| \,{=}\, 0$) to a broken symmetry state with macroscopic polar order in $A$ only ($|\alpha_1^{(0)}|  \,{>}\, 0$, $|\beta_1^{(0)}| \,{=}\, 0$).
    $A$ second transition to a state with polar order in both species ($|\alpha_1^{(0)}| \,{>}\, 0$ and $ |\beta_1^{(0)}| \,{>}\, 0$) occurs at $\sigma_t^B(\lambda) \,{\leq}\, \sigma_t^A(\lambda)$. The threshold values for the noise strengths coincide if the relative game strength $\lambda \,{=}\, 1$.
    }
\label{fig:uniformSol}
\end{figure}

The hydrodynamic equations for the velocity fields, Eqs.~\eqref{eq:Hydro_velocities}, of a spatially uniform system are given by 
\begin{subequations}
\label{eq:uniform_order_fields}
\begin{align}
    \partial_t \alpha_1 
    &= \left( \mu^A_1 - \xi |\alpha_1|^2 \right) \alpha_1 ,
    \\
    \partial_t \beta_1 
    &= \left( \mu^B_1 - \xi |\beta_1 |^2 \right) \beta_1 .
\end{align}
\end{subequations}
Hence, $ \alpha_1 \,{=}\, \beta_1  \,{=}\, 0 $ (corresponding to a disordered state) is always a trivial solution.
A nontrivial solution occurs when the terms in the brackets balance each other. 
Since $\xi \,{>}\, 0$ for all values of the control parameters (see Appendix~\ref{app:A_DerivationHydro}), such a solution for $\alpha_1$ ($\beta_1$) exists for $\mu^A_1 \,{>}\, 0$ ($\mu^B_1 \,{>}\, 0$). 
We know from the last section that the growth rate $\mu^A_1$ ($\mu^B_1$) is negative at high values of the noise strength $\sigma$ and becomes positive below the threshold value $\sigma_t^A$ ($\sigma_t^B$), which depends on the relative fitness $\lambda$, as shown in Fig.~\ref{fig:uniformSol}b.
Note that the threshold values for the noise strength are different for species $A$ and $B$ because their densities are different in steady state for games with $\lambda \,{\neq}\, 1$.
The `weaker' species (here B) has a lower density and hence is more prone to the disordering effect of noise, i.e.\ $\sigma_t^B \,{\leq}\, \sigma_t^A$.
Moreover, all results are independent of the game strength $\tau$, since the terms involving $\tau$ in $\mu_ 1^A$ and $\mu_1^ B$ (see Eqs.~\eqref{eq:MuaMub}) cancel as the densities reach their spatially uniform steady state solutions $a^{(0)}$ and $b^{(0)}$ given by Eqs.~\eqref{eq:UnifSolDensities}. The game strength $\tau$ will, however, play an important role for the dynamics of spatially non-uniform systems.

For values of the noise strength below the respective threshold, the solution $\alpha_1 \,{=}\, \beta_1 \,{=}\, 0$ is linearly unstable and additional, stable solutions appear, given by
\begin{subequations}
\label{eq:UnifSolVelocities}
\begin{align}
    \alpha_1^{(0)} 
    = \sqrt{\frac{\mu^A_1}{\xi }} \, e^{i \theta_a} \qquad \text{for}\,\, \sigma < \sigma_t^A,
    \\
    \beta_1^{(0)}
    = \sqrt{\frac{\mu^B_1}{\xi}} \, e^{i \theta_b}\qquad \text{for} \,\,\sigma < \sigma_t^B.
\end{align}
\end{subequations}
These solutions correspond to collective states which spontaneously break the rotational symmetry of the system by choosing random directions of macroscopic order $\theta_a$ and $\theta_b$, respectively. 
Figure~\ref{fig:uniformSol}a shows the absolute values (amplitudes) $|\alpha_1^{(0)}|$ and $|\beta_1^{(0)}|$ as functions of the noise strength $\sigma$ for different values of $\lambda$.
For equally competitive species, $\lambda \,{=}\, 1$, the densities are the same for both species: $a^{(0)} \,{=}\, b^{(0)} \,{=}\, 1/2$ (see Eqs.~\eqref{eq:UnifSolDensities}).
In this case the growth rates $\mu_1^A(a^{(0)},b^{(0)})$ and $\mu_1^B(a^{(0)},b^{(0)})$ are identical for all values of the noise strength $\sigma$ (see Eqs.~\eqref{eq:MuaMub}). 
Thus, the amplitudes of the velocity fields, $|\alpha_1^{(0)}|$ and $|\beta_1^{(0)}|$, are also identical as shown in the upper panel of Fig.~\ref{fig:uniformSol}. 
When $\sigma$ is decreased below $\sigma_t^A\, \,{=}\,  \sigma_t^B$, the amplitudes of the velocity fields become non-zero, which means that both species exhibit spatially uniform macroscopic order. 
For $\lambda \,{<}\, 1$, the amplitudes $|\alpha_1^{(0)}|$ and $|\beta_1^{(0)}|$ differ. 
Since a relative game strength $\lambda \,{<}\, 1$ indicates that species $A$ is `stronger' than species $B$, its stationary density (fixed point) is larger than that of species $B$: $a^{(0)} \,{>}\, b^{(0)}$.
As a consequence it has the larger growth rate, $\mu_1^A \,{>}\, \mu_1^B$, and thus polar order of $A$ is less prone to noise resulting in a larger threshold value, $\sigma_t^A \,{>}\, \sigma_t^B$, see Fig.~\ref{fig:uniformSol}b and Fig.~\ref{fig:uniformSol}a for $\lambda \,{=}\, 0.6$ and $\lambda \,{=}\, 0.2$.
This underlines an important aspect of the game interactions. These determine the relative density of the species $A$ and $B$, which directly affects the ability of these species to establish macroscopic polar order, i.e.\ the threshold values of the noise.

In summary, for our two-species system with game interactions, the spatially uniform solutions show a new aspect of macroscopic order that is absent in one-species SPP systems~\cite{Bertin06,Bertin09,Peshkov14}. 
There is not only one transition between a polar-ordered and a disordered state at some threshold value for the noise strength but two successive transitions.
The underlying reason is that the competition of the two species, as defined by the (snowdrift) game, implies that their stationary densities are different: the species with the lower fitness also has a lower density.
As the competition between alignment, favoring polar order, and noise, favoring disorder, crucially depends on the particle density, the noise thresholds for the ordering transition of the two species are different. 
The more competitive species has a higher density and, therefore, orders at a higher noise strength than the less competitive species. 
This is summarized by the phase diagram shown in Fig.~\ref{fig:uniformSol}b. 
By varying the noise strength $\sigma$ and the relative fitness $\lambda$ there are three phases: a disordered phase above $\sigma_t^A$, an intermediate phase for $\sigma_t^A \,{>}\, \sigma \,{>}\, \sigma_t^B$ where only the more competitive species, $A$, shows polar order, and a phase where both species exhibit polar order below $\sigma_t^B$.

\subsection{\label{subsec:StabilityAnalysis}Stability against spatially non-uniform perturbations} 

In this section we will test the linear stability of the spatially uniform solutions given by Eqs.~\eqref{eq:UnifSolDensities} and Eqs.~\eqref{eq:UnifSolVelocities}, against perturbations with a finite wavelength.
To this end, we introduce spatially non-uniform perturbations around the uniform solutions 
\begin{equation}
\label{eq:inhom_pert}
\begin{split}
 a(\mathbf{r}, t) &=a^{(0)} \,{+}\, \delta a(\mathbf{r}, t) , \;  \alpha_1(\mathbf{r}, t) = \alpha_1^{(0)} \,{+}\, \delta \alpha_1(\mathbf{r}, t), 
\\
b(\mathbf{r}, t) &= b^{(0)} \,{+}\, \delta b(\mathbf{r}, t) , \; \, \, \beta_1(\mathbf{r}, t) = \beta_1^{(0)} \,{+}\,\delta \beta_1(\mathbf{r}, t) .
\end{split}
\end{equation}
Assuming that the perturbations are small, we insert the perturbed fields in the full hydrodynamic equations, Eq.~\eqref{eq:Hydro_densities} and Eq.~\eqref{eq:Hydro_velocities}, and keep only terms up to linear order in the perturbation. 
Note that one has to account for both perturbations \(\delta \alpha_1\), \(\delta \beta_1\) and its complex conjugates \(\delta \alpha_1^*\), \(\delta \beta_1^*\). 
Introducing Fourier modes $\delta a (\mathbf{r}, t)  \,{=}\, \frac{1}{2 \pi} \int_{-\infty}^{\infty} \mathrm{d}  \mathbf{q} \, \mathrm{e}^{\mathrm{i} \mathbf{q} \cdot \mathbf{r}} \delta a(\mathbf{q},t)$ and analogously for all other fields, one obtains a set of linear equations for $\delta \mathbf{m}_{\mathbf{q}}  \,{=}\, (\delta a(\mathbf{q}), \delta b(\mathbf{q}), \delta \alpha_1 (\mathbf{q}), \delta \alpha_1^*(\mathbf{q}), \delta \beta_1 (\mathbf{q}), \delta \beta_1^*(\mathbf{q}) )^T$
\begin{equation}
\label{eq:Jacobian}
    \partial_{t} \, 
    \delta \mathbf{m}_{\mathbf{q}}  
    = 
    \mathbf{J}_{\mathbf{q}} \, 
    \delta \mathbf{m}_{\mathbf{q}} 
\end{equation}
with the Jacobian $\mathbf{J}_{\mathbf{q}}$. 
It is a \( 6 {\times} 6\)  matrix that depends on the wave vector \(\mathbf{q}\) and the spatially uniform solution \((a^{(0)},b^{(0)},\alpha_1^{(0)},\beta_1^{(0)})\) and therefore on all control parameters $(\lambda,\tau,\sigma)$. 
An explicit expression for $\mathbf{J}_{\mathbf{q}}$ can be found in Appendix~\ref{app:B_Jacobian}. 
For simplification, we assume that the spatially uniform solutions of the velocity fields are parallel, \(\alpha_1^{(0)} \,{\parallel}\, \beta_1^{(0)}\).
Numerical solutions of the hydrodynamic equations and agent-based simulations in the next sections will show that this assumption is justified. 
\begin{figure}[!t]
\centering
\includegraphics[width=\linewidth]{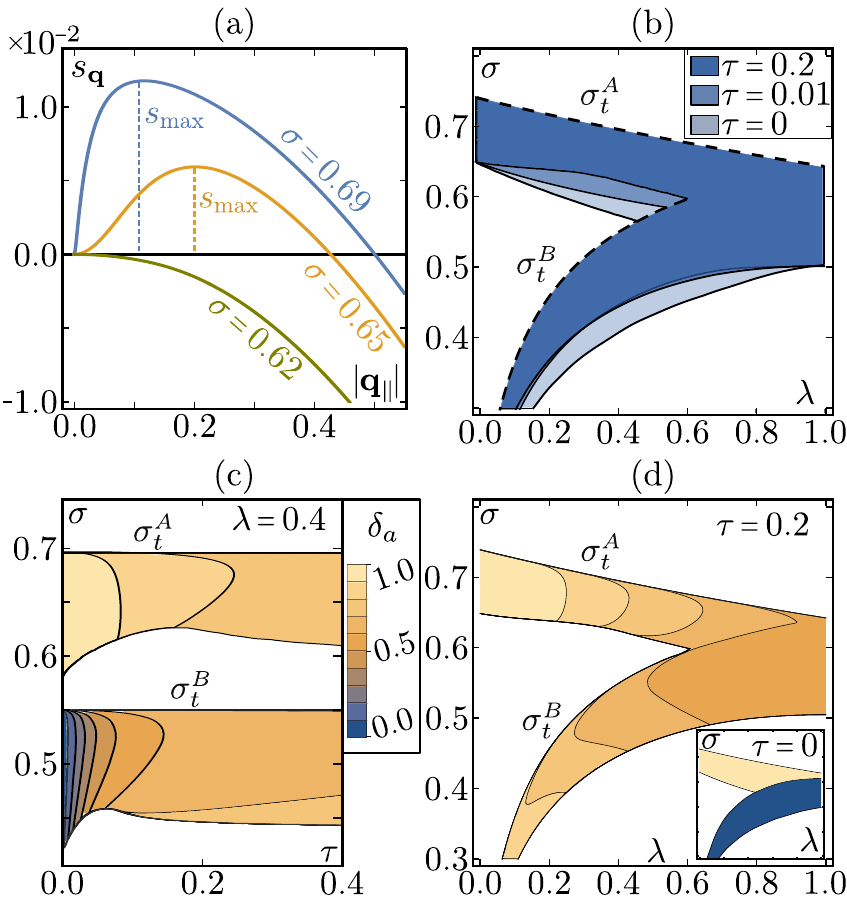}
\caption{
Stability analysis of the hydrodynamic equations Eqs.~\eqref{eq:Hydro_densities} and \eqref{eq:Hydro_velocities}.
(a) Dispersion relation of the eigenvalue with the largest real part, $s_\mathbf{q}$, for different values of the noise strength $\sigma$, and fixed game strength $\tau \,{=}\, 0.2$ and relative fitness $\lambda \,{=}\, 0.4$. $s_{\mathbf{q}}$ is plotted for wave vector $\mathbf{q}$ pointing in the direction parallel to macroscopic order, $\mathbf{q} \parallel \alpha_1, \beta_1$ (denoted by $\mathbf{q}_\parallel$). The spatially uniform state, Eqs.~\eqref{eq:UnifSolVelocities}, is unstable when the maximum of the dispersion relation, $s_\text{max}$, is larger zero. 
(b) Parameter regime (bands) of $\lambda$ (relative fitness) and $\sigma$ (noise strength) where the spatially uniform polar-ordered states are linearly unstable against spatial perturbations (i.e. $s_\text{max}>0$), for different values of the game strength $\tau$ (shades of blue) as indicated in the graph.
The two bands are bounded from above by the threshold values $\sigma_t^A$ and $\sigma_t^B$ indicating the onset of spatially uniform polar order for species $A$ and $B$, respectively (dashed lines). 
The width of the bands decreases with increasing game strength $\tau$.
White regions indicate parameter regimes where the spatially uniform ordered states are stable against spatial perturbations.
(c), (d) Linearly unstable regions in (\(\tau,\sigma\))-space and (\(\lambda,\sigma\))-space, respectively. 
The color code compares the two density field components ($\delta a$ and $\delta b$) of the eigenvector corresponding to the eigenvalue with the largest positive real part: $\delta_a \,{:=}\, \delta a /(\delta a+\delta b)$ which ranges from $0$ (the largest linear growth rate corresponds to an eigenvector with $\delta a$-component equal to zero and a non-zero $\delta b$-component) to $1$ ($\delta b$-component equal to zero and non-zero $\delta a$-component).
}
\label{fig:StabilityAnalysis}
\end{figure}

The real parts of the eigenvalues of the Jacobian $\mathbf{J}_{\mathbf{q}}$ determine the linear growth rate of the perturbations: An eigenvalue with a positive real part indicates that the corresponding uniform solution is linearly unstable and the perturbation given by the respective eigenvector of the Jacobian will grow (exponentially) in time.
Figure~\ref{fig:StabilityAnalysis}a shows examples of the real part of the eigenvalue with the largest real part, $s_\mathbf{q}$, as a function of $\mathbf{q}$ for different values of $\sigma$ and fixed $\lambda \,{=}\, 0.4$ and $\tau \,{=}\, 0.2$. 
For certain values of the noise strength $\sigma$ there are long wavelength instabilities (i.e.\ for small values of the wave vector $\mathbf{q}$) .

Our linear stability analysis shows that the spatially uniform ordered states are unstable with respect to spatial perturbations right at the onsets of both transitions to macroscopic order, $\sigma_t^A$ and $\sigma_t^B$ (Fig.~\ref{fig:StabilityAnalysis}b). 
In other words, the disordered phase ($\sigma > \sigma_t^A$) is stable, while the phases with spatially uniform polar order in species $A$ ($\sigma_t^A >\sigma > \sigma_t^B$) and uniform polar order in both species ($\sigma_t^B > \sigma$) are both unstable towards spatial perturbations for $\sigma$ close enough to the respective onset of order.
By computing the real part of the largest eigenvalue, $s_\mathbf{q}$, for different wave vectors $\mathbf{q}$, we find that $s_\mathbf{q}$ is largest for wave vectors parallel to the direction of macroscopic order ($\mathbf{q} \parallel \alpha_1^{(0)}, \beta_1^{(0)}$); i.e.\ instabilities are strongest when longitudinal.
This suggests that the ensuing patterns will form along the direction parallel to macroscopic order, $\alpha_1^{(0)}$ and $\beta_1^{(0)}$, while the system remains homogeneous along the direction perpendicular to macroscopic order. 
For $\tau \,{=}\, 0$, i.e.\ in the absence of snowdrift interactions, we recover the same unstable regions in parameter space as previously obtained from one-species SPP systems~\cite{Bertin06,Bertin09} (see Fig.~\ref{fig:StabilityAnalysis}b). 
For a finite game strength, $\tau \,{>}\, 0$, the upper boundaries of the unstable regions in parameter space are still $\sigma_t^A$ and $\sigma_t^B$ (as in the case of no snowdrift interaction). 
The width of the unstable regions, however, is smaller indicating that the snowdrift interaction has a stabilizing effect on the spatially uniform solutions.

The perturbation with the largest linear growth rate is given by the eigenvector $\delta \mathbf{m}_{\mathbf{q}}$ corresponding to the eigenvalue $s_\mathbf{q}$ with the largest positive real part (and with wave vector $\mathbf{q}$ fixed such that $s_\mathbf{q}$ is maximal); see Eq.~\eqref{eq:Jacobian}.
We compute this eigenvector as a function of the three control parameters $\sigma, \lambda$ and $\tau$ in order to see how the interplay of snowdrift game and self-propelled motion affects which perturbations have the largest growth rate.
In order to compare the growth rates of the fields of the two species, we first look at the two density field components of the eigenvector, $\delta a$ and $\delta b$.
We use the expression $\delta_a := \delta a/(\delta a+\delta b)$ to measure the direction of the eigenvector projected onto the plane spanned by the two density fields.
This value can range from $0$ ($\delta a \,{=}\, 0$ and non-zero $\delta b$-component), meaning that only perturbations in the density of species $B$ grow, to $1$ ($\delta b \,{=}\, 0$ and non-zero $\delta a$-component) meaning that only the density of species $A$ is unstable towards perturbations. 
Fig.~\ref{fig:StabilityAnalysis}c and Fig.~\ref{fig:StabilityAnalysis}d show the value of $\delta_a$ as a function of the control parameters in  ($\tau$,$\sigma$)- and ($\lambda$,$\sigma$)-space, respectively.
Analysing the results in ($\tau$,$\sigma$)-space (Fig.~\ref{fig:StabilityAnalysis}c), we can read off how the game strength, $\tau$, affects the relative growth rates of perturbations of the two densities, $a$ and $b$:
For large enough game strength, $\tau$, perturbations in the densities of both species grow equally fast ($\delta_a \rightarrow 0.5$). 
Decreasing $\tau$ changes the value of $\delta_a$ dependent on the region in parameter space:
In the unstable region with upper boundary $\sigma \,{=}\, \sigma_t^A $, perturbations in the density of species $A$ are growing faster than perturbations in the density of species $B$ such that $\delta_a \rightarrow 1$ for small enough game strength $\tau$ (see Fig.~\ref{fig:StabilityAnalysis}c). 
In other words, the perturbation with the largest growth rate changes from an even orientation in the directions of both types to an orientation only in the $a$-direction.
In the unstable region with upper boundary $\sigma \,{=}\, \sigma_t^B $, $\delta_a \,{\rightarrow}\, 0$ for small enough $\tau$, which means that only the density of species $B$ is unstable.
The inset of Fig.~\ref{fig:StabilityAnalysis}d shows the limiting case $\tau \,{=}\, 0$ in the $(\lambda,\sigma)$-plane. 
From the color code we read off that, depending on the region, either $\delta b \,{=}\, 0$ (yellow region) or $\delta a \,{=}\, 0$ (blue region). 
In other words, for two non-interacting species ($\tau \,{=}\, 0$), perturbations only grow in either species $A$ or B. 
Note that in this case, instabilities in the density fields for each species occur at the respective threshold to macroscopic order: perturbations $\delta a$ grow at $\sigma \,{<}\, \sigma_t^A $; perturbations $\delta b$ grow at $\sigma \,{<}\, \sigma_t^B$.

There are four more components to the eigenvector, encoding the perturbations of the velocity fields. An analogous analysis as above for the velocity field components parallel to macroscopic order, $\delta\alpha_1^{\parallel}$ and $\delta\beta_1^{\parallel}$, yields the same phenomenology as for the densities: 
For high values of $\tau$, perturbations in the velocity fields of both species grow equally fast ($\delta\alpha_1^{\parallel}/(\delta\alpha_1^{\parallel}+\delta\beta_1^{\parallel})\rightarrow 0.5$). 
For $\tau \rightarrow 0$, on the other hand, perturbations $\delta\alpha_1^{\parallel}$ only grow in the unstable region in the vicinity of $\sigma_t^A$, while perturbations $\delta\beta_1^{\parallel}$ only grow in the unstable region at $\sigma_t^B$. 
These results are discussed in more depth in Appendix \ref{app:B_subsec_eigenvector}.
Components perpendicular to the macroscopic order are stable to perturbations.

In summary, the main insight gained in this section is that there are regions in parameter space where spatially uniform polar ordered states are linearly unstable against spatially non-uniform perturbations. These regions are located in the immediate vicinity of the threshold values for the noise strength, $\sigma_t^A$ and $\sigma_t^B$, that mark the instability of a disordered state to states with different degrees of spatially uniform polar order. 
These threshold values of the noise strength are functions of the relative fitness, $\lambda$.
The game strength, $\tau$, affects the relative growth rates of perturbations of the two densities, $a$ and $b$, and velocities parallel to macroscopic order, $\alpha_1^\parallel$ and $\beta_1^\parallel$: 
For high values of $\tau$, perturbations in the fields of both species grow equally fast;  for decreasing $\tau$, growth rates differ between the species until for $\tau \rightarrow 0$ instabilities in the fields of $A$ and $B$ occur independently. 
In the next section we will study the spatial patterns that emerge in these unstable regions.

\subsection{\label{sec:Simulations_Hydro}Numerical solutions of the hydrodynamic equations}

The spatially uniform solutions, Eqs.~\eqref{eq:UnifSolDensities} and~\eqref{eq:UnifSolVelocities}, and their linear stability analysis (Fig.~\ref{fig:StabilityAnalysis}) suggest five distinct phases. 
Especially in the regimes where spatially uniform ordered states are unstable against spatially non-uniform perturbations (coloured regimes in Fig.~\ref{fig:StabilityAnalysis}b,c,d), we expect the formation of spatio-temporal patterns.
In order to study the nonlinear dynamics of the hydrodynamic equations, Eqs.~\eqref{eq:Hydro_densities} and~\eqref{eq:Hydro_velocities}, beyond linear stability, we solve these equations numerically.
Specifically, we use a finite difference scheme for a two-dimensional grid with periodic boundary conditions. 
We employ an explicit iterative process, the 4th order Runge-Kutta method, to determine the time evolution; for details on the implementation please see Appendix~\ref{app:C_HydroNumericSimulation}.

\subsubsection{Phase diagram and phase transitions}

\begin{figure}[tb]
\includegraphics[width=\linewidth]{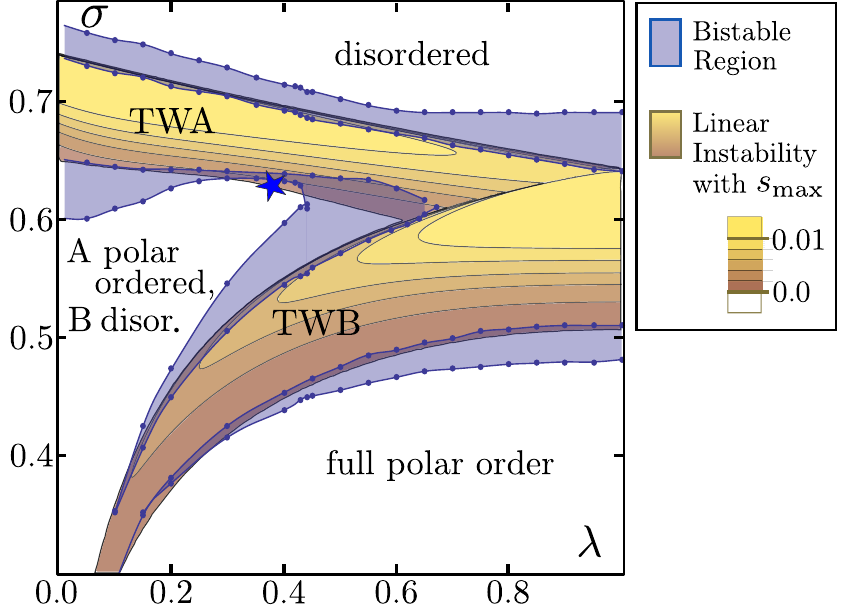}
\caption{Phase diagram in control parameter space ($\lambda, \sigma$) for game strength $\tau \,{=}\, 0.2$. Coloured regions indicate parameter regimes where one finds phases that exhibit traveling polar waves, \textit{traveling polar wave phase A} (TWA) and \textit{traveling polar wave phase B} (TWB).
White regions indicate regimes showing spatially uniform phases. Regions of linear instability are in yellow-brown (depending on the value of the maximum of the largest real part of the eigenvalue of the Jacobian, $s_{\text{max}}$). Bistable regions at the transitions between phases are coloured in blue. Blue dots are calculated numerically and indicate transitions between spatially uniform phases, bistable regions and traveling wave phases.
The blue star indicates the vicinity of a small parameter regime, where the transition between the partially polar-ordered uniform phase and the TWA phase becomes supercritical.
\label{fig:PhaseDiagram}}
\end{figure}

These numerical solutions of the hydrodynamic equations confirm the existence of five distinct phases (Fig.~\ref{fig:PhaseDiagram}).
We find three spatially uniform phases, a disordered phase, an intermediate phase where species $A$ shows polar order while species $B$ does not, and a phase where both species exhibit polar order. 
The parameter regimes determined numerically are in accordance with our analytical calculations (described in sections \ref{sec:uniform_Sol} and \ref{subsec:StabilityAnalysis}), where we found uniform solutions to be \textit{linearly} stable against non-uniform perturbations (the three white regions in Fig.~\ref{fig:StabilityAnalysis}b,c,d).

In the parameter regime where the spatially uniform solutions are linearly unstable against spatial perturbations, we observe the formation of spatial patterns. 
These patterns are characterized by density-segregated, polar-ordered bands in both species that concurrently move through the system; see Movie \textbf{movie1} in the supplementary material for an illustration (an explanation and the parameter set for the movie can be found in Appendix~\ref{app:E_Movies}).
We will refer to these traveling bands as \textit{polar waves}. 
The wave vector of the traveling patterns always points in the same direction as the macroscopic polar order, as suggested by linear stability analysis (see Section~\ref{subsec:StabilityAnalysis}).
When viewed from a co-moving frame, the patterns are uniform in the direction perpendicular to the wave vector.
Similar density-segregated traveling wave states are known from one-species SPP systems~\cite{Gregoire04, Bertin09, Caussin14}. 
Interestingly, the domains where traveling wave patterns exist, extends beyond the parameter regime where the spatially uniform polar-ordered states are unstable to spatial perturbations of finite wavelength. 
In the blue shaded regime in Fig.~\ref{fig:PhaseDiagram} the system exhibits bistabilty and subcritcal behavior as discussed in detail below; see section \ref{subsubsec:Bistability_nnd_subcriticality}. We have also investigated phase diagrams for different values of $\tau$ and found the same topology.

\begin{figure}[tb]
\includegraphics[width=\linewidth]{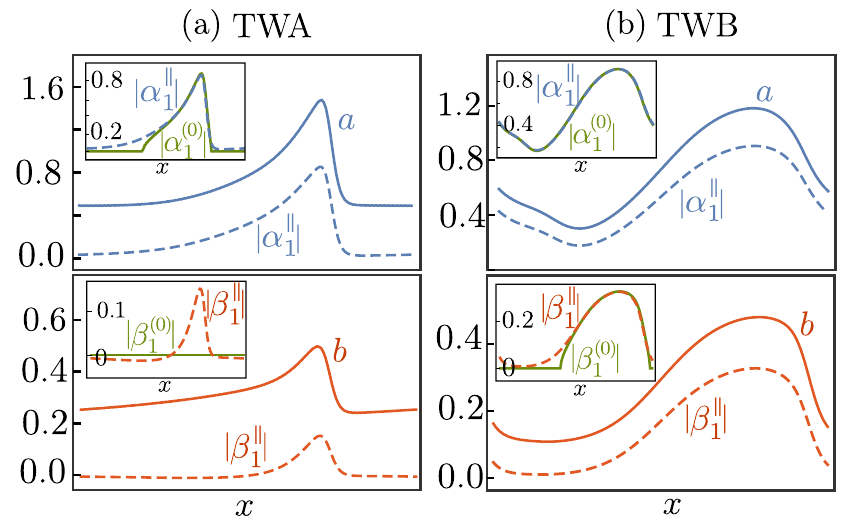}
\caption{
Typical wave profiles for the density fields (solid lines), $a$ and $b$, and the velocity fields (dashed lines),  \(|\alpha_1^{\parallel}|\) and \(|\beta_1^{\parallel}|\), in the traveling wave phases TWA (a) and TWB (b). 
Control parameters are set to relative fitness $\lambda \,{=}\, 0.4$, game strength $\tau \,{=}\, 0.2$, and noise strength $\sigma \,{=}\, 0.67$ in (a), and $\sigma \,{=}\, 0.47$ in (b). 
The insets show how the density and velocity fields are related to each other: 
At every point in space, the velocity fields $\alpha_1$ and $\beta_1$ (dashed lines in the insets) are approximately equal to the values $|\alpha_1(\mathbf{r},t)| \,{\approx}\,  |\alpha_1^{(0)}(\mathbf{r},t)| \,{=}\, \sqrt{\mu_1^A(a,b)/\xi(a)}$ and $|\beta_1(\mathbf{r},t)| \,{\approx}\,  |\beta_1^{(0)}(\mathbf{r},t)| \,{=}\, \sqrt{\mu_1^B(a,b)/\xi(b)}$ corresponding to spatially uniform solutions (see Eqs.~\eqref{eq:UnifSolVelocities}) for the respective local densities $a({\bf r},t)$ and $b({\bf r},t)$ (green solid lines in the insets). 
The waves in species $B$ in the TWA phase seem to be an exception to this observation (inset in the bottom left figure).
}
\label{fig:WaveProfiles1}
\end{figure}

We will see in the next section that the mechanism underlying pattern formation varies according to the region in the parameter space. 
A distinction is therefore made between two different phases with spatial patterns.
We refer to the traveling polar wave state emerging in the unstable region in parameter space at $\sigma_t^A$ as \textit{traveling polar wave phase A} (TWA) and the one in the unstable region at $\sigma_t^B$ as \textit{traveling polar wave phase B} (TWB).
For both of these phases, examples of the wave profiles for the density fields, $a$ and $b$, and the velocity fields, $|\alpha_1^{\parallel}|$ and $|\beta_1^{\parallel}|$, are shown in Fig.~\ref{fig:WaveProfiles1}; the superscript for the velocity fields indicates that we consider the component parallel to the direction of macroscopic order, $\alpha_1^{(0)}$and $\beta_1^{(0)}$, and wave vector, $\mathbf{q}$ (as mentioned in section \ref{subsec:StabilityAnalysis}, the stability analysis showed that instabilities are largest for $\mathbf{q} \parallel \alpha_1^{(0)}, \beta_1^{(0)}$, meaning that patterns form along the direction of macroscopic order). 
Our numerical solutions show that the polar waves for species $A$ and $B$ are strongly correlated spatially, meaning that they follow each other through the system.  
These findings are consistent with our results from linear stability analysis, where we found that instabilities always occur simultaneously for interacting species $A$ and $B$.

In the three spatially uniform phases, all fields show values that are close to their uniform fixed point values given by Eqs.~\eqref{eq:UnifSolDensities} and~\eqref{eq:UnifSolVelocities}.
Although in the TWA and TWB phase, the fields exhibit spatial patterns and deviate significantly from their uniform fixed point values at different points in space, the spatial averages $\bar{a},\bar{b},|\bar{\alpha}_1|$ and $|\bar{\beta}_1|$ are approximately also given by the corresponding uniform fixed point values.

Moreover, we find that at every point in space, the velocity fields, $|\alpha_1(\mathbf{r},t)|$ and $|\beta_1(\mathbf{r},t)|$,
are well approximated by the local values of the spatially uniform solutions, Eq.~\eqref{eq:UnifSolVelocities}. 
For local densities $a(\mathbf{r},t)$ and $b(\mathbf{r},t)$ with $\mu_1^{A/B} \,{>}\, 0$, these are given by  
$|\alpha_1^{(0)}(\mathbf{r},t)| \,{=}\, \sqrt{\mu_1^A(a,b)/\xi(a)}$ and $ |\beta_1^{(0)}(\mathbf{r},t)| \,{=}\, \sqrt{\mu_1^B(a,b)/\xi(b)}$, else we have $|\alpha_1| , |\beta_1| \,{=}\, 0$.
Hence, non-zero values for the velocity fields $\alpha_1$ and $\beta_1$ requires that the densities are large enough such that locally one is above the threshold for polar order, i.e.\ both $\mu_1^A(a,b)$ and $\mu_1^B(a,b)$ are positive (see section \ref{sec:uniform_Sol}).
The only exception are patterns of the $B$ species in the TWA phase in a regime where both $\lambda$ and $\tau$ are small enough such that the density $b$ falls below the threshold value for polar order, $\mu_1^B(a,b)$. 
Nevertheless we see order emerge in the high-density wave peak (see lower inset of Fig.~\ref{fig:WaveProfiles1}a). 
In summary, the pattern can segregate the system in regions of high density with macroscopic order (\(|\alpha_1| >0\) or $|\beta_1|>0$) and low density regions that are disordered ((\(|\alpha_1|  \,{=}\, 0\) or $|\beta_1| \,{=}\, 0$)).

\subsubsection{\label{subsubsec:Bistability_nnd_subcriticality}Bistability and subcriticality}

Previous studies on SPP systems indicate that patterns are stable even in regimes where uniform solutions are linearly stable~\cite{Thueroff14,Peshkov14}.
To identify the full regime in which pattern solutions exist, we use the following approach (`hysteresis loop'): 
We first compute the stationary solution for fixed parameters \((\lambda,\tau)\) and $\sigma > \sigma_t^A$ where there is no macroscopic order, neither in $A$ nor in $B$. 
Then, we reduce the noise strength \(\sigma\) quasi-statically in small steps, thereby sweeping through the different phases in parameter space. 
By `quasi-static' we mean that the system is given enough time to equilibrate between successive adjustments in \(\sigma\).
Afterwards, we increase the noise strength again quasi-statically back to its initial value. 
During this $\sigma$-sweep, we monitor the spatial average of the velocity fields, $|\bar{\alpha}_1|$ and $ |\bar{\beta}_1|$.
To identify the presence of spatial patterns, we use parameters that we call \textit{variation parameters} $\eta_A$ and $\eta_B$, which measure spatial variations in the densities $a$ and $b$. 
They are defined as
\begin{equation}
    \eta_A 
    := 
    \overline{\left(\frac{|\Delta a|}{\Delta x} \right)}
    , 
    \quad 
    \text{and} 
    \quad 
    \eta_B 
    := 
    \overline{\left(\frac{|\Delta b|}{\Delta x} \right)},
\end{equation}
where $\frac{|\Delta a|}{\Delta x}$ is the absolute value of the numerical spatial derivative of the density field $a$ and the bar indicates the average over all grid points (see App.~\ref{app:C_HydroNumericSimulation}). 
For $\eta_A \,{=}\, 0$ ($\eta_B \,{=}\, 0$), species $A$ ($B$) is in a spatially uniform state, while a non-zero value $\eta_A \,{>}\, 0$ ($\eta_B \,{>}\, 0$) indicates that the system exhibits some kind of spatial patterns in $A$ ($B$). 
By combining the information about average velocities and spatial variations we can infer the phase of the system, as we discuss next. 

\begin{figure}[!t]
\includegraphics[width=\linewidth]{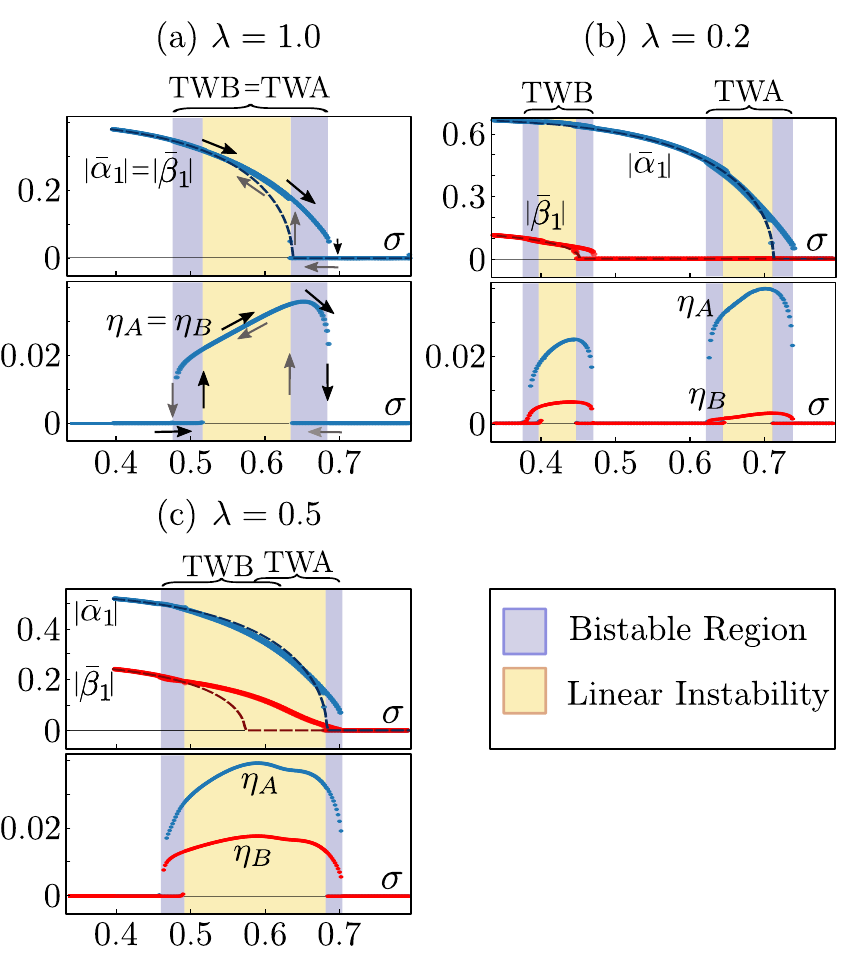}
\caption{
Bifurcations of the spatial average of the velocity fields, $|\bar{\alpha}_1|$ and $|\bar{\beta}_1|$, and the spatial variation parameters, $\eta_A$ and $\eta_B$, as a function of the noise strength \(\sigma\) for a fixed game strength \(\tau \,{=}\, 0.2\) and for different values of the relative fitness \(\lambda \): (a) $\lambda \,{=}\, 1$, (b) $\lambda \,{=}\, 0.2$, and (c) $\lambda \,{=}\, 0.5$. 
Blue and red lines consist of sets of dots that each indicate a stable state calculated numerically. 
The dashed lines indicate the analytical results for the spatially uniform amplitudes of the velocity fields, Eq.~\eqref{eq:UnifSolVelocities}; compare with Fig.~\ref{fig:uniformSol}. 
Regions shaded in blue and yellow indicate parameter regimes, in which the system is bistable or the spatially uniform states are linearly unstable with respect to spatially heterogeneous perturbations, respectively.
}
\label{fig:Bifurcation_Vel_Var1}
\end{figure}

Figure~\ref{fig:Bifurcation_Vel_Var1} shows the spatial variation parameters, $\eta_A$ and $\eta_B$, and spatial averages of the velocity fields, $|\bar{\alpha}_1|$ and $|\bar{\beta}_1|$, obtained from $\sigma$-sweeps for different values of the relative fitness $\lambda$ and a fixed game strength $\tau \,{=}\, 0.2$. 
A general observation from our numerical analysis is that phases with traveling polar waves exist in parameter regimes wider than the domains predicted by the linear stability analysis of spatially uniform states; compare the yellow- and blue-shaded regimes in Fig.~\ref{fig:Bifurcation_Vel_Var1}. 
While in the yellow-shaded parameter regime only the traveling waves are stable solutions, in the blue-shaded parameter regime both these waves and spatially uniform states are stable solutions, i.e.\ the system shows a bistable dynamics and subcritical behavior.

For a neutral game ($\lambda \,{=}\, 1$), species $A$ and $B$ show identical behavior (Fig.~\ref{fig:Bifurcation_Vel_Var1}a): Quasi-statically decreasing (grey arrows) and then increasing (black arrows) the noise strength $\sigma$ one observes a hysteresis loop.
Both, the variation parameters, $\eta_A$ and $\eta_B$, and the average velocities, $|\bar{\alpha}_1|$ and $|\bar{\beta}_1|$, show discontinuous changes (up and down arrows) at the boundaries of the bistable regime at high noise values (right blue-shaded region). 
Hence, in the bistable regime, the disordered phase and the traveling wave phase (TWA) are possible metastable states.
In contrast, in the bistable regime at low noise strength (left blue-shaded region) only the spatial variation parameter exhibits a hysteresis loop while the average velocities change continuously. 
Moreover, the spatial averages of the numerically determined values for the velocities closely follow the results calculated analytically for a spatially uniform system (dashed lines in Fig.~\ref{fig:Bifurcation_Vel_Var1}a)); compare Eq.~\eqref{eq:UnifSolVelocities}.
This has two implications. First, a spatially uniform polar-ordered state and a traveling wave state are metastable solutions (in the left bistable regime). Second, the spatially uniform and spatially non-uniform state have the same average velocities. We therefore call this transition quasi-continuous.

For small enough values of the relative fitness $\lambda$ ($\lambda \,{=}\, 0.2$ in Fig.~\ref{fig:Bifurcation_Vel_Var1}b), one observes all of the five different phases (compare also Fig.~\ref{fig:PhaseDiagram}) with bistable regions as well as hysteresis loops at each transition.
In addition to the disordered phase at high noise strength and the spatially uniform polar-ordered phase at low noise strength, in which both species show polar order ($|\alpha_1|, |\beta_1|>0$), there is now an intermediate phase, in which only species $A$ shows polar order, but species $B$ does not ($|\alpha_1|>0, |\beta_1|=0$). 
Similar as for a neutral game, the spatial mean of both velocities is fairly well approximated by the analytical results we obtained for a spatially uniform system (dashed lines in Fig.~\ref{fig:Bifurcation_Vel_Var1}b), even in the regime where the system actually shows phase separation into polar bands of high density and disordered spatial regions of low density.

For large enough values of the relative fitness $\lambda$ ($\lambda \,{=}\, 0.5$ in Fig.~\ref{fig:Bifurcation_Vel_Var1}c), the two traveling polar wave phases, TWA and TWB, merge; compare also Fig.~\ref{fig:PhaseDiagram}.
The phase behavior is similar as in the neutral case with the difference, that the travelling polar waves in species $A$ and $B$ differ in shape and degree of polarization (see the different values of $\eta_A$, $\eta_B$ and $|\alpha_1|$, $|\beta_1|$ in Fig.~\ref{fig:Bifurcation_Vel_Var1}c).

Last, in a small parameter regime (close vicinity around the blue star in Fig.~\ref{fig:PhaseDiagram}), the transition between the uniform partially polar-ordered phase and the TWA phase becomes \textit{supercritical}, i.e.\ we do not find jumps in the variation parameters and average velocities, but a continuous change between zero and non-zero values. 
Moreover, in this regime the dynamics become extremely slow, so that the system takes a long time to reach its spatially non-uniform, stationary state, indicating critical slowing down.

\subsubsection{\label{subsec:GameInducedPatternFormation}Game-induced pattern formation }

In the last section we found that the two traveling wave phases, TWA and TWB, exhibit coupled traveling polar wave patterns.
In a next step, we now investigate the relationship between the interaction mediated by the snowdrift game and the formation and coupling of polar wave patterns. 
This will shed light on the differences between the TWA and TWB phases and on the underlying mechanism of pattern formation. 

The linear stability analysis performed in Section~\ref{subsec:StabilityAnalysis} already indicated a possible role of the game interaction in pattern formation.
There we found that for finite game strength ($\tau \,{>}\, 0$) spatial perturbations in the density and velocity fields of species $A$ and $B$ grow simultaneously, while in the absence of game interactions ($\tau \,{=}\, 0$) instabilities in the fields of $A$ and $B$ occur independently. 
This suggests that the games interaction is responsible for the coupling between the polar wave patterns of species $A$ and $B$.

\begin{figure}[t]
\centering
\includegraphics[width=\linewidth]{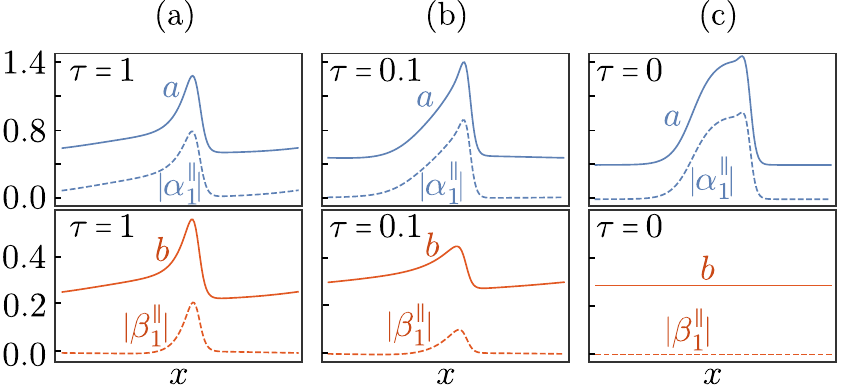}
\caption{
Wave profiles for noise strength $\sigma \,{=}\, 0.68$, relative fitness $\lambda \,{=}\, 0.4$ and different game strengths $\tau$ as indicated in the graphs. The system is in the TWA phase. 
(a) For $\tau \,{=}\, 1$, species $A$ and species $B$ exhibit polar waves in the density fields,  $a$ and $b$, with polar-ordered wave peaks (non-zero velocity fields $|\alpha_1^{\parallel}|$ and $|\beta_1^{\parallel}|$).
(b) For $\tau \,{=}\, 0.1$, we observe an increase of wave amplitudes in $a$ and $|\alpha_1^{\parallel}|$ as well as a decrease in wave amplitudes in $b$ and $|\beta_1^{\parallel}|$. This effect is stronger the smaller the value of the game strength $\tau$ is.
(c) For $\tau \,{=}\, 0$, the polar wave pattern in $B$ has completely disappeared and species $B$ is in a disordered, spatially uniform state.
}
\label{fig:waveProfiles}
\end{figure}

To pinpoint this idea, we systematically study the formation of polar wave patterns as a function of the game strength $\tau$. 
Figure~\ref{fig:waveProfiles} shows profiles of the densities, $a$ and $b$, and the velocity fields, $|\alpha_1^{\parallel}|$ and $|\beta_1^{\parallel}|$, for different values of the game strength $\tau$. 
The other two control parameters are set to \(\sigma \,{=}\, 0.68\) and \(\lambda \,{=}\, 0.4\), such that the system is in the TWA phase.  
For $\tau \,{=}\, 1$, traveling polar waves emerge in both species, as expected (Fig.~\ref{fig:waveProfiles}a).
For smaller values of the game strength $\tau$, we observe a decrease of wave amplitudes in \(b\) and \(|\beta_1^{\parallel}|\) but an increase of wave amplitudes in \(a\) and \(|\alpha_1^{\parallel}|\) (Fig.~\ref{fig:waveProfiles}b).
In general, we find that the smaller the value of \(\tau\), the lower the amplitude of waves in species $B$ and the higher the amplitudes in species $A$. 
In the absence of a snowdrift interaction ($\tau \,{=}\, 0$), the wave pattern in species $B$ completely disappears and species $B$ is in a disordered, spatially uniform state with \(b  \,{=}\, \frac{\lambda }{1+\lambda}  \,{\approx}\, 0.29\) and \(|\beta_1^{\parallel}| \,{=}\, 0\) throughout the system (Fig.~\ref{fig:waveProfiles}c). 
We conclude that the pattern in $B$ observed for $\tau  \,{>}\, 0$ in the TWA phase is due to the snowdrift interaction.
This behaviour is consistent with our analysis of the direction of the linear instability as a function of $\tau$ in Section~\ref{subsec:StabilityAnalysis} (see Fig.~\ref{fig:StabilityAnalysis}c). 
There we found that the $b$- and $\beta_1^{\parallel}$-components of the eigenvector pointing in the direction of the instability gradually decreased with the game strength $\tau$, such that the $a$- and $\alpha_1^{\parallel}$-components remained as the only non-zero contributions for $\tau \rightarrow 0$.

Our interpretation of these observations is the following. 
The alignment interactions between the particles of species $A$ cause polar wave patterns in the TWA phase.
This is consistent with previous studies on SPP systems, that find traveling polar waves at the onset of ordered motion.

For a finite game strength, $\tau  \,{>}\, 0$, the game interaction induces a shift of the density of the $B$ species towards the value $b  \,{=}\, \lambda \, a$, given by the fixed point of the snowdrift game (Eq.~\eqref{eq:UnifSolDensities}).
The magnitude of this shift (flow in the phase space of the game) depends the game strength $\tau$. 
The stronger the game interaction, the stronger the mutual influence of the two species. 
For large $\tau$, one expects an adiabatic limit where $a$ and $b$ are strictly correlated by $b/a  \,{=}\, \lambda$; compare Figs.~\ref{fig:waveProfiles}a-c.
We call this mechanism \textit{game-induced pattern formation}. 

There is an analogue mechanism in the TWB phase. 
Our simulations show that with decreasing game strength $\tau$, the pattern amplitudes in \(a\) and \(|\alpha_1^{\parallel}|\) decrease while they increase in \(b\) and \(|\beta_1^{\parallel}|\).
For vanishing game strength ($\tau \rightarrow 0$), patterns in species $A$ fully disappear and one is left with spatially uniform polar order in species $A$ and polar wave patterns in species $B$.
Hence, in the TWB phase, species $B$ induces the traveling wave pattern in species $A$. 

In parameter regions where the TWA and TWB phases overlap, both species form polar waves independently, due to alignment interactions. 
Game interactions cause synchronization of these wave patterns.
Interestingly, when we tune the game strength $\tau$ to very small values, in a parameter regime where the densities of both species are high enough to exhibit pattern formation, we observe ``oscillating'' solutions where a faster moving traveling wave periodically passes a slower wave.
Our explanation for this phenomenon is that the active properties responsible for the wave formation dictate different wave speeds for both species and the game strength is too weak to synchronize both patterns. 

\newpage 

\section{\label{sec:Agent-based}Agent-based simulation}

As a complementary approach and to test the robustness of the results obtained with the kinetic Boltzmann approach, we implemented agent-based simulations of a system of two species that exhibit intra-species alignment and inter-species interactions mediated by the snowdrift game.
For a single species, previous agent-based simulations showed that aligning collisions can lead to the formation of macroscopic order when the density of particles is high enough or orientational noise is weak~\cite{Vicsek95, Chate08, Bertin09}. 
In addition, at the onset of macroscopic order, particles were observed to form polar-ordered high-density waves rather than spatially uniform polar order \cite{Chate08, Gregoire04}.

\subsection{Agent-based model}

Following Bertin et al.~\cite{Bertin09}, we use a generalized Vicsek model~\cite{Vicsek95} for self-propelled particles and extend it to two species, which interact through a snowdrift game.
We consider a system of $N$ particles on a two-dimensional square plane of area $L \,{\times}\, L$ with periodic boundary conditions.
Each particle, located at position $\mathbf{r}_i$, is either of species A or B, and moves with a velocity $\mathbf{v}_i \,{=}\, v_0\, \mathbf{e}_{\theta}^{i}$; the speed $v_0$ is assumed to be constant and the direction vector reads $\mathbf{e}_{\theta}^{i} \,{=}\, (\cos \theta_i , \sin \theta_i)$, where $\theta_i$ denotes the angle enclosed with the $x$-axis. 

The positions of all particles are updated in discrete time steps $dt$  
\begin{equation}
\label{eq:PosUpdateAgentBased}
    \mathbf{r}_{j}^{t+dt} 
    = 
    \mathbf{r}_{j}^{t} + 
    \mathbf{v}_{j}^{t} \,dt. 
\end{equation}
Particles move ballistically unless they change their direction $\mathbf{e}_{\theta}^{i}$ in random `diffusion' or `collision' events.
A `collision' event is defined in the spirit of a Vicsek model~\cite{Vicsek95}:
Each particle aligns its direction to the average direction of all particles of the same species
in its \textit{alignment vicinity} $\mathcal{A}_{j}^{t}$ at time $t$. 
$\mathcal{A}_{j}^{t}$ is defined as the disk around $\mathbf{r}_{j}^{t}$ with radius $d_{\text{align}}$, the alignment range, corresponding to $d_0$ in the kinetic model.
Independent of alignment, each particle is subject to `diffusion' events, in which its direction changes by a random amount $\eta$. We assume that $\eta$ is a random variable that is uncorrelated in time and space and given by a flat distribution over the interval $(-h\pi,h\pi]$ with $h  \,{\in}\,  [0,1]$. 
The standard deviation $\sigma$ of a flat distribution is then given by $\sigma \,{=}\, h \pi/\sqrt{3}$. 
We will refer to $\sigma$ as the noise strength.
The direction of each particle at time $t \,{+}\, dt$ is therefore given by
\begin{equation} 
\label{eq:DirUpdateAgentBased}
	\theta_{j}^{t+dt} 
	= 
	\textrm{arg}
	\biggl( 
	\sum_{k\in \mathcal{A}_{j}^{t}}^{} 
	e^{i\theta_{k}{t}} 
	\biggr) 
    + \eta^{t}_{j}
	\, .
\end{equation}

Previous studies have shown that if the noise strength $\sigma$ is sufficiently low, species begin to form macroscopic polar order, i.e.\ particles move collectively over distances much larger than their alignment range $d_{\text{align}}$~\cite{Vicsek95, Chate08}
In order to measure polar order of each species we define the polar order parameters
\begin{equation}
	\alpha_{1}^t
	=
	\frac{1}{N_{A}^{t}} 
	\Bigl| \sum_{k=1}^{N_{A}^{t}} e^{i\theta_{k}^{t}} \Bigr|,~ \; \text{and} \quad
	\beta_{1}^t
	=
	\frac{1}{N_{B}^{t}} 
	\Bigl| \sum_{k=1}^{N_{B}^{t}} e^{i\theta_{k}^{t}} \Bigr|,
\end{equation}
where $N_{A}^{t}$ and $N_{B}^{t}$ denote the total particle number of species A and B, respectively, at time $t$;
$\alpha_{1}^t$ and $\beta_{1}^t$ range from $0$ (no order) to $1$ (perfect alignment of all particles).
The temporal averages of these polar order parameters are given by
\begin{equation}
	\langle \alpha_1 \rangle = \frac{1}{T} \sum\limits_{t=1}^{T}\alpha_{1}^t,~ \; \text{and} \quad
	\langle \beta_1 \rangle = \frac{1}{T} \sum\limits_{t=1}^{T}\beta_{1}^t.
\end{equation}

In addition to intra-species alignment interactions, particles also play an inter-species snowdrift game each time they `encounter' each other.
Analogously to the alignment vicinity, we define a \textit{snowdrift game vicinity} $\mathcal{G}_{j}^{t}$ of particle $j$ at time $t$ by a disk around particle $j$ with radius $d_{\text{game}}$, which we will refer to in the following as the \textit{game range}.

All particles $N_j^t$ within this range are assumed to participate in the game, similar as in an urn model \cite{Frey.2010}.
The corresponding particle numbers for species A and B are denoted by $N_{A,j}^{t}$ and $N_{B,j}^{t}$, respectively. 
Consider the case that particle $j$ belongs to species A:
This A-particle will be at a disadvantage in the competition with the B-particles if the relative abundance, $N_{B,j}^{t}/N_{A,j}^{t}$, lies below a threshold $\lambda$, which defines their relative fitness. 
We assume, in analogy to the replicator equations Eqs.~\eqref{eq:ReplicatorEquations}, that a disadvantaged particle will switch species, i.e.\ change strategy (here $A \,{\to}\, B$) with a rate $\tau ( a_{j}^t \lambda \,{-}\, b_{j}^t )$. Here, 
$a_{j}^t \,{=}\, N^{t}_{A,j} / N_j^t$ and 
$b_{j}^t \,{=}\, N^{t}_{B,j} / N_j^t$ denote the particle densities of species A and B, respectively (within the game range of particle $j$).
On the other hand, if a particle is at an advantage in the game it plays, i.e.~the relative abundance is greater than the threshold $\lambda$, we assume that it keeps its strategy. An analogous argument holds if $j$ happens to a B-particle.

Taken together, we define the probability for switching species as 
\begin{equation} 
\label{eq:SwitchProb}
    p_{j}^t= 
	\begin{cases}
	\tau(a_{j}^t \lambda {-} b_{j}^t) dt 
	& \textrm{if $j {\in} \text{A}$ 
	and $\tau(a_{j}^t \lambda {-} b_{j}^t)>0$}  \\
	\tau(b_{j}^t {-} a_{j}^t \lambda)dt 
	& \textrm{if $j {\in} \text{B}$ 
	and $\tau(a_{j}^t \lambda {-} b_{j}^t)<0$}\\
	\hfil 0 & \textrm{otherwise.} 
	\end{cases}
\end{equation}
As in Sec.~\ref{sec:Model}, $\tau \,{\geq}\, 0$ is interpreted as the game strength: With increasing $\tau$ the strength of the snowdrift interaction compared to the alignment interaction increases. 
Because the species are antisymmetric under relabeling A and B, we can restrict the relative fitness to $\lambda  \,{\in}\,  [0,1]$, i.e.\ species A is the `fitter' species.
We assume that particles which change species lose their directional information and move in a random direction afterwards until they align with partners of their new species. 
This assumption, which we refer to as \textit{`switch randomization'}, prevents transmission of directional information between species, such that all inter-species interactions are purely due to the snowdrift game. 

To be able to identify spatial structure as observed in previous agent-based SPP models~\cite{Bertin09,Chate08,Vicsek95} we divide the system into $C$ channels perpendicular to the direction of macroscopic order.
Each channel has area $L^2/C$.
We then calculate the densities $a_{c}^t$ and $b_{c}^t$ and polar order parameters $\alpha_{1,c}^t$ and $\beta_{1,c}^t$ for each channel $c$.
The same partitioning is made parallel to the direction of macroscopic order.
If waves are present in the system, we expect a difference in the magnitude of the order parameter for different channels, whereas for an uniformly ordered phase the order parameters for each of the channels should be almost identical.
We measure this difference with the velocity variation parameter $\nu_{A}^t$, defined as twice the standard deviation of polar order parameter $\alpha_{1,c}^t$ over all channels: 
\begin{equation}
    \label{eq:velocity_variation}
	\nu_{A}^t=2\sqrt{\frac{1}{C}\sum_{c=1}^{C} \left( \alpha_{1,c}^t -  \langle \alpha_{1}^t \rangle \right)^2}.
\end{equation} 
The factor of $2$ is introduced such that $\nu_{A}^t  \,{\in}\,  [0,1]$. 
The velocity variation $\nu_{B}^t$ for species B is defined analogously.
To see if waves are present over an extended period of time, we use the differences between the time averages $\Delta \nu_A \,{=}\, \langle \nu_{A,\perp} \rangle \,{-}\, \langle \nu_{A,\parallel} \rangle$, with channels perpendicular and parallel to the direction of macroscopic order, respectively.
Further details can be found in Appendix~\ref{app:D_Agent-Based}.

\subsection{\label{sec:AgentBased_results}Results of the agent-based simulation}

Of the eight model parameters ($\lambda$, $\tau$, $\rho$, $\sigma$, $v_0$, $d_{\text{align}}$, $d_{\text{game}}$, and $L$) we focus mainly on analyzing the role of two: (i)
the noise strength $\sigma$, and (ii) the relative fitness $\lambda$.
The noise strength $\sigma$ is used as a control parameter to distinguish between the different phases, with the density kept fixed at $\rho \,{=}\, N/L^2 \,{=}\, 1.5$.
Varying the relative fitness $\lambda$ tunes the ratio of species B to A.
The remaining parameter values are listed in  Appendix~\ref{app:D_Agent-Based}.
We have tested that the same phenomenology found for these specific parameter values is also observed in a broader parameter range, similar to the kinetic Boltzmann approach~\cite{MichaelMasterthesis}. 

\begin{figure}[bt]
\centering
\includegraphics[width=\linewidth]{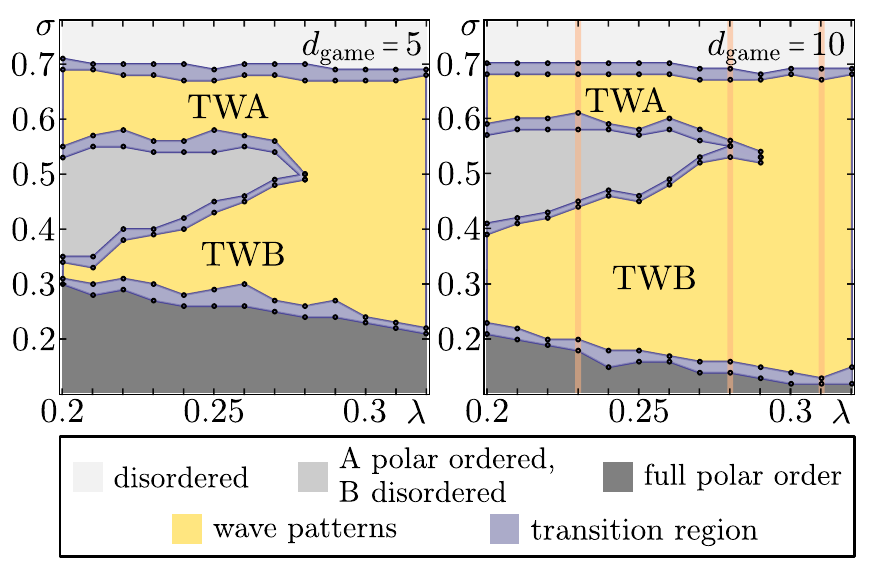}
\caption{Regimes of different macroscopic order and spatio-temporal patterns obtained from agent-based simulations for game strength $\tau \,{=}\, 0.4$ and two distinct values of the game theory interaction radius $d_{\text{game}}$ as indicated in the graph.
Yellow areas mark parameter regimes with traveling polar wave patterns, where TWA and TWB mark the regions where species A and B, respectively, promote wave formation.
These two phases merge with increasing values of the relative fitness $\lambda$.
Grey shaded areas indicate parameter regimes where the system is in either a disordered or uniformly polar-ordered state, as indicated in the figure legend. 
For all transitions between phases there are transition regions in which both phases --- a uniform phase (either ordered or disordered, depending on the transition) and a polar wave pattern ---  can dynamically change (light blue regions). 
Note that, as explained at the end of Section~\ref{sec:AgentBased_results}, due to demixing of species inside the waves, the relative size of the parameter regime covered by the TWA and merged wave phase widens with increasing $d_{\text{game}}$.
The vertical orange lines in the right panel indicate the three slices in parameter space displayed in Fig.~\ref{fig:AgentBasedPhaseDiag_OPs}.
}
\label{fig:AgentBasedPhaseDiag}
\end{figure}

In our agent-based simulations for the `active matter game' we observe a similar phenomenology as in the kinetic Boltzmann approach: 
Depending on the parameters, especially the game strength $\tau$, relative fitness $\lambda$, and noise strength $\sigma$, we observe up to five different phases. 
For neutral games ($\lambda \,{=}\, 1$), we observe, similar to previous studies~\cite{Bertin09, Chate08, Vicsek95}, three different phases: a disordered, a traveling polar wave, and a uniformly polar-ordered phase.
Figure~\ref{fig:AgentBasedPhaseDiag} shows the diagram for a game strength $\tau \,{=}\, 0.4$, and two distinct game ranges $d_{\text{game}}  \,{\in}\,  \{5,10\}$.
The topology of the diagram is similar to the one obtained from the hydrodynamic equations (Fig.~\ref{fig:PhaseDiagram}), and it is independent of the specific value for the game range \footnote{This is the case as long as the game range is not chosen too small, i.e.~that the average number of particles participating in an interaction is insufficient to roughly achieve the desired ratio given by $\lambda$.}. In the following discussion, the specific parameter values given refer to the case $d_{\text{game}} \,{=}\, 10$.

\begin{figure}[tb]
\centering
\includegraphics[width=\linewidth]{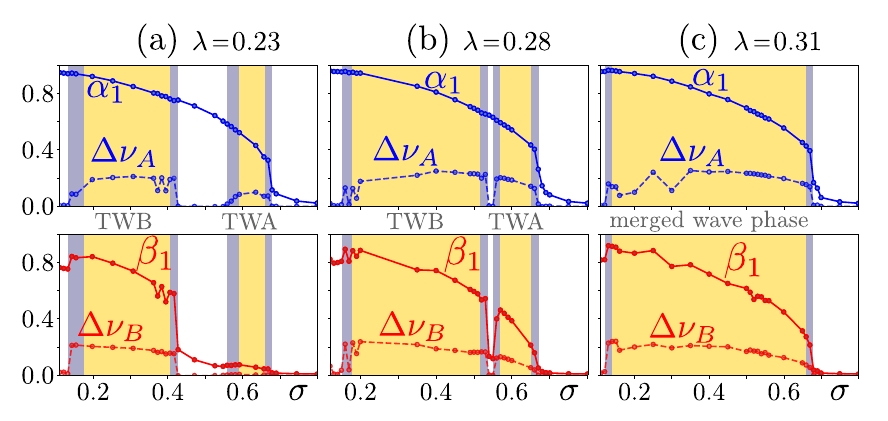}
\caption{
Polar order (solid line) and velocity variation (dashed line) against noise strength $\sigma$ for species A (top row) and species B (bottom row) and a set of different values for the relative fitness $\lambda$ (a-c), indicated in the graph.
There are parameter regimes with spatially uniform states (white regions, corresponding to grey shaded areas in Fig. \ref{fig:AgentBasedPhaseDiag}), wave patterns (yellow shaded region), and bistable behavior where both wave patterns and uniform (dis-)order (blue shaded region) stochastically alternate.
We observe three qualitatively different regimes.
For $\lambda \,{=}\, 0.23$ (a) there are distinct TWA and TWB phases, 
separated by an intermediate phase. 
The TWA phase shows non-zero polar order for species A ($\alpha_1 \,{>}\, 0$) and finite velocity variations ($\Delta \nu_A \,{>}\, 0$), while there is neither significant polar order in species B ($\beta_1 \,{\approx}\, 0$) nor velocity variations $\Delta \nu_B$. 
In contrast, the TWB phase is characterized by polar order as well as velocity variation in both species A and B.
The case $\lambda \,{=}\, 0.28$ (b) is phenomenologically similar to (a), but in the TWA phase there is now also polar order ($\beta_1 \,{>}\, 0$) and finite velocity variations ($\Delta \nu_B \,{>}\, 0$) for species B.  
The intermediate phase in which species A is uniformly ordered and species B disordered (center white strip) is much smaller than in (a).
For $\lambda=0.31$ (c), the  TWA and TWB phases merge to one broad, single travelling wave phase which exhibits polar order and finite velocity variation in both species.
}
\label{fig:AgentBasedPhaseDiag_OPs}
\end{figure}

The agent-based simulations show that above some threshold value for the noise strength ($\sigma \,{ \gtrsim }\, 0.7$) both species are in a spatially uniform, disordered state, with the polar order parameters, $\alpha_1$ and $\beta_1$, and the velocity variation $\Delta \nu$ for both species approximately zero; see
upper-most light grey area in Fig.~\ref{fig:AgentBasedPhaseDiag} and Fig.~\ref{fig:AgentBasedPhaseDiag_OPs}, as well as Fig.~\ref{fig:5_phases_all}a. 
Below this threshold noise value, 
species A shows polar waves independent of the value for $\lambda$, as can be inferred from Fig.~\ref{fig:AgentBasedPhaseDiag_OPs}a where $\alpha_1$ and $\Delta \nu_A$ are shown as functions of $\sigma$.
In analogy to the corresponding phase observed in the kinetic Boltzmann approach, we call this phase \textit{traveling wave phase A} (TWA).
How these polar waves in species A affect the degree of order in species B depends on the strength of the relative fitness $\lambda$. 
Within the polar waves, the density of A is enhanced, which in turn implies an increased density for species B due to the game interaction; recall that at the mean-field level we have $b = \lambda \cdot a$.
Whether this induces polar order depends on whether the density of species B within the band exceeds the threshold for polar order (in the absence of the game).
Indeed, we find that for low values of the relative fitness, $\lambda \lesssim 0.26$, there is only a density wave in species B but no (or only weak) polar order (see Fig.~\ref{fig:AgentBasedPhaseDiag_OPs}a and Fig.~\ref{fig:5_phases_all}b).
For larger values of $\lambda$ (Fig.~\ref{fig:AgentBasedPhaseDiag_OPs}b,c), however, the density of species B becomes large enough within the polar bands of species A such that the threshold for polar order is crossed and species B exhibits polar bands as well.  
Since this polar order is mediated by the polar waves of species A through the game interaction and not through the alignment interaction of species B per se, it is a game-induced phenomenon similar as in the hydrodynamic approach (Sec. \ref{subsec:GameInducedPatternFormation}). 
A movie displaying the dynamics in the TWA phase can be found in the supplementary material (movie2).

Upon further decreasing the noise strength $\sigma$, there is a parameter domain for $\lambda \lesssim 0.29$ where we find an intermediate phase (see the grey-shaded region in Fig.~\ref{fig:AgentBasedPhaseDiag} enclosed between the TWA and TWB phases). 
In this region of the diagram, there are no longer any wave patterns in species A or B; see central white regions in Figs.~\ref{fig:AgentBasedPhaseDiag_OPs}a,b.
Furthermore, while the polar order of species A is maintained ($\alpha_1 \,{>}\, 0$), species B no longer exhibits polar order ($\beta_1 \,{\approx}\, 0$).  
In other words, species A is now in a uniformly polar-ordered state, while species B is in a disordered state; see also Fig.~\ref{fig:5_phases_all}c.

\begin{figure}[t]
\centering
\includegraphics[width=\linewidth]{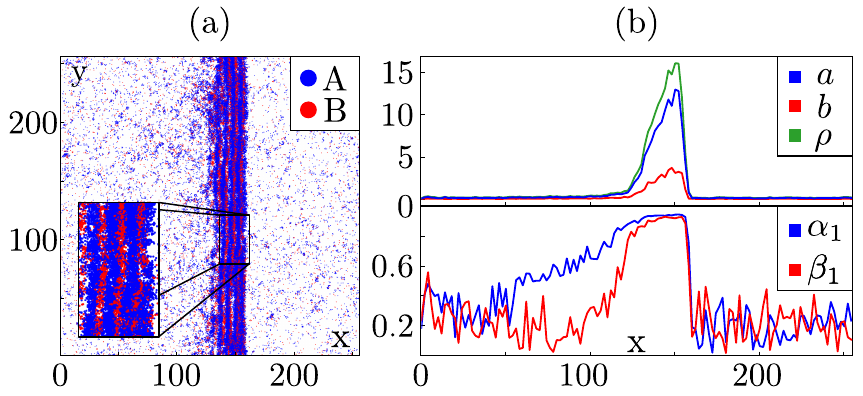}
\caption{
Density and velocity profiles in the TWB phase for noise strength $\sigma \,{=}\, 0.32$ and relative fitness $\lambda \,{=}\, 0.26$. (a) Snapshot from the agent-based simulation with particle positions of species A and B marked in blue and red, respectively. Inset: blow-up showing species demixing inside the wave. 
(b) Corresponding wave profile as a function of position $x$ averaged over the $y$-direction. 
For the analysis we partitioned the system into $256$ channels along the $x$-axis, counted the particles per channel and evaluated the channel densities $a$, $b$, and $\rho$ (upper plot) and polar order $\alpha_1$, $\beta_1$ (lower plot). There is a clear correlation between the density wave and polar order; outside the wave the system shows a low-density disordered phase.
}
\label{fig:WaveBandsAndProfile}
\end{figure}

Upon decreasing the noise strength $\sigma$ even further, we observe wave patterns again.
Now, both species A and B display traveling waves with polar order; see Figs.~\ref{fig:AgentBasedPhaseDiag_OPs}a,b and Fig.~\ref{fig:5_phases_all}d.
Species B forms polar waves by itself due to alignment interactions.  
The polar waves in species A, however, have a different origin: Without the game interaction species A would already be in a uniform polar-ordered phase. 
The role of the interaction mediated by the snowdrift game is to spatially modulate the density of species A such that it mirrors the wave pattern of species B; for spatial profiles of the density and the polar order see Fig.~\ref{fig:WaveBandsAndProfile}b. 
We call this phase \textit{traveling wave phase B} (TWB); the dynamics are displayed in movie3.
At even lower noise strength, the waves disappear again, and both species exhibit uniform polar order (dark grey area in Fig.~\ref{fig:AgentBasedPhaseDiag} and movie4). 
Transitions between different phases do not occur at exact parameter values, instead there is a region with stochastic switching, where the system alternates between the two adjacent phases.

While the diagrams obtained from the agent-based simulations (Fig.~\ref{fig:AgentBasedPhaseDiag}) and using the hydrodynamic theory (Fig.~\ref{fig:PhaseDiagram}) are qualitatively fairly similar, the agent-based simulation reveal interesting additional phenomena which are absent in the Boltzmann approach.
For phases in which both species show long-range polar order, i.e.\ polar waves or uniform polar order, we observe a demixing of species A and B, meaning clusters or alternating bands of only one species or the other species.
Figure~\ref{fig:WaveBandsAndProfile}a illustrates this demixing in the TWB phase for polar waves parallel to the direction of wave propagation. 
Note that the bands are wider in species A , as it has a fitness advantage compared to species B ($\lambda \,{=}\, 0.26$).

The size of these demixed regions grows linearly with increasing game range $d_{\text{game}}$, as is illustrated in Fig.~\ref{fig:r_delta_vs_dgame}. 
If demixing occurs, the ensuing density ratio $b/a$ becomes larger than the relative fitness $\lambda$.
Both demixing and the density shift occur only if the relative motion between particles is slow (e.g.\ in the polar-ordered phases) compared to the rate $\tau$ of snowdrift game interactions. 
In the agent-based simulations used in this study, this requirement is met only for the traveling wave phase induced by species B (TWB) and the uniformly polar ordered phase, the only phases in which both species A and B display sufficient polar order. 
While the demixing manifests itself as bands in the TWB phase, the species form clusters in the uniformly polar ordered phase; see Fig.~\ref{fig:5_phases_all}e.
The observed demixing of species is an interesting feature of the snowdrift game and occurs even in the absence of alignment interactions (see left panel of Fig.~\ref{fig:r_delta_vs_dgame}), as long as the diffusive mixing of the system is sufficiently slow.

We attribute the formation of such clusters and bands to the \textit{finite} interaction range of the snowdrift game.
In a well-mixed population the density ratio is given by $b/a=\lambda$ and, hence, all particles have equal fitness.
Upon demixing within the interaction range of the game, however, the overall fitness can be increased as alike species cluster together. 
This tendency for demixing dominates over `entropic' mixing effects due to random particle motion, i.e.~for slow movement of particles relative to each other (in comparison to the game strength).

In the agent-based approach we observe that the TWB phase is stable over a much wider range of noise strength $\sigma$ than in the hydrodynamic approach, and becomes even larger with increasing relative fitness levels $\lambda$; cf. Fig.~\ref{fig:AgentBasedPhaseDiag} and Fig.~\ref{fig:PhaseDiagram}.
We hypothesize that this stabilization of travelling waves is due to the interplay between demixing and the loss of directional information during switch randomization: 
Particles that change species at the rear end of a wave and thereby lose their directional information counteract the formation of uniform polar order.

\section{Discussion and conclusions}\label{sec:DiscussionAndConclusions}

We investigated an active matter system consisting of two different types of self-propelled particles interacting both mechanically and through birth-death processes described by a game. 
For the mechanical interaction we used standard alignment of the particles' velocities implemented either as pairwise collisions with half-angle alignment (following Bertin et al.~\cite{Bertin06})  or alignment in a certain neighborhood (in the spirit of a Vicsek model~\cite{Vicsek95}).
In addition, we considered competition between the different particle types (`species') mediated through a snowdrift game, again played either upon collision or in some neighborhood. 
Since these interactions can change the (local) density of particles ---
a critical parameter for symmetry-breaking and pattern-formation in active matter --- they are an essential and interesting extension of classical active matter models. 
We would even suggest that they may be seen as a simple classic example of a more general class of interactions that affect the density of the self-propelled particles rather than their orientation. 

We have derived hydrodynamic equations for the densities and velocities of the particles, starting from a kinetic Boltzmann equation for polar particles that we extended by the replicator dynamics of the snowdrift game.
Using linear stability analysis we have studied the system's dynamics as a function of two key control parameters, the noise amplitude of the polar alignment interactions and the relative fitness of the two species in the snowdrift game.
These studies revealed different regimes of macroscopic polar order and pattern formation for one or both species. 
Below a critical noise strength, we observe that particles of the species with the higher fitness form polar order and evolve into polar traveling wave patterns. 
Interestingly, these patterns---mediated by snow drift interactions between the two species---induce polar travelling waves in the other species as well. 
Furthermore, we find that in a regime of even lower noise values, the species with the lower fitness forms polar wave patterns, and these patterns again induce polar traveling waves in the other species. 
We referred to this phenomenon as \textit{game-induced pattern formation}.
Although we have discovered this phenomenon for the specific case of polar active particles interacting through a snowdrift game, we would argue that it is a fairly general phenomenon that applies to a broad class of active systems where the \textit{local} density of particles is affected by (nonlinear) reaction kinetics.  
We have complemented the kinetic approach with agent-based simulations in which interactions between particles take place in some extended neighbourhood rather than in collisions. 
These simulations qualitatively confirm the phase diagram and the various regimes of macroscopic order and pattern formation, including the game-induced patterns.
This shows that our findings are robust and do not depend on the particular implementation of the active matter system.

Moreover, our numerical simulations of the hydrodynamic theory and the agent-based model consistently show that the transitions between the different types of collective behavior are characterised by discontinuous changes in the velocity and spatial density variation. 
In particular, they also exhibit hysteresis and bistability.
This shows that the ensuing phase transitions are generically first order and the formation of patterns is subcritical.
Only by fine-tuning parameters could we find a regime where the transition becomes continuous.
Finally, our agent-based simulations show the formation of species separated clusters (disks and bands) in the ordered phases due to the finite range of game interactions. 
The simulations suggest that these spatial structures can lead to a stabilization of game-induced wave pattern leading to a broader parameter regime of these waves. 

From a broader perspective, this study suggests a possibly general mechanism in active matter systems where patterns in one species induce patterns in another species.
Here, since we focused on two species with qualitatively identical pattern phenomenology (traveling waves), the patterns induced by game interactions are qualitatively the same (traveling waves).
More generally, it would be interesting to ask what could happen if the individual species showed qualitatively different types of collective behavior and different pattern phenomenology.
For instance, different symmetries of intra-species alignment, e.g. due to different shapes of the species' agents, could lead to density regimes of patterns with different symmetries for the two species (e.g.\ polar clusters and nematic patterns)~\cite{Baer_etal:2020}. 
Based on our study, patterns in one species induce density variations in the second species (game-induced patterns).
These density variations in the second species could then locally enter regimes of pattern formation and induce patterns in the second species. 
As a result there would be patterns of different symmetries, which undergo mutual feedbacks and possibly lead to a rich pattern phenomenology with novel collective states in active matter.

Our study could also, perhaps only on a conceptual level, relate directly to `real life' systems, where motile organisms undergo chemical interactions with each other. Examples for biological systems whose spatiotemporal dynamics are shaped by the interplay of motility and chemical interactions include motile bacteria that compete for shared resources~\cite{Gude:2020} and bacterial systems that interact through chemotaxis and other types of extracellular signals.
For a direct comparison with these systems, it will certainly be necessary to consider a more complex interaction scenario than the snowdrift game considered here and, for example, explicitly include the spatiotemporal dynamics of the extracellular chemicals and more complex chemical reaction kinetics \cite{CremerFrey19}. 
Other systems, where we expect similar phenomenology, may include systems with different chemical species engineered such that these species are self-propelled and show chemical reactions. 

\begin{acknowledgments}
We would like to thank Lorenz Huber for stimulating discussions, and Fridtjof Brauns and Laeschkir W\"urthner for their help in implementing the numerical simulations. 
This research was supported by the Excellence Cluster ORIGINS, which is funded by the Deutsche Forschungsgemeinschaft (DFG, German Research Foundation) under Germany's Excellence Strategy EXC-2094-390783311, and the funding initiative ``What is life''of the Volkswagenstiftung.
\end{acknowledgments}

\appendix

\section{\label{app:A_DerivationHydro}Derivation of the hydrodynamic equations}

The kinetic Boltzmann Eqs.~(\ref{eq:Boltzmann}) set the mathematical framework for the investigation of our active particle system with snowdrift game interaction.
Following~\cite{Bertin09,Peshkov14}, we use these equations as a starting point to derive hydrodynamic equations that describe the dynamics of the density and polar order at least in a parameter regime close to the onset of macroscopic polar order.\\

\subsection{\label{app:Boltzmann}Extended kinetic Boltzmann equations}
As discussed in Section \ref{sec:Model}, we model the dynamics of two species $A$ and $B$ with self-propulsion and self-diffusion, that undergo intraspecies polar alignment, and interspecies game interactions. All agents are assumed disc-shaped with diameter $d_0$ and velocity $v_0$. We therefore assume the following dynamics for the one-particle probability density functions $\alpha \left( \mathbf{r},\theta,t\right)$ and $\beta \left( \mathbf{r},\theta,t\right)$, for the species $A$ and $B$, respectively:
\begin{widetext}
\begin{subequations}
\label{eq:appBoltzmann}
\begin{align}
\label{eq:appBoltzmannA}
\partial_t \alpha \left( \mathbf{r},\theta,t\right) &+  v_0\left(\mathbf{e}_{\theta} \cdot \nabla \right) \alpha \left( \mathbf{r},\theta,t\right)
 =  I_{\text{diff}} \left[\alpha\right] + I_{\text{coll}} \left[\alpha,\alpha\right] + I_{\text{game}}^A \left[\alpha,\beta\right] \\
\label{eq:appBoltzmannB}
\partial_t \beta \left( \mathbf{r},\theta,t\right) &+  v_0\left(\mathbf{e}_{\theta} \cdot \nabla \right) \beta \left( \mathbf{r},\theta,t\right) 
 =  I_{\text{diff}} \left[\beta\right] + I_{\text{coll}} \left[\beta,\beta\right] + I_{\text{game}}^B \left[\beta,\alpha\right] 
\end{align}
\end{subequations}
with the terms describing self-diffusion, $ I_{\text{diff}} \left[f\right] $, and alignment interaction, $I_{\text{coll}} \left[f,f\right] $
\begin{subequations}
\label{eq:appBoltzamnnTerms}
\begin{align}
    I_{\text{diff}} \left[f\right] 
    &=
    -\omega f \left(\theta\right) +\omega \int_{-\pi}^\pi d\theta' \int_{-\infty}^\infty d\eta_0 \,  f\left(\theta'\right)   \
    \mathcal{P}_0\left(\eta_0\right)   \, \delta_{2\pi}\left(\theta' + \eta_0 -\theta\right) ,
    \\
    I_{\text{coll}} \left[f,f\right] 
    &=
    -f\left(\theta\right) \int_{-\pi}^\pi d\theta' \, \mathcal{R}_{\theta,\theta'}
    f\left(\theta'\right)   
    + \int_{-\pi}^\pi d\theta_1 \int_{-\pi}^\pi d\theta_2  \, \mathcal{R}_{\theta_1,\theta_2} f\left(\theta_1\right) f\left(\theta_2\right)  \int_{-\infty}^\infty d\eta \, P\left(\eta\right)  \, 
    \delta_{2\pi}
    \left( 
    \tfrac{\theta_1+\theta_2}{2} 
    {+} \eta 
    {-} \theta
    \right)  ,
\end{align} 
\end{subequations}
\end{widetext}
and the terms for the game interaction
\begin{subequations}
\begin{align}
\label{eq:appBoltzamnnTerms_game}
    I_{\text{game}}^{A}  \left[\alpha,\beta\right] 
    &=
    \tau \left(  b - \lambda a \right)  b  \,\alpha\left(\theta\right) ,
    \\
    I_{\text{game}}^{B}  \left[\alpha,\beta\right] 
    &=
    \tau \left( \lambda a - b \right) a \,\beta\left(\theta\right)\,.
\end{align} 
\end{subequations}
The collision kernel $\mathcal{R}_{\theta_1,\theta_2} \,{=}\, 4  \, d_0 v_0 \left|\sin\left(\tfrac12 (\theta_1-\theta_2) \right)\right|$ characterizes the collisions between the agents and 
\begin{equation}
    \delta_{2\pi}\left(x\right) 
    := \sum_{m = -\infty}^{+\infty} \delta(x + 2 \pi m).
\end{equation}
denotes the $2 \pi$ periodic delta distribution. 
We assume that both random variables, $\eta_0$ and $\eta$, describing diffusion and collision noise, respectively, are Gaussian distributed:
\begin{align}
    \mathcal{P}_0(\eta_0) 
    &= \frac{1}{\sqrt{2 \pi \sigma_0^2}} \,  
    \exp 
    \biggl[ 
    - \frac{\eta_0^2}{2\sigma_0^2}
     \biggr] , 
    \\
    \mathcal{P}(\eta) 
    &= \frac{1}{\sqrt{2 \pi \sigma^2}} \, 
    \exp 
    \biggl[ 
    - \frac{\eta^2}{2\sigma^2}
    \biggr] . 
\end{align} 
For a more detailed derivation of $I_{\text{diff}}$ and $I_{\text{coll}}$, we refer to Ref.~\cite{Bertin09}. 
By using the rescaling
\begin{equation}
\begin{split}
\tilde{t} :=& \, t \cdot \omega, \qquad
\tilde{\mathbf{r}} := \,\mathbf{r}\cdot \omega  \, v_{0} ^{-1},   \\ 
\tilde{\tau} :=& \,\tau \cdot \omega \, d_0^{-2} v_0^{-2}, \qquad
\tilde{f}:= \, f \cdot d_0 v_0 \omega^{-1} .
\end{split}
\end{equation}
we can set $\omega$, $v_0$, and $d_0$ equal to $1$ without loss of generality. 
In the following, we will omit the tilde and just write $t$, $\mathbf{r}$, $\tau$, and $f$ to simplify notation.

\subsection{\label{app:BoltzmannFourier}Fourier transformed Boltzmann terms}

As a first step, we perform the Fourier transformation of the extended kinetic Boltzmann equations \eqref{eq:appBoltzmann} in $\theta$ with the Fourier representation 
\begin{align} 
	f \left(\mathbf{r}, \theta, t\right) 
	&= \frac{1}{2 \pi} \sum\limits^{\infty}_{k= - \infty} e^{-ik\theta}  \, \hat{f}_k (\mathbf{r},t) 
    , \\
    \hat{f}_k( \mathbf{r}, t) 
    &= \int\limits^{\pi}_{-\pi} d \theta ~ e^{ik\theta} \,  f \left(\mathbf{r}, \theta, t\right) ,
\end{align}
where $f$ stands for $\alpha$ or $\beta$. With this convention for the Fourier transformation the following identities hold: $\int_{-\pi}^{\pi} e^{ik\theta} = 2 \pi \delta_{k,0}$ and $\sum_{k= - \infty}^{\infty} e^{-ik\theta} = 2 \pi \delta (\theta)$. In the following, we omit spatial and temporal dependencies, and, if not specified differently, all integrals over $\theta$ have boundaries $-\pi$ to $\pi$, all integrals over the noise have boundaries $\pm\infty$, and sums over wave numbers $k$ run from $-\infty$ to $\infty$. 
In order to calculate the Fourier transform of the convection term, we use $\mathbf{e}_\theta = \left( \cos\theta,~  \sin\theta  \right)^T$. In Fourier space, the convection term reads
\begin{align} 
    &\int \! d \theta  ~e^{ik\theta}   \, 
    \mathbf{e}_{\theta} \cdot 
    \partial_\mathbf{r} \alpha(\theta) 
    = \int d \theta ~e^{ik\theta} 
    \Bigl[
    \cos\theta \, \partial_x + \sin\theta \, \partial_y
    \Bigr] 
    \alpha(\theta) 
    \nonumber \\ 
    &\qquad =
    \int \!\mathrm{d} \theta \, 
    \mathrm{e}^{\mathrm{i} k \theta} \, 
    \tfrac{1}{2} 
    \left[
    \left(
    \mathrm{e}^{\mathrm{i} \theta} {+}
    \mathrm{e}^{-\mathrm{i} \theta}
    \right) 
    \partial_{x}
    -\mathrm{i} 
    \left(
    \mathrm{e}^{\mathrm{i} \theta} {-} 
    \mathrm{e}^{-\mathrm{i} \theta}
    \right)\partial_{y}\right] 
    \alpha( \theta)
    \nonumber \\
	&\qquad 
	=\tfrac{1}{2} \partial_x \left( \hat{\alpha}_{k+1} + \hat{\alpha}_{k-1} \right) - \tfrac{i}{2} \partial_y \left( \hat{\alpha}_{k+1} - \hat{\alpha}_{k-1} \right) 
	\nonumber \\
	&\qquad 
	=\tfrac{1}{2} 
	\bigl( \nabla \hat{\alpha}_{k-1} + \nabla^{*} \hat{\alpha}_{k+1} 
	\bigr),
\end{align}
where $\nabla \,{:=}\, \partial_x {+} i\partial_y$ and $\nabla^{*} \,{:=}\, \partial_x {-} i\partial_y$.
The Fourier transform of the diffusion term can be calculated as
\begin{align}
    &\int  \! d  \theta ~ e^{ik\theta} I_{\text{diff}} \left[ \alpha \right] 
    = 
    - \int \! d \theta ~ e^{ik\theta}  \alpha (\theta) 
	\nonumber \\
	&\quad+  
	\int \! d \theta ~ e^{ik\theta} 
	\int \! d \theta' \!
	\int \! d \eta 
	~ \mathcal{P}_0 (\eta)  \, 
	\alpha( \theta') \,  
	\delta_{2 \pi} 
	( \theta {-} \theta' {-} \eta) 
	\nonumber \\
	&\quad= 
	- \hat{\alpha}_k +  \int d \theta' \int  \! d  \eta ~ e^{ik(\theta' + \eta)} \mathcal{P}_0 (\eta)  \, \alpha (\theta') \nonumber \\
	&\quad= 
	- \hat{\alpha}_k + \hat{\alpha}_k \hat{\mathcal{P}}_{0,k} 
	=
	( \hat{\mathcal{P}}_{0,k} -1) \hat{\alpha}_k
\end{align}
with $\hat{\mathcal{P}}_{0,k}=$ exp$(-k^2\sigma_{0}^{2}/2)$ the $k$-th Fourier mode of the Gaussian distribution $\mathcal{P}_0 (\eta) $.
The diffusion term consists of a gain and a loss term, given by the first and the second term in Eq.~\eqref{eq:appBoltzamnnTerms}b, respectively. We denote the Fourier transforms of the gain and the loss terms as $(I)$ and $(II)$, respectively,
$\int d  \theta ~ e^{ik\theta} I_{\text{coll}} \left[ \alpha, \beta \right] = (I) + (II)$:
\begin{align} 
	(I)=& - \int d \theta \int\limits_{\theta-\pi}^{\theta+\pi} d \theta'~ e^{ik\theta} ~ \mathcal{R}_{\theta, \theta'}    \,  \alpha(\theta) \,  \beta( \theta') \nonumber \\
    &= - \int d\theta \int\limits_{\theta -\pi}^{\theta+\pi} d \theta'~ e^{ik\theta} \mathcal{R}_{\theta, \theta'} \sum\limits_{q,p}  \frac{\hat{\alpha}_q \hat{\beta}_p }{(2 \pi)^2}  e^{-iq\theta}  e^{-ip\theta'} 
    \nonumber \\
	&= -\int d \theta \int d \vartheta \sum\limits_{q,p} \frac{\hat{\alpha}_q \hat{\beta}_p}{(2 \pi)^2} \mathcal{R}_{|\vartheta|} \,  e^{i\theta(k-q-p)} e^{-ip\vartheta} \nonumber \\
	&= -\frac{1}{2\pi} \int d\vartheta ~ \sum\limits_{q,p}   \hat{\alpha}_q \hat{\beta}_p \mathcal{R}_{|\vartheta|}  \, \delta_{q,k-p}  \, e^{-ip\vartheta} \nonumber \\
	&= -\frac{1}{2\pi} \sum\limits_{p}   \hat{\alpha}_{k-p} \hat{\beta}_p \int d \vartheta ~ \cos (p\vartheta) \mathcal{R}_{|\vartheta|} .
\end{align}

Analogously, $(II)$ can be calculated as
\begin{widetext}
\begin{equation} 
\begin{split}
	(II) 
	=& \int d\theta ~ e^{ik\theta} \int d \theta_1 \int\limits_{\theta_1-\pi}^{\theta_1+\pi} d\theta_2 \int  \! d  \eta ~ \alpha (\theta_1) \beta ( \theta_2)
      \mathcal{R} _{\theta_1,\theta_2} \mathcal{P}(\eta)  \, \delta_{2\pi} 
      \left( 
      \theta {-} \tfrac{\theta_1 {+} \theta_2}{2} {-} \eta
      \right)
    \\
    =&  
    \int d\theta \int d\theta_1  \int\limits_{\theta_1-\pi}^{\theta_1+\pi} d\theta_2  
    \int d\eta ~ \sum\limits_{q,p} e^{ik\theta} e^{-iq\theta_1} e^{-ip\theta_2} 
    \frac{\hat{\alpha}_q \hat{\beta}_p }{(2 \pi)^2} \mathcal{R}_{\theta_1,\theta_2} \mathcal{P}(\eta)  \, \delta_{2\pi} 
    \left( 
    \theta {-} \tfrac{\theta_1 {+} \theta_2}{2} {-} \eta
    \right)
    \\
	=& 
	\int d\theta_1 
	\int d \vartheta ~ \sum\limits_{q,p} e^{i \theta_1 (k-q-p)} e^{i\vartheta\left(k/2-p\right)} \frac{\hat{\alpha}_q \hat{\beta}_p }{(2 \pi)^2}  \mathcal{R}_{|\vartheta|} \hat{\mathcal{P}}_k
	=
	\frac{1}{2 \pi}  \int d \vartheta ~ \sum\limits_{q,p} \delta_{q,k-p} e^{i\vartheta \left(k/2-p\right)}  \hat{\alpha}_q \hat{\beta}_p \mathcal{R}_{|\vartheta|} \hat{\mathcal{P}}_k
    \\
    =&\frac{1}{2 \pi} \sum\limits_p \hat{\alpha}_{k-p} \hat{\beta}_p \hat{\mathcal{P}}_k 
    \int d \vartheta ~ 
    \cos \Bigl( \bigl(\tfrac{k}{2}-p\bigr) \vartheta \Bigr) \mathcal{R}_{|\vartheta|},
\end{split}
\end{equation}
\end{widetext}

In total, we get for the collision term:
\begin{align} 
    \int  d  \theta ~ e^{ik\theta} I_{\text{coll}} \left[ \alpha, \beta \right] 
    = \sum\limits_p \hat{\alpha}_{k-p} \hat{\beta}_p \mathcal{I}_{p,k},
\end{align}
with
\begin{align}
  \mathcal{I}_{p,k} 
  := \int \frac{d\vartheta}{2 \pi} \mathcal{R}_{|\vartheta |} 
  \Bigl[
  \hat{\mathcal{P}}_k  \, 
  \cos
  \bigl( 
  (\tfrac{k}{2}-p) \vartheta 
  \bigr) - 
  \cos (p \vartheta) 
  \Bigr].
\end{align}

The last remaining part to complete the Fourier representation of Eq.~\eqref{eq:appBoltzmann} is $I_{\text{game}}$:
\begin{equation}
\int d \theta ~e^{ik\theta} I^{A}_{\text{game}} = \tau (b - \lambda a) b\,\hat{\alpha}_k 
\end{equation}
and $I^{B}_{\text{game}}$ analogously. 
The Fourier transformation of the equation for species B, Eq.~\eqref{eq:appBoltzmannB} works analogous to the transformation for species A shown here.
In summary, the Fourier representations of Eqs.~\eqref{eq:appBoltzmann} read
\begin{subequations} 
\label{eq:appFourierBoltzmann}
\begin{align}
	\partial_t \hat{\alpha}_k + \frac{1}{2} &\left( \nabla \hat{\alpha}_{k-1} + \nabla^{*} \hat{\alpha}_{k+1} \right) 
	\\
	&= ( \hat{\mathcal{P}}_{0,k} -1) \hat{\alpha}_k 
    + \sum\limits_p \hat{\alpha}_{k-p} \hat{\alpha}_p \mathcal{I}_{p,k} +\tau (b - \lambda a) b\,\hat{\alpha}_k 
    \nonumber \\
	\partial_t \hat{\beta}_k + \frac{1}{2}& \left( \nabla \hat{\beta}_{k-1} + \nabla^{*} \hat{\beta}_{k+1} \right) 
	\\
	&= ( \hat{\mathcal{P}}_{0,k} -1) \hat{\beta}_k 
    + \sum\limits_p \hat{\beta}_{k-p} \hat{\beta}_p \mathcal{I}_{p,k} - \tau( b - \lambda a ) a \hat{\beta}_k .
    \nonumber 
\end{align}
\end{subequations}
where we used the fact that the 0-th Fourier mode is the local density of the respective species, since
\begin{align}
    \alpha_0(\mathbf{r},t) 
    &= \int_{-\pi}^{\pi} \mathrm{d} \theta \,e^{i k \cdot 0}\alpha(\mathbf{r},\theta,t) 
    \\
    &= \int_{-\pi}^{\pi} \mathrm{d} \theta \,\alpha(\mathbf{r},\theta,t)= a(\mathbf{r},t)
\end{align}
and $\beta_0(\mathbf{r},t) = b(\mathbf{r},t)$  analogously.

\subsection{\label{app:Hierarchy}Uniform disordered solution}
One solution of~\eqref{eq:appFourierBoltzmann} can be identified by using the fact that $(\hat{\mathcal{P}}_{0,k}-1)=0$ and $\mathcal{I}_{n,k}=0$ for $k=0$ for all $n$. 
Assuming spatial homogeneity, we set $\nabla \alpha_{k}  \,{=}\, \nabla \beta_{k}  \,{=}\, 0 $. The kinetic Boltzmann equations for $k \,{=}\, 0$ then read 
\begin{equation}
\label{eq:appReplicatorEq}
\partial_t a  = \tau \left(  b - \lambda a \right) a \, b
,\quad
\partial_t b = - \tau \left(  b -  \lambda a \right)  a \, b.
\end{equation}
Substituting the control parameter $\bar{\rho}  \,{=}\,  a + b$, we get the three solutions
\begin{subnumcases}{(a^{(0)}, b^{(0)}) =}
  \hspace{6.6 mm} (0,\bar{\rho})\\
    \hspace{6.6 mm} (\bar{\rho},0)  \\
    (\bar{\rho} \, \tfrac{ 1}{1+\lambda},\bar{\rho}  \, \tfrac{\lambda}{1+\lambda}) \label{eq:appHomSol3}
\end{subnumcases}
and $\alpha_k  \,{=}\,  \beta_k  \,{=}\,  0$ for all $k \,{>}\, 0$. Eqs.~(\ref{eq:appReplicatorEq}) are the replicator equations described in Section~\ref{sec:Model}. As argued there, considering the dynamics of the equations , only the third solution is linearly stable against perturbations.

In order to investigate the linear stability of this solution, we linearize the equations \eqref{eq:appFourierBoltzmann} around the solution Eq.~\eqref{eq:appHomSol3} and calculate the linear growth rate of small perturbations $\delta \alpha_k$ and $\delta \beta_k$. The linearized dynamics of $\delta \alpha_k$ and $\delta \beta_k$ is given by (for $k \,{>}\, 0$)
\begin{align}
	\partial \delta \alpha_k 
	&= \left[ \hat{\mathcal{P}}_{0,k} -1 + (\mathcal{I}_{k,k}+\mathcal{I}_{0,k}) \, \bar{\rho} \,  \frac{1}{1+\lambda} \right] \delta \alpha_k  \nonumber \\
	&= \mu_k^A(a^{(0)},b^{(0)}) \delta \alpha_k ,\\
	\partial \delta \beta_k 
	&= \left[ \hat{\mathcal{P}}_{0,k} -1 + (\mathcal{I}_{k,k}+\mathcal{I}_{0,k}) \, \bar{\rho} \,  \frac{\lambda}{1+\lambda}  \right] \delta \beta_k \nonumber \\
	&= \mu_k^B(a^{(0)},b^{(0)}) \delta \beta_k ,
\end{align}
with $\mu_k^A$ and $\mu_k^B$ defined as in Eqs.~(\ref{eq:MuaMub}). The solution Eq.~\eqref{eq:appHomSol3} is therefore unstable when the linear growth rates $\mu_k^A$ or  $\mu_k^B$ have positive real parts. Because $\hat{\mathcal{P}}_{0,k}$ is the Fourier transformed distribution of the noise, we have $\hat{\mathcal{P}}_{0,k}  \,{\leq}\,  1 $, thus $\hat{\mathcal{P}}_{0,k} -1  \,{\leq}\,  0$. The linear growth rates $\mu_k^A$ and  $\mu_k^B$ can therefore only be positive when $\mathcal{I}_{k,k}+\mathcal{I}_{0,k} \,{>}\, 0$. 
Calculating \(\left(\mathcal{I}_{k,k}+\mathcal{I}_{0,k}\right)  \) for different values of \(k \,{>}\, 0\) shows that it is positive only for \(k \,{=}\, 1\) (for collision noise values \(\sigma <1\)) and negative for all \(k \,{>}\, 1\). Hence, small perturbations of the modes \(\alpha_k,\beta_k\) for \(k \,{>}\, 1\) will decay and only perturbations in $\alpha_1$ and $\beta_1$ may grow (if $\mu_k^A \,{>}\, 0$ or $\mu_k^B \,{>}\, 0$, respectively). 

The definition of the Fourier series yields for the first mode, $k \,{=}\, 1$:
\begin{align}
    \alpha_1 (\mathbf{r}, t) 
    &=  \int_{-\pi}^\pi  d\theta  \, e^{i\theta} \alpha( \mathbf{r},\theta, t) , \\
    \beta_1 (\mathbf{r}, t) 
    &=  \int_{-\pi}^\pi  d\theta  \, e^{i\theta} \beta( \mathbf{r},\theta, t) .
\end{align}
The first modes can be related to the polar order fields \(\mathbf{P}_A\) and  \(\mathbf{P}_B\), which measure the polar order of the system, through
\begin{align}
    a\,\mathbf{P}_A   
    = \left( \begin{array}{c} \text{Re}(\alpha_1 ) \\ \text{Im}(\alpha_1) \end{array}\right)
    \, , \quad
    b\, \mathbf{P}_B   
    = \left( \begin{array}{c} \text{Re}(\beta_1) \\ \text{Im}(\beta_1) \end{array}\right).
\end{align}
Hence, the real and imaginary part of $\alpha_1$ and $\beta_1$ are the $x$- and $y$- components of the overall polar order of the system at a given point in space and time. Since we have set the particle velocity $v_0$ to one, the polar order fields are identical to the velocity fields of both species. In the following, we will refer to $\alpha_1$ and $\beta_1$ as \textit{velocity fields} of the species $A$ and $B$, respectively. This means that when $\mu_k^A$ or  $\mu_k^B$ have positive real parts, the absolute value of the polar order of species $A$ or $B$, respectively, will grow. Indeed, as discussed in section \ref{sec:uniform_Sol}, we find critical values for $\sigma$ (as functions of $\lambda$) where either the real part of $\mu_k^A$ or $\mu_k^B$ is zero, indicating a transition from zero to nonzero polar order.

\subsection{Scaling ansatz}

Following~\cite{Bertin06,Bertin09,Peshkov14} we employ a scaling ansatz to derive hydrodynamic equations for the density, $a$ and $b$, and velocity fields $\alpha_1$ and $\beta_1$, of the species $A$ and $B$, respectively:
Specifically, we introduce the small parameter $\varepsilon$ and make the ansatz
\begin{equation} 
\begin{split}
	a = \mathcal{O} (\varepsilon)~, \quad  \alpha_k =& \mathcal{O} (\varepsilon^{|k|}) \quad \forall k \neq 0 \\
	 b = \mathcal{O} (\varepsilon)~, \quad  \beta_k =& \mathcal{O} (\varepsilon^{|k|}) \quad \forall k \neq 0
\end{split}
\end{equation}
Furthermore, we assume the scaling 
\begin{equation}
	\partial_t ~ \sim ~ \partial_x ~ \sim ~ \partial_y ~ \sim ~ \varepsilon.
\end{equation}
We can now expand the Fourier representation of the kinetic Boltzmann equations \eqref{eq:appFourierBoltzmann} up to the lowest order which gives non-trivial solutions, which is $\varepsilon^3$ in our case and truncate higher orders. The remaining terms are:
\begin{widetext}
\begin{subequations} 
\begin{align}
\label{eq:f0}
&k=0: && \partial_t a + \tfrac{1}{2} ( \nabla \alpha_{1}^* + \nabla^{*} \alpha_{1})=\tau (b - \lambda a) a \,  b \\
\label{eq:f1}
&k=1: &&
	\partial_t \alpha_{1} + \tfrac{1}{2} ( \nabla \alpha_{0} +  \nabla^{*} \alpha_2)=(\mathcal{P}_{0,1} -1)\alpha_{1} + (\mathcal{I}_{0,1}+\mathcal{I}_{1,1})a \,  \alpha_{1} 
	+ (\mathcal{I}_{2,1} + \mathcal{I}_{-1,1})\alpha_{1}^* \alpha_2 + \tau ( b - \lambda a) \alpha_{1} b \\
\label{eq:f2}
&k=2: &&
	\partial_t \alpha_2 + \tfrac{1}{2} \nabla \alpha_{1} = (\mathcal{P}_{0,2}-1) \alpha_2 +(\mathcal{I}_{0,2} + \mathcal{I}_{2,2}) a  \, \alpha_2 + \mathcal{I}_{1,2}  \, \beta_{1}^{2} 
\end{align}
\end{subequations}
We only show the equations for $a$, $\alpha_{1}$ and $\alpha_{2}$ here. The equations for $b$ and $\beta_{1}$ are analogue. Because we only take into account $\mathcal{O}(\epsilon^3)$ in our final equations for $a$, $b$, $\alpha_1$, and $\beta_1$, we can drop the terms $\partial_t \alpha_2$ in (\ref{eq:f2}) and solve \eqref{eq:f2} for $\alpha_2$. By defining 
\begin{equation} 
	\nu(a) = -\frac{1}{4[P_{0,2}-1 + (\mathcal{I}_{0,2} + \mathcal{I}_{2,2})a]}
\end{equation}  
we can write $
	\alpha_2=4 \, \nu(a)\mathcal{I}_{1,2} \, \beta_{1}^{2} -2 \, \nu(a) \nabla \alpha_{1}$.
Substituting this into (\ref{eq:f1}), we get
\begin{equation} 
\begin{split}
	\partial_t \alpha_{1}=& -\tfrac{1}{2}\nabla a - 2 \, \nabla^{*} \nu(a) \left( \mathcal{I}_{1,2}  \, \alpha^{2}_{1}-\tfrac{1}{2} \nabla \alpha_{1} \right) 
    -2 \,  \nu(a)  \left(2\mathcal{I}_{1,2}  \, \alpha_{1} \nabla^{*}\alpha_{1}-\tfrac{1}{2} \nabla^{*} \nabla \alpha_{1} \right) 
	+(\mathcal{P}_{0,1}-1)\alpha_{1}  \\
    &+(\mathcal{I}_{0,1}+\mathcal{I}_{1,1}) a \,  \alpha_{1} + 4 \, \nu(a)\mathcal{I}_{1,2}(\mathcal{I}_{2,1} +\mathcal{I}_{-1,1})\alpha_{1}^{*}\alpha_{2}^{2} 
	- 2  \, \nu(a)(\mathcal{I}_{2,1}+\mathcal{I}_{1,1}) \alpha_{1}^*\nabla \alpha_{1} + \tau ( b - \lambda a) b  \, \alpha_{1}
\end{split}
\end{equation}
Neglecting order $\mathcal{O}(\varepsilon^4)$ and rearranging the equation, we arrive at
\begin{equation} 
\begin{split}
\partial_t \alpha_{1}=&\big [(\mathcal{P}_{0,1}-1) + (\mathcal{I}_{0,1}+\mathcal{I}_{1,1})a  + \tau ( b - \lambda a) b + 4(\mathcal{I}_{2,1}+\mathcal{I}_{-1,1}) |\alpha_{1}|^2 \nu(a)\mathcal{I}_{1,2}  \big ] \alpha_{1} \\
	&- \tfrac{1}{2} \nabla a + \nu(a) \nabla^{*} \nabla \alpha_{1} - 4\nu(a) \mathcal{I}_{1,2} \alpha_{1} \nabla^{*} \alpha_{1} 
 -2 \nu(a) (\mathcal{I}_{2,1}+\mathcal{I}_{-1,1}) \alpha_{1}^* \nabla \alpha_{1}.
\end{split}
\end{equation}
\end{widetext}
Following the same steps for $\beta_1$ and with the definitions 
\begin{subequations}
\begin{align}
	\mu_1^A  
	&= \, \left(\mathcal{P}_{0,1} {-} 1\right) {+} \left(\mathcal{I}_{0,1} {+} \mathcal{I}_{1,1}\right) a {+} \tau\left(  b {-} \lambda a \right)  b  ,\quad \\
    \mu_1^B  
    &= \, \left(\mathcal{P}_{0,1} {-} 1\right) {+} \left(\mathcal{I}_{0,1} {+} \mathcal{I}_{1,1}\right) b {+} \tau(  \lambda a {-} b) a \\
    \nu(x) &= - \tfrac{1}{4} \left[ \mathcal{P}_{0,2} -1 +(\mathcal{I}_{0,2}+\mathcal{I}_{2,2}) x\right]^{-1}  , \\
    \gamma(x) 
    &= 4  \, \nu(x) \mathcal{I}_{1,2} 
        \\
    \kappa(x)
    &= 2  \, \nu(x)(\mathcal{I}_{2,1}+\mathcal{I}_{-1,1}) , \\
    \xi(x) &= -4 (\mathcal{I}_{2,1} + \mathcal{I}_{-1,1}) \nu(x) \mathcal{I}_{1,2} ,
\end{align}
\end{subequations}
where $x$ stands for either $a$ or $b$ and
\begin{subequations}
\begin{align}
    \nabla 
    &= \partial_x + i \partial_y ,\quad   
    \nabla^* = \partial_x - i \partial_y,
    \\
    \mathcal{P}_{0,k} 
    &= \int \! d \eta \, e^{i k\eta_0} \, \mathcal{P}_0 (\eta_0) 
    = \exp \bigl[ -\tfrac12 \sigma_0^2 k^2 \bigr],
    \\
    \mathcal{P}_{k} 
    &= \int \!d \eta \, e^{i k\eta} \, \mathcal{P} (\eta) 
    = \exp \bigl[ -\tfrac12 \sigma^2 k^2 \bigr],
    \\
    \mathcal{I}_{n,k} 
    &= \int \! d \psi \, \mathcal{R}_{|\psi |} 
    \left[ 
    \mathcal{P}_k 
    \cos \bigl((n {-}\tfrac{k}{2})\psi\bigr) {-}
    \cos (n \psi) 
    \right].
\end{align}
\end{subequations}
we get the final set of hydrodynamic equations for the densities $a$ and $b$ and the polar order fields $\alpha_1$ and $\beta_1$
\begin{subequations}
\label{eq:appHydro}
\begin{align}
    \partial_t a 
    &=
    -\frac{1}{2} \left( \nabla  \alpha_{1}^* + \nabla^*  \alpha_{1} \right) + \tau(  b - \lambda a ) b \, a 
    \\
    \partial_t b 
    &=
    -\frac{1}{2} \left( \nabla  \beta_{1}^* + \nabla^*  \beta_{1} \right) + \tau(  \lambda a -b) a \, b 
    \\
    \partial_t \alpha_{1} 
    &=
    \left( \mu_1^A(a,b) - \xi(a) |\alpha_{1} |^2 \right) \alpha_{1} + \nu(a) \, \nabla^* \nabla \alpha_{1} \\
    &- \gamma(a) \, \alpha_{1} \nabla^* \alpha_{1} - \kappa(a)\, \alpha_{1}^* \nabla \alpha_{1} - \tfrac{1}{2} \nabla a
    \\
    \partial_t \beta_{1} 
    &=
    \left( \mu_1^B(a,b) - \xi(b) |u |^2 \right) \beta_{1}+ \nu(b) \, \nabla^* \nabla \beta_{1} \\
    &- \gamma(b) \, \beta_1 \nabla^* \beta_1 - \kappa(b)\, \beta_{1}^* \nabla \beta_1 - \tfrac{1}{2} \nabla b.
\end{align}
\end{subequations}
The set of coupled partial differential equations \eqref{eq:appHydro} describes the evolution of the two densities $a$ and $b$ and the corresponding polar order fields $\alpha_1$ and $\beta_1$ for the species $A$ and $B$, respectively. When we set the game parameter \(\tau=0\), we recover (for each species) the hydrodynamic equations derived in \cite{Bertin06,Bertin09}.
 
The fact that the snowdrift interaction term only appears in the term linear in the velocity fields in Eqs.~\eqref{eq:appHydro} is a result of the truncation scheme: 
The snowdrift terms in Eqs.~(\ref{eq:appFourierBoltzmann}) have the form $\tau (b - \lambda a) b\,\hat{\alpha}_k$ in the dynamic equation for $\alpha_k$ and $-\tau (b - \lambda a) a\,\hat{\beta}_k$ in the equation for $\beta_k$. 
Since we assume a scaling $ a = \mathcal{O} (\varepsilon), b = \mathcal{O} (\varepsilon),\alpha_k = \mathcal{O} (\varepsilon^{|k|})  ,   \beta_k = \mathcal{O} (\varepsilon^{|k|}) $ for all $ k \neq 0$, the snowdrift term is of order $\epsilon^3$ for $k=0$ and $k=1$. For all $k \,{>}\, 1$, the snowdrift term has a order of at least $\epsilon^4$ and is therefore cut off in the truncation scheme. 
Hence, the only snowdrift terms possible are the ones linear in the density and velocity fields, hence $\pm\tau(  b - \lambda a ) b \, a$, $\tau(  b - \lambda a ) b \,\alpha_1$ and $\tau(  b - \lambda a ) a \,\beta_1$, which all appear in the hydrodynamic Eqs.~(\ref{eq:appHydro}).

\section{\label{app:B_Jacobian}Jacobian of the hydrodynamic equations}

To analyze the stability of our solutions against spatial perturbations, we employ a plane wave ansatz:
\begin{subequations}
\begin{align}
\label{eq:SpaceFT}
	f(\mathbf{r},t) 
	&= \frac{1}{2 \pi} \int\limits_{-\infty}^{\infty} d\mathbf{q} ~ e^{i \mathbf{qr}} f(\mathbf{q}), \\
	f(\mathbf{q}) 
	&= \frac{1}{2 \pi} \int\limits_{- \infty}^{\infty} d\mathbf{r} ~ e^{-i\mathbf{qr}} f(\mathbf{r}), \\
	f^{*}(\mathbf{r},t)
	&= \frac{1}{2 \pi} \int\limits_{-\infty}^{\infty} d\mathbf{q} ~ e^{-i \mathbf{qr}} f^{*}(\mathbf{q}), 
	\\ 
	f^{*}(\mathbf{q}) 
	&= \frac{1}{2 \pi} \int\limits_{- \infty}^{\infty} d\mathbf{r} ~ e^{i\mathbf{qr}} f^{*}(\mathbf{r}),
\end{align}
\end{subequations}
where $f$ stands for $\delta a$, $\delta b$, $\delta \alpha_{1}$ and $\delta \beta_1$. 
Linearizing the hydrodynamic Eqs.~(\ref{eq:Hydro_densities}) and Eqs.~(\ref{eq:Hydro_velocities}) around the solutions given by Eqs.~(\ref{eq:UnifSolDensities}) and Eqs.~(\ref{eq:UnifSolVelocities}) and substituting \eqref{eq:SpaceFT}, we arrive at the following linear equations for $\delta a(\mathbf{q})$, $\delta b(\mathbf{q})$, $\delta \alpha_{1}(\mathbf{q})$, $\delta \alpha_{1}^{*}(-\mathbf{q})$, $\delta \beta_{1}(\mathbf{q})$ and $\delta \beta_{1}^{*}(-\mathbf{q})$:
\begin{widetext}
\begin{subequations}
\label{eq:HydroMatrixSystem}
\begin{align}
	\partial_t \delta a(\mathbf{q}) =& \left[ \tau ( b - 2 \lambda a ) b \right] \delta a(\mathbf{q}) + \left[ \tau (2 b - \lambda a )a \right] \delta b(\mathbf{q}) 
	+ \left[ - \tfrac{1}{2} (i q_x +q_y ) \right] \delta \alpha_{1}(\mathbf{q}) + \left[- \tfrac{1}{2}( i q_x - q_y)\right] \delta \alpha_{1}^{*}(-\mathbf{q})
	\\
	\partial_t \delta b(\mathbf{q}) =& \left[ -\tau ( b - 2 \lambda a ) b \right] \delta a(\mathbf{q}) + \left[ - \tau (2 b - \lambda a )a \right] \delta b(\mathbf{q}) 
    + \left[ - \tfrac{1}{2} (i q_x +q_y ) \right] \delta \beta_{1}(\mathbf{q}) + \left[ -\tfrac{1}{2}( i q_x - q_y)\right] \delta \beta_{1}^{*}(-\mathbf{q})
	\\
	\begin{split}
	\partial_t \delta \alpha_{1}(\mathbf{q}) =& \left[ \left( \frac{\partial \mu_1^A }{\partial a} - \frac{\partial \xi}{\partial a } |\alpha_{1}|^2 \right) \alpha_{1} - \frac{1}{2} ( i q_x - q_y ) \right] \delta a (\mathbf{q}) + \left[ \frac{\partial \mu_1^A}{\partial b } \alpha_{1} \right] \delta b(\mathbf{q})
	\\
	&+ \left[ \mu_1^A - 2 \xi |\alpha_{1}|^2 - \nu( q_{x}^{2} + q_{y}^{2}) - \gamma \alpha_{1} (i q_x + q_y ) - \kappa \alpha_{1}^{*}(i q_x -q_y) \right] \delta \alpha_{1}(\mathbf{q}) + \left[ - \xi \alpha_{1}^2 \right] \delta \alpha_{1}^{*}(-\mathbf{q})
	\end{split}
	\\
	\begin{split}
	\partial_t \delta \alpha_{1}^{*}(-\mathbf{q}) =& \left[  \left( \frac{\partial \mu_1^A }{\partial a} - \frac{\partial \xi}{\partial a } |\alpha_{1}|^2 \right) \alpha_{1}^{*} - \frac{1}{2} ( i q_x + q_y ) \right] \delta a (\mathbf{q}) + \left[ \frac{\partial \mu_1^A}{\partial b } \alpha_{1}^{*} \right] \delta b(\mathbf{q}) 
	\\
	&+ \left[ - \xi (\alpha_{1}^{*})^2 \right] \delta \alpha_{1}(\mathbf{q}) + \left[ \mu_1^A - 2 \xi |\alpha_{1}|^2 - \nu( q_{x}^{2} + q_{y}^{2}) - \gamma \alpha_{1}^{*} (i q_x - q_y ) - \kappa \alpha_{1}(i q_x +q_y) \right] \delta \alpha_{1}^{*}(-\mathbf{q})
	\end{split}
	\\
	\begin{split}
	\partial_t \delta \beta_{1}(\mathbf{q}) =& \left[ \frac{\partial \mu_1^B}{\partial a} \beta_{1} \right] \delta a(\mathbf{q}) +  \left[ \left( \frac{\partial \mu_1^B }{\partial b} - \frac{\partial \xi}{\partial b } |\beta_{1}|^2 \right) \beta_{1} - \frac{1}{2} ( i q_x - q_y ) \right] \delta b(\mathbf{q}) 
	\\
	&+ \left[ \mu_1^B - 2 \xi |\beta_{1}|^2 - \nu( q_{x}^{2} + q_{y}^{2}) - \gamma \alpha_{1} (i q_x + q_y ) - \kappa \beta_{1}^{*}(i q_x -q_y) \right] \delta \beta_{1}(\mathbf{q}) + \left[ - \xi \beta_{1}^2 \right] \delta \beta_{1}^{*}(-\mathbf{q})
	\end{split}
	\\
	\begin{split}
	\partial_t \delta \beta_{1}^{*}(-\mathbf{q}) =& \left[ \frac{\partial \mu_1^B}{\partial a } \beta_{1}^{*} \right] \delta a(\mathbf{q}) + \left[ \left( \frac{\partial \mu_1^B }{\partial b} - \frac{\partial \xi}{\partial b } |\beta_{1}|^2 \right) \beta_{1}^{*} - \frac{1}{2} ( i q_x + q_y ) \right] \delta b(\mathbf{q}) 
	\\
	&+ \left[ - \xi (\beta_{1}^{*})^2 \right] \delta \beta_{1}(\mathbf{q}) + \left[ \mu_1^B - 2 \xi |\beta_{1}|^2 - \nu( q_{x}^{2} + q_{y}^{2}) - \gamma \beta_{1}^{*} (i q_x - q_y ) - \kappa \beta_{1}(i q_x +q_y) \right] \delta \beta_{1}^{*}(-\mathbf{q})
	\end{split}
\end{align}
\end{subequations}
This system of linear equations for the Fourier modes  $\delta \mathbf{m}_{\mathbf{q}}  \,{=}\, (\delta a(\mathbf{q}), \delta b(\mathbf{q}), \delta \alpha_1 (\mathbf{q}), \delta \alpha_1^*(\mathbf{q}), \delta \beta_1 (\mathbf{q}), \delta \beta_1^*(\mathbf{q}) )^T$ can be written as 
\begin{equation}
\label{eq:JacobianAPP}
    \partial_{t} \, 
    \delta \mathbf{m}_{\mathbf{q}}  
    = 
    \mathbf{J}_{\mathbf{q}} \, 
    \delta \mathbf{m}_{\mathbf{q}} 
\end{equation}
where we introduced the Jacobian matrix $\mathbf{J}_{\mathbf{q}}$, which can be read off from Eq.~\eqref{eq:HydroMatrixSystem}. 
As an example, for the homogeneous, isotropic fixed point with $\alpha_{1} = \alpha_{1}^{*} = \beta_{1} = \beta_{1}^{*} = 0$, 
$a=\bar{\rho} /(1 + \lambda)$ and $b = \bar{\rho} \lambda/(1 + \lambda)$, $\mathbf{J}$ is given by:
\begin{equation}
	\mathbf{J}_{\mathbf{q}}=
	\begin{pmatrix}
	- \rho^2 \tau \lambda^{2}/(1+\lambda)^2 & \rho^2 \tau \lambda/(1+\lambda)^2 & -(iq_x+q_y)/2 & (q_y-iq_x)/2 & 0 & 0 \\
	 \rho^2 \tau \lambda^{2}/(1+\lambda)^2 & -\rho^2 \tau \lambda/(1+\lambda)^2 & 0 & 0 & -(iq_x+q_y)/2 & (q_y-iq_x)/2 \\
	-(iq_x-q_y)/2 & 0 & \mu_1^A- \nu q^2& 0 & 0 & 0  \\
	-(iq_x+q_y)/2 & 0 & 0& \mu_1^A- \nu q^2& 0 & 0  \\
	0 & -(iq_x-q_y)/2 & 0 & 0 & \mu_1^B- \nu q^2& 0  \\
	0 & -(iq_x+q_y)/2 & 0 & 0 & 0&  \mu_1^B- \nu q^2
	\end{pmatrix}
\end{equation}
\end{widetext}
with $q^2=q_{x}^2+q_{y}^2$. The eigenvalues of this matrix encode the stability of the fixed point: If for some vector $\mathbf{q}$, the Jacobian $\mathbf{J}_{\mathbf{q}}$ has an eigenvalue with positive real part, wave-like perturbations with the wave-vector $\mathbf{q}$ will grow exponentially, indicating the formation of patterns. On the other hand, if all eigenvalues have negative real part, for all $\mathbf{q}$, spatial perturbations decay and the spatially uniform state is (linearly) stable.

\subsection{\label{app:B_subsec_eigenvector}Velocity components of the eigenvector}

\begin{figure}[htb]
    \centering
    \includegraphics[width=0.6\columnwidth]{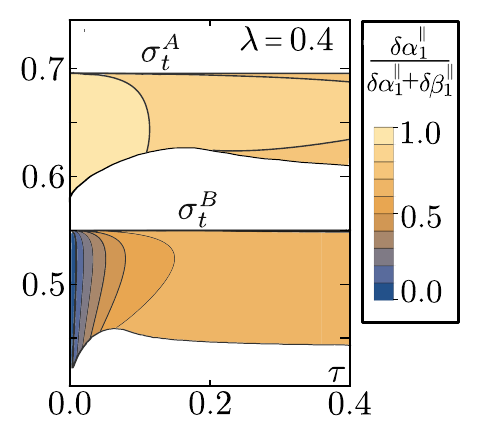}
    \caption{Stability analysis of the hydrodynamic equations in (\(\tau,\sigma\))-space. The color code measures how fast perturbations in the velocity field of species $A$ ($\delta \alpha_1$) grow compared to perturbations in the density of species $B$ ($\delta \beta_1$): $\delta \alpha_1/(\delta \alpha_1+\delta \beta_1)$ which ranges from $0$ (only perturbations $\delta \beta_1$ grow) to $1$ (only perturbations $\delta \alpha_1$ grow).}
    \label{fig:appStabilityAnalysisVelocities}
\end{figure}
In Sec.~\ref{subsec:StabilityAnalysis} we compute $\delta a/(\delta a+\delta b)$, as a function of $\tau$ (see Fig.~\ref{fig:StabilityAnalysis}c) in order to see how the snowdrift game affects which density perturbations grow in time. 
There are four more components to the eigenvector, $\delta\alpha_1$, $\delta\alpha_1^*$, $\delta\beta_1$ and  $\delta\beta_1^*$ , encoding the perturbations of the velocity fields. An analogue analysis as in Sec.\ref{subsec:StabilityAnalysis} for the velocity field components parallel to macroscopic order, $\delta\alpha_1^{\parallel}$ and $\delta\beta_1^{\parallel}$, yields the same phenomenology as for the densities and is shown in Fig.~\ref{fig:appStabilityAnalysisVelocities}: 
For high values of $\tau$, perturbations in the velocity fields of both species grow equally fast ($\delta\alpha_1^{\parallel}/(\delta\alpha_1^{\parallel}+\delta\beta_1^{\parallel})\rightarrow 0.5$). For $\tau \rightarrow 0$ on the other hand, perturbations $\delta\alpha_1^{\parallel}$ only grow in the unstable region in the vicinity of $\sigma_t^A$, while perturbations $\delta\beta_1^{\parallel}$ only grow in the unstable region at $\sigma_t^B$.
The components perpendicular to the macroscopic order, $\delta\alpha_1^{\perp}$ and $\delta\beta_1^{\perp}$, are always zero. Hence, perturbations of the velocity field components perpendicular to macroscopic order decay in time.

\section{\label{app:C_HydroNumericSimulation} Numerical methods - hydrodynamic simulation}

To investigate the full non-linear dynamics of the hydrodynamic Eqs.~(\ref{eq:Hydro_densities}) and Eqs.~(\ref{eq:Hydro_velocities}), we numerically implemented a finite-difference approximation (4th order Runge-Kutta method). Specifically, we model our system by a 2D-grid with periodic boundary conditions and replace all spatial derivatives by their finite difference approximations. 
\begin{align*}
f'(z) &\approx \frac{1}{2h} \left[ f(z+h) - f(z-h)\right] 
\\
f''(z) &\approx \frac{1}{h^2} \left[ f(z+h) -2 f(z) + f(z-h)\right]
\end{align*}
with $h$ a small number.
These approximations are the same for partial derivatives. In our case $f$ represents the density fields $a$ and $b$ or the velocity fields $\alpha_1$ and $\beta_1$ (then $f$ is a complex function) and $z$ is a space variable.
To solve our equations by finite-differences, we start by defining a square grid in space with grid points
\begin{align*}
x_i &= ih
\\
y_j &= jh , \qquad i,j \in {1,2,3,...,N}.
\end{align*}
The fields are then represented by time dependent $N \times N$ matrices 
\begin{align*}
a_{i,j} (t)&:= a (t,x_i,y_j) \\
b_{i,j} (t) &:= b (t,x_i,y_j)  \\
\alpha_{1;i,j} (t)&:= \alpha_1 (t,x_i,y_j) \\
\beta_{1;i,j} (t) &:= \beta_1 (t,x_i,y_j) 
\end{align*}
with complex entries in the case of $\alpha_1$ and $\beta_1$. Using the approximations for spatial derivatives we get the following replacement rules for the field equations
\begin{align*}
a(t,x,y) &\rightarrow a_{i,j}(t) \\
\partial_x a(t,x,y) &\rightarrow \frac{1}{2h} \left[ a_{i+1,j}(t) - a_{i-1,j}(t)\right] \\
\partial_y a(t,x,y) &\rightarrow \frac{1}{2h} \left[ a_{i,j+1}(t) - a_{i,j-1}(t)\right] \\
\partial_x^2 a(t,x,y) &\rightarrow \frac{1}{h^2} \left[ a_{i+1,j}(t) - 2 a_{i,j}(t) + a_{i-1,j}(t)\right] \\
\partial_y^2 a(t,x,y) &\rightarrow \frac{1}{h^2} \left[ a_{i,j+1}(t) - 2 a_{i,j}(t) + a_{i,j-1}(t)\right]
\end{align*}
and the corresponding rules for $b$, $\alpha_1$ and $\beta_1$. Overall we get $4 N^2$ ODEs (one ODE for every entry of the four $N\times N$ matrices) of the general form 
\begin{equation*}
    \partial_t a_{i,j} 
    = G(a_{k,l},b_{k,l},\alpha_{1;k,l},\beta_{1;k,l})
\end{equation*}
with $k,l \in {1,2,3,...,N}$.
We assume now that we know the state of the system at a time \(t=0\). To iteratively solve for later times, the 4th order Runge-Kutta (RK4) method worked very well for our system. The general concept of RK4 is the following: Consider the initial value problem 
\begin{equation*}
\dot{f} = F(f), \qquad f(t=0) = f_0
\end{equation*}
where the initial value \(f_0\) and the function \(F\) are given. As a first step we discretise the time
\begin{equation*}
t_n = \Delta t \, n, \qquad n \in {1,2,3, ...,M}
\end{equation*}
with \(\Delta t\) a small number. Hence, time dependent functions are disctretised as  \(f_n := f(t_n)\). The RK4 method now states that \(f_{n+1} = f(t_{n+1})\) can be approximated by
\begin{equation}
f_{n+1} =f_{n}+\frac{\Delta t}{6}\left(k_{1}+2 k_{2}+2 k_{3}+k_{4}\right)
\end{equation}
with
\begin{align*}
k_{1} &= F\left( f_{n}\right) \\
k_{2} &= F\left( f_{n}+\tfrac12 k_{1} \right) \\ 
k_{3} &= F\left( f_{n}+\tfrac12 k_{2}\right) \\ 
k_{4} &= F\left(f_{n}+k_{3}\right) .
\end{align*}
In our case, $f$ is an array of the four relevant fields
\begin{equation}
f_n = \left(\begin{array} {c}
{a_{n,i,j}} \\ {b_{n,i,j}} \\ {\alpha_{1;n,i,j}} \\ {\beta_{1;n,i,j}}
\end{array} \right) := 
\left(\begin{array} {c}
{a_{i,j}(t_n)} \\ {b_{i,j}(t_n)} \\{\alpha_{1;i,j}(t_n)} \\ {\beta_{1;i,j}(t_n)}
\end{array} \right)
\end{equation}
and $F = F(f)$ is given by the hydrodynamic equations (\ref{eq:Hydro_densities},\ref{eq:Hydro_velocities}). We chose initial conditions $({a_{0,i,j}}$, ${b_{0,i,j}}$, ${\alpha_{1;0,i,j}}$, ${\beta_{1;0,i,j}})$  for the fields at every grid point $(i,j)$ and  find the solution for $t= \Delta t$, $2 \Delta t$, $3 \Delta t$, ... via the RK4 method outlined above.

We chose periodic boundary conditions for our system. In order to discretise the fields we had to select the size of the 2D grid (the matrices) and the spacing between grid points, $h$. 
For our simulations we chose
\begin{equation*}
\mathrm{gridsize} = 160 \times 160, \quad h = 0.4, \quad \Delta t = 0.25*h^2\,.
\end{equation*}
A finer discretization in time and space did not change the outcome of the calculated stationary states of the system.
For the simulation results discussed in section~\ref{sec:Simulations_Hydro}, the simulation started at homogeneous fields $({a_{0,i,j}}$, ${b_{0,i,j}}$, ${\alpha_{1;0,i,j}}$, ${\beta_{1;0,i,j}})$ = ($\rho/2$, $\rho/2$, $0$ ,$ 0)$ for fixed control parameters $\rho=1$, $\sigma$, $\tau$, $\lambda$ with small spatial random fluctuations.

We tested the validity of our Runge-Kutta implementation by using COMSOL Multiphysics finite element analysis software and found very well agreement with our implementation.

\section{\label{app:D_Agent-Based} Agent-based simulations}

\subsection{\label{app:D_Agent-Based_Data}Implementation and parameters}

The agent-based simulation is designed, as described in Sec.~\ref{sec:Agent-based}, as a system of particles in a 2D-box of size $L \times L$ with periodic boundary conditions. 
Each of the $N$ particles in the system has a position $\mathbf{r}$, velocity $\mathbf{v}_0=v_0\mathbf{e}(\theta)$, species type $S=\{ A, B \}$, and fitness $f$.
The simulation is set up in discrete time steps and interactions are evaluated using a modified Verlet algorithm~\cite{BoxAlgo}. 
The update for each time step $dt$ is performed as follows:
First, the game theoretical interspecies interaction is evaluated and the particle species are updated according to Eq.~\eqref{eq:SwitchProb}.
In the next step, the intraspecies collisions are processed, the particle directions adjusted, and noise $\eta$ is added according to Eq.~\eqref{eq:DirUpdateAgentBased}. 
Then, the particle positions are updated according to Eq.~\eqref{eq:PosUpdateAgentBased}. 
As a last step, a snapshot of the system is  saved into a hdf5 database for later analysis. 
These steps are repeated $sim\_steps=T/dt$ times, where $T$ is the total time of the simulation. 
Where possible, these steps were implemented utilizing parallel programming.

Parameters used for all shown data are given by $dt=1$, $L=256$, $N=98304$, $d_{\text{align}}=1$, $v_0=0.5$ (except for Fig.~\ref{fig:r_delta_vs_dgame}: $v_0=0.015$), $T=10^5$ and Table~\ref{tab:ParametersAgentBased}.

\begin{table}
\caption{\label{tab:ParametersAgentBased}Parameters of the agent-based simulation. The notation $[x,y,z]$ indicates the parameter ranging from $x$ to $y$ with a step-size of $z$, e.g. $[0.3,0.6,0.1]$ denotes the set $\{0.3,0.4,0.5,0.6\}$.}
\begin{ruledtabular}
\begin{tabular}{ccccc}
  Figure/Movie & $\lambda$ & $\sigma$ &  $d_{\text{game}}$\\
  Fig.~\ref{fig:AgentBasedPhaseDiag}a & $[0.2,0.32,0.1]$ & $[0.05,0.8,0.01]$ & 5\\
  Fig.~\ref{fig:AgentBasedPhaseDiag}b & $[0.2,0.32,0.1]$ & $[0.05,0.8,0.01]$ & 10 \\
  Fig.~\ref{fig:WaveBandsAndProfile} & 0.26 & 0.32 & 5 \\ 
  Fig.~\ref{fig:5_phases_all}a & 0.23 & 0.71 & 10 \\
  Fig.~\ref{fig:5_phases_all}b, movie2 (TWA) & 0.23 & 0.65 & 10 \\
  Fig.~\ref{fig:5_phases_all}c & 0.23 & 0.56 & 10 \\
  Fig.~\ref{fig:5_phases_all}d, movie3 (TWB) & 0.23 & 0.40 & 10 \\
  Fig.~\ref{fig:5_phases_all}e, movie4 (ordered) & 0.23 & 0.17& 10 \\
  Fig.~\ref{fig:r_delta_vs_dgame}a & 0.4 & 0.5 & 10 \\
  Fig.~\ref{fig:r_delta_vs_dgame}b & $\{0.2,0.4,0.5\}$ & 0.5 & 10 
\end{tabular}
\end{ruledtabular}
\end{table}

The hdf5 database generated by the simulation was analyzed using Python scripts; the movies were generated with ffmpeg.

\subsection{\label{app:D_Agent-Based_nu}Detailed explanation of the velocity variation parameter}

\begin{figure}[ht]
    \centering
    \includegraphics{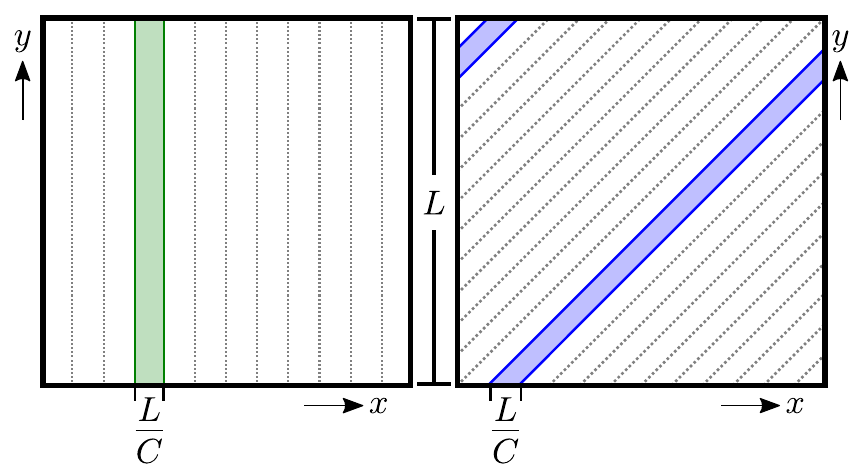}
    \caption{Examples of different channel subdivisions. Systems of size $L \times L$ with channels perpendicular to the $x$-axis (left) and diagonal (right) are shown. Due to periodic boundary conditions channels not aligned with either $x$- or $y$-axis may be split in two parts. Channels always have area $L^2/C$. }
    \label{fig:channels}
\end{figure}

In order to identify polar wave patterns in the system, we introduce the velocity variation parameter $\nu$ (Eq.~\eqref{eq:velocity_variation}). 
We therefore choose to divide the system into $C$ channels along the average direction of particle propagation and perpendicular to it (see Fig.~\ref{fig:channels}).

Each particle in the system is assigned to the corresponding channel based on its location. 
The channel densities $a_{c}^{t}$ and $b_{c}^{t}$ are calculated at time $t$ for each channel $c$ as well as the corresponding channel polar order parameter $\alpha_{1,c}^{t}$ and $\beta_{1,c}^{t}$.
The velocity variation $nu_{A}^t$ is then defined as twice the standard deviation of polar order parameter $\alpha_{1,c}^t$ over all channels: 
\begin{equation}
	\nu_{A}^t=2\sqrt{\frac{1}{C}\sum_{c=1}^{C} \left( \alpha_{1,c}^t -  \langle \alpha_{1}^t \rangle \right)^2}.
\end{equation} 
The factor of $2$ is introduced such that $\nu_{A}^t \in [0,1]$. 
The velocity variation $\nu_{B}^t$ for species $B$ is defined accordingly. 
The temporal averages are denoted by $\langle \nu_{A} \rangle$ and $\langle \nu_{B} \rangle$.

If waves are present in the system, the difference between $\langle \nu_{A,\perp} \rangle$
and $\langle \nu_{A,\parallel} \rangle$ for channels perpendicular and parallel to the average direction of motion is nonzero, while for systems without spatial structure they cancel each other.
We denote this by 
\begin{equation}
	\Delta \nu_{A}=\langle \nu_{A,\perp} \rangle - \langle \nu_{A,\parallel} \rangle .
\end{equation} 

The channel subdivisions and corresponding densities $a_{c}^{t}$ and $b_{c}^{t}$ and polar order parameters $\alpha_{1,c}^{t}$ and $\beta_{1,c}^{t}$ are also used to generate the wave profile plots such as Fig.~\ref{fig:WaveBandsAndProfile}b.
 
\subsection{Five phases}
\label{app:D_Agent-Based_5phases}

For small enough values of relative fitness $\lambda$, we observe the five distinct phases (disorder in Fig.~\ref{fig:5_phases_all}a, TWA (see Fig.~\ref{fig:5_phases_all}b), partial order (only one species polar ordered, Fig.~\ref{fig:5_phases_all}c), TWB  (Fig.~\ref{fig:5_phases_all}d), and full polar order (Fig.~\ref{fig:5_phases_all}e) in the agent-based simulations. 
The travelling wave phases induced by species $A$ and $B$ (TWA and TWB), as well as the phase of full polar order are shown in the attached movies (see Section~\ref{app:E_Movies}).

\subsection{Size and consequences of the demixing patterns}
\label{app:D_Agent-Based_Demixing}
A distinct feature of the agent-based simulation is the formation of demixing patterns in environments with slow relative particle movement, compared to the game strength $\tau$. 
These patterns are a consequence of the range-based interactions in the implementation of the snowdrift game and can also occur in the absence of an alignment interaction, as long as the movement of particles is slow ($v_0/\tau\ll 1$, s.t. the number of games played by the same set of agents in a row is much larger than one.)
This can be seen from the demixing patterns in Fig.~\ref{fig:r_delta_vs_dgame}a, which shows a snapshot of a system where instead of self-propelled particles we simulated particles that show slow diffusive motion.
The size of these patterns is proportional to the game range $d_\text{game}$ (see Fig.~\ref{fig:r_delta_vs_dgame}b). On the other hand, the spatially averaged densities of A and B do not depend on the game range $d_\text{game}$.
For simulations with alignment interaction, the system can also display demixing patterns in the form of bands (see Fig.~\ref{fig:5_phases_all}d).

\begin{widetext}

\begin{figure}[htb]
    \centering
    \includegraphics{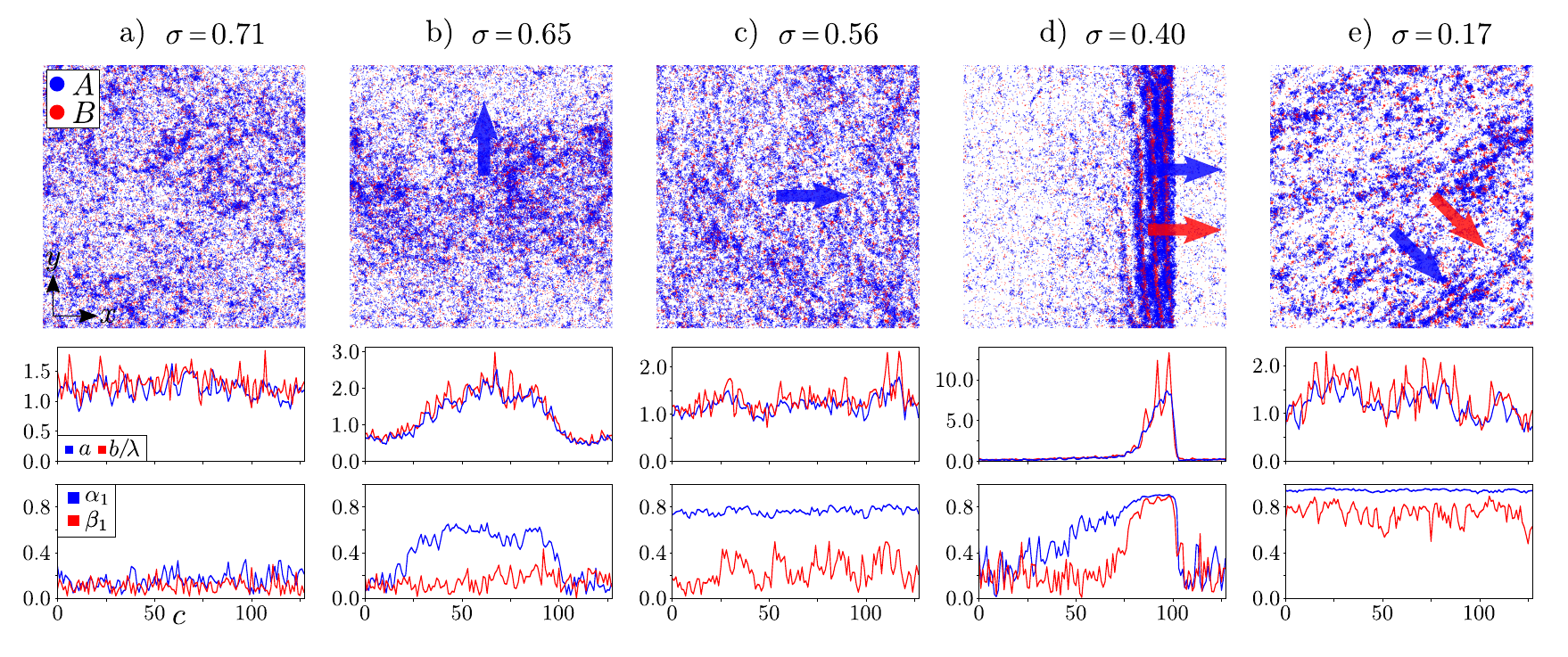}
    \caption{
    Snapshots (top row), density (middle row) and polar order (bottom row) of a set of different noise strength $\sigma$ as indicated in the graph; the relative fitness is fixed to $\lambda  \,{=}\, 0.23$ and the interaction range to $d_\text{game}  \,{=}\, 10$. (Remaining parameters can be found in Table~\ref{tab:ParametersAgentBased}.)
    The profiles for the densities and the polar order are displayed along the direction of polar order as described in Appendix~\ref{app:D_Agent-Based_nu}. 
    In the density plot (middle row), the density of species B is adjusted by a factor of inverse relative fitness $\lambda^{-1}$ to emphasize the relation between the snowdrift game and local densities of both species; note that at the mean-field level one has $a  \,{=}\, b / \lambda$.
    (a) \textit{Disordered phase} ($\sigma  \,{=}\, 0.71$): The system neither shows any spatial patterns (middle panel) nor  polar order (bottom panel).
    (b) \textit{TWA phase} ($\sigma \,{=}\, 0.65$): Species A forms polar ordered waves (direction of order indicated by blue arrow). Species B forms a density wave mirroring species A but does not show polar order.
    (c) \textit{Uniform polar order of species A }$  (\sigma \,{=}\, 0.56$): The system is spatially uniform with no spatial  patterns visible. Species A shows uniform polar order, while species B is still disordered.
    (d)  \textit{TWB phase} ($\sigma \,{=}\, 0.40$): Both species A and B show polar waves (direction of order indicated for each species by an arrow in its respective color). While species B only shows order inside the wave, species A still has order in the wake of the wave, evident by the high polar observable in the region behind the wave. 
    (e) \textit{Uniform polar order} ($\sigma \,{=}\, 0.17$): Both species show polar order over the whole system, no wave pattern is visible.
    }
    \label{fig:5_phases_all}
\end{figure}

\begin{figure}[!h]
\centering
\includegraphics{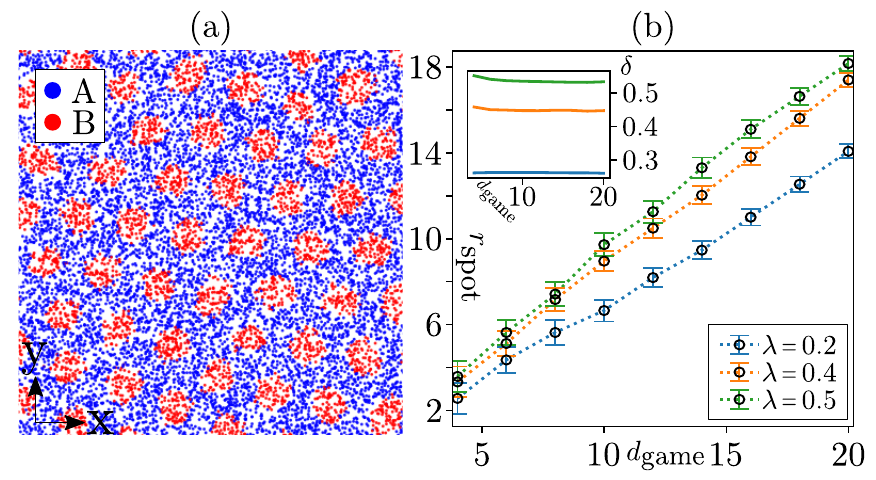}
    \caption{Size dependence of demixing patterns on $d_\text{game}$. 
    (a) Snapshot of a system with slow diffusive motion ($v_0/\tau\ll 1$) [$\lambda \,{=}\, 0.4$, $d_\text{game}=10$, $v_0 \,{=}\, 0.015$].
    Demixing into clusters of species B surrounded by species A is clearly visible. 
    (b) Average radius of spots $r_\text{spot}$ as a function of $d_\text{game}$ for different values of relative fitness $\lambda$.
    While the dependence is clearly linear, for larger values of relative fitness $\lambda$, the radius $r_\text{spot}$ grows faster for increasing game range $d_\text{game}$.
    Inset: independence of the ratio between species $\delta=b/a$ on $d_\text{game}.$
    }
    \label{fig:r_delta_vs_dgame}
\end{figure}

\end{widetext}

\cleardoublepage\newpage

\section{Movie Descriptions}
\label{app:E_Movies}

\textbf{movie1.mp4:} Simulations of the hydrodynamic equations show the formation of spatial patterns. These spatial patterns are characterised by density-segregated polar ordered bands in both species that move concurrently through the system. When viewed from a comoving frame, the patterns are uniform in the direction perpendicular to the wave vector. The movie shows the profiles of the two density fields $a$ and $b$ over time. Control parameters are set to $\sigma = 0.68$, $\lambda = 0.4$ and $\tau = 1.0$, such that the system is in its traveling wave phase $A$ (TWA). 

Movies 2-4 show simulations of the agent-based approach. Parameter values for movie2-movie4 are given in Table~\ref{tab:ParametersAgentBased}.

\textbf{movie2.mp4:} The formation of a traveling wave phase induced by species $A$ (TWA). The wave is clearly visible as a broad density pattern and travels in positive y-direction. The dominance of species $A$ can be inferred from the relative abundance of species $A$ compared to $B$. Only species A shows polar order inside the wave.

\textbf{movie3.mp4:} The formation of a traveling wave phase induced by species $B$ (TWB). The sharply contrasted wave travels in positive x-direction with polar order in both species. Species $B$ is in a parameter regime where it would display wave pattern on its own. Species $A$ on the other hand would be uniformly ordered in this parameter regime and the observed wave patterns are induced by the snowdrift interaction between the species. Furthermore there is clearly visible demixing, as the species form alternating bands inside the wave due to the finite range of the snowdrift interaction.

\textbf{movie4.mp4:} During the formation of order the two species initially cluster together and start to display uniform polar order in both species $A$ and $B$. The species then spread out over the whole system while maintaining high polar order. In the end small clusters with internal demixing can be observed moving thorugh the system parallel to each other with high polar order. 

\cleardoublepage
\newpage


\begin{thebibliography}{61}%
\makeatletter
\providecommand \@ifxundefined [1]{%
 \@ifx{#1\undefined}
}%
\providecommand \@ifnum [1]{%
 \ifnum #1\expandafter \@firstoftwo
 \else \expandafter \@secondoftwo
 \fi
}%
\providecommand \@ifx [1]{%
 \ifx #1\expandafter \@firstoftwo
 \else \expandafter \@secondoftwo
 \fi
}%
\providecommand \natexlab [1]{#1}%
\providecommand \enquote  [1]{``#1''}%
\providecommand \bibnamefont  [1]{#1}%
\providecommand \bibfnamefont [1]{#1}%
\providecommand \citenamefont [1]{#1}%
\providecommand \href@noop [0]{\@secondoftwo}%
\providecommand \href [0]{\begingroup \@sanitize@url \@href}%
\providecommand \@href[1]{\@@startlink{#1}\@@href}%
\providecommand \@@href[1]{\endgroup#1\@@endlink}%
\providecommand \@sanitize@url [0]{\catcode `\\12\catcode `\$12\catcode
  `\&12\catcode `\#12\catcode `\^12\catcode `\_12\catcode `\%12\relax}%
\providecommand \@@startlink[1]{}%
\providecommand \@@endlink[0]{}%
\providecommand \url  [0]{\begingroup\@sanitize@url \@url }%
\providecommand \@url [1]{\endgroup\@href {#1}{\urlprefix }}%
\providecommand \urlprefix  [0]{URL }%
\providecommand \Eprint [0]{\href }%
\providecommand \doibase [0]{https://doi.org/}%
\providecommand \selectlanguage [0]{\@gobble}%
\providecommand \bibinfo  [0]{\@secondoftwo}%
\providecommand \bibfield  [0]{\@secondoftwo}%
\providecommand \translation [1]{[#1]}%
\providecommand \BibitemOpen [0]{}%
\providecommand \bibitemStop [0]{}%
\providecommand \bibitemNoStop [0]{.\EOS\space}%
\providecommand \EOS [0]{\spacefactor3000\relax}%
\providecommand \BibitemShut  [1]{\csname bibitem#1\endcsname}%
\let\auto@bib@innerbib\@empty
\bibitem [{\citenamefont {Cremer}\ \emph
  {et~al.}(2019{\natexlab{a}})\citenamefont {Cremer}, \citenamefont
  {Melbinger}, \citenamefont {Wienand}, \citenamefont {Henri­quez},
  \citenamefont {Jung},\ and\ \citenamefont {Frey}}]{CremerFrey19}%
  \BibitemOpen
  \bibfield  {author} {\bibinfo {author} {\bibfnamefont {J.}~\bibnamefont
  {Cremer}}, \bibinfo {author} {\bibfnamefont {A.}~\bibnamefont {Melbinger}},
  \bibinfo {author} {\bibfnamefont {K.}~\bibnamefont {Wienand}}, \bibinfo
  {author} {\bibfnamefont {T.}~\bibnamefont {Henriquez}}, \bibinfo {author}
  {\bibfnamefont {H.}~\bibnamefont {Jung}},\ and\ \bibinfo {author}
  {\bibfnamefont {E.}~\bibnamefont {Frey}},\ }\bibfield  {title} {\bibinfo
  {title} {Cooperation in microbial populations: Theory and experimental model
  systems},\ }\href {10.1016/j.jmb.2019.09.023} {\bibfield  {journal} {\bibinfo
   {journal} {Journal of Molecular Biology}\ }\textbf {\bibinfo {volume} {431}}
  (\bibinfo {year} {2019}{\natexlab{a}})}\BibitemShut {NoStop}%
\bibitem [{\citenamefont {Dobramysl}\ \emph {et~al.}(2018)\citenamefont
  {Dobramysl}, \citenamefont {Mobilia}, \citenamefont {Pleimling},\ and\
  \citenamefont {T\"auber}}]{Dobramysl:2018}%
  \BibitemOpen
  \bibfield  {author} {\bibinfo {author} {\bibfnamefont {U.}~\bibnamefont
  {Dobramysl}}, \bibinfo {author} {\bibfnamefont {M.}~\bibnamefont {Mobilia}},
  \bibinfo {author} {\bibfnamefont {M.}~\bibnamefont {Pleimling}},\ and\
  \bibinfo {author} {\bibfnamefont {U.~C.}\ \bibnamefont {T\"auber}},\
  }\bibfield  {title} {\bibinfo {title} {{Stochastic population dynamics in
  spatially extended predator-prey systems}},\ }\href
  {https://doi.org/10.1088/1751-8121/aa95c7} {\bibfield  {journal} {\bibinfo
  {journal} {Journal of Physics A: Mathematical and Theoretical}\ }\textbf
  {\bibinfo {volume} {51}},\ \bibinfo {pages} {063001} (\bibinfo {year}
  {2018})}\BibitemShut {NoStop}%
\bibitem [{\citenamefont {Friedman}\ and\ \citenamefont
  {Gore}(2017)}]{friedman_ecological_2017}%
  \BibitemOpen
  \bibfield  {author} {\bibinfo {author} {\bibfnamefont {J.}~\bibnamefont
  {Friedman}}\ and\ \bibinfo {author} {\bibfnamefont {J.}~\bibnamefont
  {Gore}},\ }\bibfield  {title} {\bibinfo {title} {Ecological systems biology:
  The dynamics of interacting populations},\ }\href
  {https://doi.org/10.1016/j.coisb.2016.12.001} {\bibfield  {journal} {\bibinfo
   {journal} {Current Opinion in Systems Biology}\ }\textbf {\bibinfo {volume}
  {1}},\ \bibinfo {pages} {114} (\bibinfo {year} {2017})}\BibitemShut {NoStop}%
\bibitem [{\citenamefont {{Maynard-Smith}}(1982)}]{Maynard}%
  \BibitemOpen
  \bibfield  {author} {\bibinfo {author} {\bibfnamefont {J.}~\bibnamefont
  {{Maynard-Smith}}},\ }\href@noop {} {\textit {\bibinfo {title} {Evolution and
  the Theory of Games}}}\ (\bibinfo  {publisher} {Cambridge University Press},\
  \bibinfo {address} {Cambridge},\ \bibinfo {year} {1982})\BibitemShut
  {NoStop}%
\bibitem [{\citenamefont {Hofbauer}\ and\ \citenamefont
  {Sigmund}()}]{Book:HofbauerSigmund98}%
  \BibitemOpen
  \bibfield  {author} {\bibinfo {author} {\bibfnamefont {J.}~\bibnamefont
  {Hofbauer}}\ and\ \bibinfo {author} {\bibfnamefont {K.}~\bibnamefont
  {Sigmund}},\ }\href@noop {} {\textit {\bibinfo {title} {{Evolutionary Games and
  Population Dynamics}}}}\ (\bibinfo  {publisher} {Cambridge University Press},\
  \bibinfo {address} {Cambridge},\ \bibinfo {year} {1998}) \BibitemShut {NoStop}%
\bibitem [{\citenamefont {Frey}(2010)}]{Frey.2010}%
  \BibitemOpen
  \bibfield  {author} {\bibinfo {author} {\bibfnamefont {E.}~\bibnamefont
  {Frey}},\ }\bibfield  {title} {\bibinfo {title} {{Evolutionary game theory:
  Theoretical concepts and applications to microbial communities}},\ }\href
  {10.1016/j.physa.2010.02.047} {\bibfield  {journal} {\bibinfo  {journal}
  {Physica A: Statistical Mechanics and its Applications}\ }\textbf {\bibinfo
  {volume} {389}},\ \bibinfo {pages} {4265} (\bibinfo {year}
  {2010})}\BibitemShut {NoStop}%
\bibitem [{\citenamefont {Reichenbach}\ \emph {et~al.}(2007)\citenamefont
  {Reichenbach}, \citenamefont {Mobilia},\ and\ \citenamefont
  {Frey}}]{Reichenbach.Frey.2007}%
  \BibitemOpen
  \bibfield  {author} {\bibinfo {author} {\bibfnamefont {T.}~\bibnamefont
  {Reichenbach}}, \bibinfo {author} {\bibfnamefont {M.}~\bibnamefont
  {Mobilia}},\ and\ \bibinfo {author} {\bibfnamefont {E.}~\bibnamefont
  {Frey}},\ }\bibfield  {title} {\bibinfo {title} {{Mobility promotes and
  jeopardizes biodiversity in rock-paper-scissors games}},\ }\href
  {https://doi.org/10.1038/nature06095} {\bibfield  {journal} {\bibinfo
  {journal} {Nature}\ }\textbf {\bibinfo {volume} {448}},\ \bibinfo {pages}
  {1046} (\bibinfo {year} {2007})},\ \Eprint {https://arxiv.org/abs/0709.0217}
  {0709.0217} \BibitemShut {NoStop}%
\bibitem [{\citenamefont {Gelimson}\ \emph {et~al.}(2013)\citenamefont
  {Gelimson}, \citenamefont {Cremer},\ and\ \citenamefont
  {Frey}}]{Gelimson.Frey.2013}%
  \BibitemOpen
  \bibfield  {author} {\bibinfo {author} {\bibfnamefont {A.}~\bibnamefont
  {Gelimson}}, \bibinfo {author} {\bibfnamefont {J.}~\bibnamefont {Cremer}},\
  and\ \bibinfo {author} {\bibfnamefont {E.}~\bibnamefont {Frey}},\ }\bibfield
  {title} {\bibinfo {title} {Mobility, fitness collection, and the breakdown of
  cooperation},\ }\href {10.1103/physreve.87.042711} {\bibfield  {journal}
  {\bibinfo  {journal} {Physical Review E}\ }\textbf {\bibinfo {volume} {87}},\
  \bibinfo {pages} {042711} (\bibinfo {year} {2013})}\BibitemShut {NoStop}%
\bibitem [{\citenamefont {Mobilia}\ \emph {et~al.}(2006)\citenamefont
  {Mobilia}, \citenamefont {Georgiev},\ and\ \citenamefont
  {T\"auber}}]{Mobilia:2006}%
  \BibitemOpen
  \bibfield  {author} {\bibinfo {author} {\bibfnamefont {M.}~\bibnamefont
  {Mobilia}}, \bibinfo {author} {\bibfnamefont {I.~T.}\ \bibnamefont
  {Georgiev}},\ and\ \bibinfo {author} {\bibfnamefont {U.~C.}\ \bibnamefont
  {T\"auber}},\ }\bibfield  {title} {\bibinfo {title} {Phase transitions and
  spatio-temporal fluctuations in stochastic lattice lotka-volterra models},\
  }\href {http://dx.doi.org/10.1007/s10955-006-9146-3} {\bibfield  {journal}
  {\bibinfo  {journal} {Journal of Statistical Physics}\ }\textbf {\bibinfo
  {volume} {128}},\ \bibinfo {pages} {447} (\bibinfo {year}
  {2006})}\BibitemShut {NoStop}%
\bibitem [{\citenamefont {Nowak}\ and\ \citenamefont
  {May}(1992)}]{Nowak.May.1992}%
  \BibitemOpen
  \bibfield  {author} {\bibinfo {author} {\bibfnamefont {M.~A.}\ \bibnamefont
  {Nowak}}\ and\ \bibinfo {author} {\bibfnamefont {R.~M.}\ \bibnamefont
  {May}},\ }\bibfield  {title} {\bibinfo {title} {Evolutionary games and
  spatial chaos},\ }\href {https://doi.org/10.1038/359826a0} {\bibfield
  {journal} {\bibinfo  {journal} {Nature}\ }\textbf {\bibinfo {volume} {359}},\
  \bibinfo {pages} {826} (\bibinfo {year} {1992})}\BibitemShut {NoStop}%
\bibitem [{\citenamefont {Nowak}\ \emph {et~al.}(1994)\citenamefont {Nowak},
  \citenamefont {Bonhoeffer},\ and\ \citenamefont {May}}]{Nowak.May.1994}%
  \BibitemOpen
  \bibfield  {author} {\bibinfo {author} {\bibfnamefont {M.~A.}\ \bibnamefont
  {Nowak}}, \bibinfo {author} {\bibfnamefont {S.}~\bibnamefont {Bonhoeffer}},\
  and\ \bibinfo {author} {\bibfnamefont {R.~M.}\ \bibnamefont {May}},\
  }\bibfield  {title} {\bibinfo {title} {{Spatial games and the maintenance of
  cooperation}},\ }\href {https://doi.org/10.1073/pnas.91.11.4877} {\bibfield
  {journal} {\bibinfo  {journal} {Proceedings of the National Academy of
  Sciences}\ }\textbf {\bibinfo {volume} {91}},\ \bibinfo {pages} {4877}
  (\bibinfo {year} {1994})}\BibitemShut {NoStop}%
\bibitem [{\citenamefont {Nakamuru}\ \emph {et~al.}(1997)\citenamefont
  {Nakamuru}, \citenamefont {Matsuda},\ and\ \citenamefont
  {Iwasa}}]{Nakamuru:1997}%
  \BibitemOpen
  \bibfield  {author} {\bibinfo {author} {\bibfnamefont {M.}~\bibnamefont
  {Nakamuru}}, \bibinfo {author} {\bibfnamefont {H.}~\bibnamefont {Matsuda}},\
  and\ \bibinfo {author} {\bibfnamefont {Y.}~\bibnamefont {Iwasa}},\ }\bibfield
   {title} {\bibinfo {title} {The evolution of cooperation in a
  lattice-structured population},\ }\href
  {https://doi.org/10.1006/jtbi.1996.0243} {\bibfield  {journal} {\bibinfo
  {journal} {Journal of Theoretical Biology}\ }\textbf {\bibinfo {volume}
  {184}},\ \bibinfo {pages} {65} (\bibinfo {year} {1997})}\BibitemShut
  {NoStop}%
\bibitem [{\citenamefont {Helbing}\ and\ \citenamefont
  {Yu}(2008)}]{Helbing:2008}%
  \BibitemOpen
  \bibfield  {author} {\bibinfo {author} {\bibfnamefont {D.}~\bibnamefont
  {Helbing}}\ and\ \bibinfo {author} {\bibfnamefont {W.}~\bibnamefont {Yu}},\
  }\bibfield  {title} {\bibinfo {title} {Migration as a mechanism to promote
  cooperation},\ }\href {https://doi.org/10.1142/s0219525908001866} {\bibfield
  {journal} {\bibinfo  {journal} {Adv. Compl. Syst.}\ }\textbf {\bibinfo
  {volume} {11}},\ \bibinfo {pages} {641} (\bibinfo {year} {2008})}\BibitemShut
  {NoStop}%
\bibitem [{\citenamefont {Szab\'o}\ and\ \citenamefont
  {Hauert}(2002)}]{Szabo:2002}%
  \BibitemOpen
  \bibfield  {author} {\bibinfo {author} {\bibfnamefont {G.}~\bibnamefont
  {Szab\'o}}\ and\ \bibinfo {author} {\bibfnamefont {C.}~\bibnamefont
  {Hauert}},\ }\bibfield  {title} {\bibinfo {title} {Phase transitions and
  volunteering in spatial public goods games},\ }\href
  {https://doi.org/10.1103/physrevlett.89.118101} {\bibfield  {journal}
  {\bibinfo  {journal} {Physical Review Letters}\ }\textbf {\bibinfo {volume}
  {89}},\ \bibinfo {pages} {118101} (\bibinfo {year} {2002})}\BibitemShut
  {NoStop}%
\bibitem [{\citenamefont {Ramaswamy}(2010)}]{RamaswamyReview10}%
  \BibitemOpen
  \bibfield  {author} {\bibinfo {author} {\bibfnamefont {S.}~\bibnamefont
  {Ramaswamy}},\ }\bibfield  {title} {\bibinfo {title} {The mechanics and
  statistics of active matter},\ }\href
  {https://doi.org/10.1146/annurev-conmatphys-070909-104101} {\bibfield
  {journal} {\bibinfo  {journal} {Annual Review of Condensed Matter Physics}\
  }\textbf {\bibinfo {volume} {1}},\ \bibinfo {pages} {323} (\bibinfo {year}
  {2010})}\BibitemShut {NoStop}%
\bibitem [{\citenamefont {Vicsek}\ and\ \citenamefont
  {Zafeiris}(2012)}]{VicsekReview12}%
  \BibitemOpen
  \bibfield  {author} {\bibinfo {author} {\bibfnamefont {T.}~\bibnamefont
  {Vicsek}}\ and\ \bibinfo {author} {\bibfnamefont {A.}~\bibnamefont
  {Zafeiris}},\ }\bibfield  {title} {\bibinfo {title} {Collective motion},\
  }\href {10.1016/j.physrep.2012.03.004} {\bibfield  {journal} {\bibinfo
  {journal} {Physics Reports}\ }\textbf {\bibinfo {volume} {517}},\ \bibinfo
  {pages} {71 } (\bibinfo {year} {2012})}\BibitemShut {NoStop}%
\bibitem [{\citenamefont {Marchetti}\ \emph {et~al.}(2013)\citenamefont
  {Marchetti}, \citenamefont {Joanny}, \citenamefont {Ramaswamy}, \citenamefont
  {Liverpool}, \citenamefont {Prost}, \citenamefont {Rao},\ and\ \citenamefont
  {Simha}}]{MarchettiReview12}%
  \BibitemOpen
  \bibfield  {author} {\bibinfo {author} {\bibfnamefont {M.~C.}\ \bibnamefont
  {Marchetti}}, \bibinfo {author} {\bibfnamefont {J.~F.}\ \bibnamefont
  {Joanny}}, \bibinfo {author} {\bibfnamefont {S.}~\bibnamefont {Ramaswamy}},
  \bibinfo {author} {\bibfnamefont {T.~B.}\ \bibnamefont {Liverpool}}, \bibinfo
  {author} {\bibfnamefont {J.}~\bibnamefont {Prost}}, \bibinfo {author}
  {\bibfnamefont {M.}~\bibnamefont {Rao}},\ and\ \bibinfo {author}
  {\bibfnamefont {R.~A.}\ \bibnamefont {Simha}},\ }\bibfield  {title} {\bibinfo
  {title} {Hydrodynamics of soft active matter},\ }\href
  {10.1103/RevModPhys.85.1143} {\bibfield  {journal} {\bibinfo  {journal} {Rev.
  Mod. Phys.}\ }\textbf {\bibinfo {volume} {85}},\ \bibinfo {pages} {1143}
  (\bibinfo {year} {2013})}\BibitemShut {NoStop}%
\bibitem [{\citenamefont {Cates}\ and\ \citenamefont
  {Tailleur}(2014)}]{Cates_Tailleur:2015}%
  \BibitemOpen
  \bibfield  {author} {\bibinfo {author} {\bibfnamefont {M.~E.}\ \bibnamefont
  {Cates}}\ and\ \bibinfo {author} {\bibfnamefont {J.}~\bibnamefont
  {Tailleur}},\ }\bibfield  {title} {\bibinfo {title} {{Motility-Induced Phase
  Separation}},\ }\href
  {https://doi.org/10.1146/annurev-conmatphys-031214-014710} {\bibfield
  {journal} {\bibinfo  {journal} {Annual Review of Condensed Matter Physics}\
  }\textbf {\bibinfo {volume} {6}},\ \bibinfo {pages} {1} (\bibinfo {year}
  {2014})}\BibitemShut {NoStop}%
\bibitem [{\citenamefont {Chat{\'e}}(2020)}]{ChateReview20}%
  \BibitemOpen
  \bibfield  {author} {\bibinfo {author} {\bibfnamefont {H.}~\bibnamefont
  {Chat{\'e}}},\ }\bibfield  {title} {\bibinfo {title} {Dry aligning dilute
  active matter},\ }\href
  {https://doi.org/10.1146/annurev-conmatphys-031119-050752} {\bibfield
  {journal} {\bibinfo  {journal} {Annual Review of Condensed Matter Physics}\
  }\textbf {\bibinfo {volume} {11}},\ \bibinfo {pages} {189} (\bibinfo {year}
  {2020})}\BibitemShut {NoStop}%
\bibitem [{\citenamefont {Osborne}(2004)}]{Osborne_game_theory}%
  \BibitemOpen
  \bibfield  {author} {\bibinfo {author} {\bibfnamefont {M.}~\bibnamefont
  {Osborne}},\ }\href@noop {} {\emph {\bibinfo {title} {An Introduction to Game
  Theory}}}\ (\bibinfo  {publisher} {Oxford University Press},\ \bibinfo {year}
  {2004})\BibitemShut {NoStop}%
\bibitem [{\citenamefont {Gore}\ \emph {et~al.}(2009)\citenamefont {Gore},
  \citenamefont {Youk},\ and\ \citenamefont {Oudenaarden}}]{Gore09}%
  \BibitemOpen
  \bibfield  {author} {\bibinfo {author} {\bibfnamefont {J.}~\bibnamefont
  {Gore}}, \bibinfo {author} {\bibfnamefont {H.}~\bibnamefont {Youk}},\ and\
  \bibinfo {author} {\bibfnamefont {A.}~\bibnamefont {Oudenaarden}},\
  }\bibfield  {title} {\bibinfo {title} {Snowdrift game dynamics and
  facultative cheating in yeast},\ }\href {https://doi.org/10.1038/nature07921}
  {\bibfield  {journal} {\bibinfo  {journal} {Nature}\ }\textbf {\bibinfo
  {volume} {459}},\ \bibinfo {pages} {253} (\bibinfo {year}
  {2009})}\BibitemShut {NoStop}%
\bibitem [{\citenamefont {Korolev}\ \emph {et~al.}(2010)\citenamefont
  {Korolev}, \citenamefont {Avlund}, \citenamefont {Hallatschek},\ and\
  \citenamefont {Nelson}}]{Korolev:2010}%
  \BibitemOpen
  \bibfield  {author} {\bibinfo {author} {\bibfnamefont {K.~S.}\ \bibnamefont
  {Korolev}}, \bibinfo {author} {\bibfnamefont {M.}~\bibnamefont {Avlund}},
  \bibinfo {author} {\bibfnamefont {O.}~\bibnamefont {Hallatschek}},\ and\
  \bibinfo {author} {\bibfnamefont {D.~R.}\ \bibnamefont {Nelson}},\ }\bibfield
   {title} {\bibinfo {title} {{Genetic demixing and evolution in linear
  stepping stone models}},\ }\href {https://doi.org/10.1103/revmodphys.82.1691}
  {\bibfield  {journal} {\bibinfo  {journal} {Reviews of Modern Physics}\
  }\textbf {\bibinfo {volume} {82}},\ \bibinfo {pages} {1691} (\bibinfo {year}
  {2010})}\BibitemShut {NoStop}%
\bibitem [{\citenamefont {Drescher}\ \emph {et~al.}(2014)\citenamefont
  {Drescher}, \citenamefont {Nadell}, \citenamefont {Stone}, \citenamefont
  {Wingreen},\ and\ \citenamefont {Bassler}}]{Drescher:2014}%
  \BibitemOpen
  \bibfield  {author} {\bibinfo {author} {\bibfnamefont {K.}~\bibnamefont
  {Drescher}}, \bibinfo {author} {\bibfnamefont {C.}~\bibnamefont {Nadell}},
  \bibinfo {author} {\bibfnamefont {H.}~\bibnamefont {Stone}}, \bibinfo
  {author} {\bibfnamefont {N.}~\bibnamefont {Wingreen}},\ and\ \bibinfo
  {author} {\bibfnamefont {B.}~\bibnamefont {Bassler}},\ }\bibfield  {title}
  {\bibinfo {title} {{Solutions to the Public Goods Dilemma in Bacterial
  Biofilms}},\ }\href {https://doi.org/10.1016/j.cub.2013.10.030} {\bibfield
  {journal} {\bibinfo  {journal} {Current Biology}\ }\textbf {\bibinfo {volume}
  {24}},\ \bibinfo {pages} {50} (\bibinfo {year} {2014})}\BibitemShut {NoStop}%
\bibitem [{\citenamefont {Kayser}\ \emph {et~al.}(2018)\citenamefont {Kayser},
  \citenamefont {Schreck}, \citenamefont {Yu}, \citenamefont {Gralka},\ and\
  \citenamefont {Hallatschek}}]{KayserReview18}%
  \BibitemOpen
  \bibfield  {author} {\bibinfo {author} {\bibfnamefont {J.}~\bibnamefont
  {Kayser}}, \bibinfo {author} {\bibfnamefont {C.}~\bibnamefont {Schreck}},
  \bibinfo {author} {\bibfnamefont {Q.}~\bibnamefont {Yu}}, \bibinfo {author}
  {\bibfnamefont {M.}~\bibnamefont {Gralka}},\ and\ \bibinfo {author}
  {\bibfnamefont {O.}~\bibnamefont {Hallatschek}},\ }\bibfield  {title}
  {\bibinfo {title} {Emergence of evolutionary driving forces in
  pattern-forming microbial populations},\ }\href {10.1098/rstb.2017.0106}
  {\bibfield  {journal} {\bibinfo  {journal} {Phil. Trans. R. Soc. B}\ }\textbf
  {\bibinfo {volume} {373}} (\bibinfo {year} {2018})}\BibitemShut {NoStop}%
\bibitem [{\citenamefont {Excoffier}\ \emph {et~al.}(2009)\citenamefont
  {Excoffier}, \citenamefont {Foll},\ and\ \citenamefont
  {Petit}}]{Excoffier.Petit.2009}%
  \BibitemOpen
  \bibfield  {author} {\bibinfo {author} {\bibfnamefont {L.}~\bibnamefont
  {Excoffier}}, \bibinfo {author} {\bibfnamefont {M.}~\bibnamefont {Foll}},\
  and\ \bibinfo {author} {\bibfnamefont {R.~J.}\ \bibnamefont {Petit}},\
  }\bibfield  {title} {\bibinfo {title} {{Genetic Consequences of Range
  Expansions}},\ }\href
  {https://doi.org/10.1146/annurev.ecolsys.39.110707.173414} {\bibfield
  {journal} {\bibinfo  {journal} {Annual Review of Ecology, Evolution, and
  Systematics}\ }\textbf {\bibinfo {volume} {40}},\ \bibinfo {pages} {481}
  (\bibinfo {year} {2009})}\BibitemShut {NoStop}%
\bibitem [{\citenamefont {Yanni}\ \emph {et~al.}(2019)\citenamefont {Yanni},
  \citenamefont {M{\'a}rquez-Zacar{\'\i}as}, \citenamefont {Yunker},\ and\
  \citenamefont {Ratcliff}}]{Yanni:2019ip}%
  \BibitemOpen
  \bibfield  {author} {\bibinfo {author} {\bibfnamefont {D.}~\bibnamefont
  {Yanni}}, \bibinfo {author} {\bibfnamefont {P.}~\bibnamefont
  {M{\'a}rquez-Zacar{\'\i}as}}, \bibinfo {author} {\bibfnamefont {P.~J.}\
  \bibnamefont {Yunker}},\ and\ \bibinfo {author} {\bibfnamefont {W.~C.}\
  \bibnamefont {Ratcliff}},\ }\bibfield  {title} {\bibinfo {title} {{Drivers of
  Spatial Structure in Social Microbial Communities}},\ }\href
  {https://doi.org/10.1016/j.cub.2019.03.068} {\bibfield  {journal} {\bibinfo
  {journal} {Current biology}\ }\textbf {\bibinfo {volume} {29}},\ \bibinfo
  {pages} {R545} (\bibinfo {year} {2019})}\BibitemShut {NoStop}%
\bibitem [{\citenamefont {Hamilton}(1964)}]{Hamilton:1964}%
  \BibitemOpen
  \bibfield  {author} {\bibinfo {author} {\bibfnamefont {W.~D.}\ \bibnamefont
  {Hamilton}},\ }\bibfield  {title} {\bibinfo {title} {The genetical evolution
  of social behaviour. {I+II}},\ }\href
  {https://doi.org/10.1016/0022-5193(64)90038-4} {\bibfield  {journal}
  {\bibinfo  {journal} {Journal of Theoretical Biology}\ }\textbf {\bibinfo
  {volume} {7}},\ \bibinfo {pages} {1} (\bibinfo {year} {1964})}\BibitemShut
  {NoStop}%
\bibitem [{\citenamefont {Killingback}\ \emph {et~al.}(1999)\citenamefont
  {Killingback}, \citenamefont {Doebeli},\ and\ \citenamefont
  {Knowlton}}]{Killingback99}%
  \BibitemOpen
  \bibfield  {author} {\bibinfo {author} {\bibfnamefont {T.}~\bibnamefont
  {Killingback}}, \bibinfo {author} {\bibfnamefont {M.}~\bibnamefont
  {Doebeli}},\ and\ \bibinfo {author} {\bibfnamefont {N.}~\bibnamefont
  {Knowlton}},\ }\bibfield  {title} {\bibinfo {title} {Variable investment, the
  continuous prisoner's dilemma, and the origin of cooperation},\ }\href
  {10.1098/rspb.1999.0838} {\bibfield  {journal} {\bibinfo  {journal}
  {Proceedings Biological sciences}\ }\textbf {\bibinfo {volume} {266}},\
  \bibinfo {pages} {1723} (\bibinfo {year} {1999})}\BibitemShut {NoStop}%
\bibitem [{\citenamefont {Fu}\ \emph {et~al.}(2010)\citenamefont {Fu},
  \citenamefont {Nowak},\ and\ \citenamefont {Hauert}}]{Fu.Hauert.2010}%
  \BibitemOpen
  \bibfield  {author} {\bibinfo {author} {\bibfnamefont {F.}~\bibnamefont
  {Fu}}, \bibinfo {author} {\bibfnamefont {M.~A.}\ \bibnamefont {Nowak}},\ and\
  \bibinfo {author} {\bibfnamefont {C.}~\bibnamefont {Hauert}},\ }\bibfield
  {title} {\bibinfo {title} {{Invasion and expansion of cooperators in lattice
  populations: Prisoner's dilemma vs. snowdrift games}},\ }\href
  {https://doi.org/10.1016/j.jtbi.2010.06.042} {\bibfield  {journal} {\bibinfo
  {journal} {Journal of Theoretical Biology}\ }\textbf {\bibinfo {volume}
  {266}},\ \bibinfo {pages} {358} (\bibinfo {year} {2010})}\BibitemShut
  {NoStop}%
\bibitem [{\citenamefont {Helbing}\ and\ \citenamefont
  {Yu}(2009)}]{Helbing:2009}%
  \BibitemOpen
  \bibfield  {author} {\bibinfo {author} {\bibfnamefont {D.}~\bibnamefont
  {Helbing}}\ and\ \bibinfo {author} {\bibfnamefont {W.}~\bibnamefont {Yu}},\
  }\bibfield  {title} {\bibinfo {title} {The outbreak of cooperation among
  success-driven individuals under noisy conditions},\ }\href
  {https://doi.org/10.1073/pnas.0811503106} {\bibfield  {journal} {\bibinfo
  {journal} {Proc. Natl. Acad. Sci. U.S.A.}\ }\textbf {\bibinfo {volume}
  {106}},\ \bibinfo {pages} {3680} (\bibinfo {year} {2009})}\BibitemShut
  {NoStop}%
\bibitem [{\citenamefont {Langer}\ \emph {et~al.}(2008)\citenamefont {Langer},
  \citenamefont {Nowak},\ and\ \citenamefont {Hauert}}]{Langer:2008}%
  \BibitemOpen
  \bibfield  {author} {\bibinfo {author} {\bibfnamefont {P.}~\bibnamefont
  {Langer}}, \bibinfo {author} {\bibfnamefont {M.~A.}\ \bibnamefont {Nowak}},\
  and\ \bibinfo {author} {\bibfnamefont {C.}~\bibnamefont {Hauert}},\
  }\bibfield  {title} {\bibinfo {title} {{Spatial invasion of cooperation}},\
  }\href {https://doi.org/10.1016/j.jtbi.2007.11.002} {\bibfield  {journal}
  {\bibinfo  {journal} {Journal of Theoretical Biology}\ }\textbf {\bibinfo
  {volume} {250}},\ \bibinfo {pages} {634} (\bibinfo {year}
  {2008})}\BibitemShut {NoStop}%
\bibitem [{\citenamefont {Bauer}\ and\ \citenamefont
  {Frey}(2018{\natexlab{a}})}]{Bauer.Frey.2018}%
  \BibitemOpen
  \bibfield  {author} {\bibinfo {author} {\bibfnamefont {M.}~\bibnamefont
  {Bauer}}\ and\ \bibinfo {author} {\bibfnamefont {E.}~\bibnamefont {Frey}},\
  }\bibfield  {title} {\bibinfo {title} {Delayed adaptation in stochastic
  metapopulation models},\ }\href {https://doi.org/10.1209/0295-5075/122/68002}
  {\bibfield  {journal} {\bibinfo  {journal} {EPL (Europhysics Letters)}\
  }\textbf {\bibinfo {volume} {122}},\ \bibinfo {pages} {68002} (\bibinfo
  {year} {2018}{\natexlab{a}})}\BibitemShut {NoStop}%
\bibitem [{\citenamefont {Bauer}\ and\ \citenamefont
  {Frey}(2018{\natexlab{b}})}]{Bauer.Frey.2018y3b}%
  \BibitemOpen
  \bibfield  {author} {\bibinfo {author} {\bibfnamefont {M.}~\bibnamefont
  {Bauer}}\ and\ \bibinfo {author} {\bibfnamefont {E.}~\bibnamefont {Frey}},\
  }\bibfield  {title} {\bibinfo {title} {Multiple scales in metapopulations of
  public goods producers},\ }\href {10.1103/physreve.97.042307} {\bibfield
  {journal} {\bibinfo  {journal} {Physical Review E}\ }\textbf {\bibinfo
  {volume} {97}},\ \bibinfo {pages} {042307} (\bibinfo {year}
  {2018}{\natexlab{b}})}\BibitemShut {NoStop}%
\bibitem [{\citenamefont {Bauer}\ and\ \citenamefont
  {Frey}(2018{\natexlab{c}})}]{Bauer.Frey.2018qp}%
  \BibitemOpen
  \bibfield  {author} {\bibinfo {author} {\bibfnamefont {M.}~\bibnamefont
  {Bauer}}\ and\ \bibinfo {author} {\bibfnamefont {E.}~\bibnamefont {Frey}},\
  }\bibfield  {title} {\bibinfo {title} {Delays in fitness adjustment can lead
  to coexistence of hierarchically interacting species},\ }\href
  {10.1103/physrevlett.121.268101} {\bibfield  {journal} {\bibinfo  {journal}
  {Physical Review Letters}\ }\textbf {\bibinfo {volume} {121}},\ \bibinfo
  {pages} {268101} (\bibinfo {year} {2018}{\natexlab{c}})}\BibitemShut
  {NoStop}%
\bibitem [{\citenamefont {Hauert}\ and\ \citenamefont
  {Doebeli}(2004)}]{Hauert:2004}%
  \BibitemOpen
  \bibfield  {author} {\bibinfo {author} {\bibfnamefont {C.}~\bibnamefont
  {Hauert}}\ and\ \bibinfo {author} {\bibfnamefont {M.}~\bibnamefont
  {Doebeli}},\ }\bibfield  {title} {\bibinfo {title} {{Spatial structure often
  inhibits the evolution of cooperation in the snowdrift game}},\ }\href
  {https://doi.org/10.1038/nature02360} {\bibfield  {journal} {\bibinfo
  {journal} {Nature}\ }\textbf {\bibinfo {volume} {428}},\ \bibinfo {pages}
  {643} (\bibinfo {year} {2004})}\BibitemShut {NoStop}%
\bibitem [{\citenamefont {Cremer}\ \emph
  {et~al.}(2019{\natexlab{b}})\citenamefont {Cremer}, \citenamefont {Honda},
  \citenamefont {Tang}, \citenamefont {Wong-Ng}, \citenamefont {Vergassola},\
  and\ \citenamefont {Hwa}}]{Cremer:2019}%
  \BibitemOpen
  \bibfield  {author} {\bibinfo {author} {\bibfnamefont {J.}~\bibnamefont
  {Cremer}}, \bibinfo {author} {\bibfnamefont {T.}~\bibnamefont {Honda}},
  \bibinfo {author} {\bibfnamefont {Y.}~\bibnamefont {Tang}}, \bibinfo {author}
  {\bibfnamefont {J.}~\bibnamefont {Wong-Ng}}, \bibinfo {author} {\bibfnamefont
  {M.}~\bibnamefont {Vergassola}},\ and\ \bibinfo {author} {\bibfnamefont
  {T.}~\bibnamefont {Hwa}},\ }\bibfield  {title} {\bibinfo {title} {{Chemotaxis
  as a navigation strategy to boost range expansion}},\ }\href
  {https://doi.org/10.1038/s41586-019-1733-y} {\bibfield  {journal} {\bibinfo
  {journal} {Nature}\ }\textbf {\bibinfo {volume} {575}},\ \bibinfo {pages}
  {658} (\bibinfo {year} {2019}{\natexlab{b}})}\BibitemShut {NoStop}%
\bibitem [{\citenamefont {Jeckel}\ \emph {et~al.}(2019)\citenamefont {Jeckel},
  \citenamefont {Jelli}, \citenamefont {Hartmann}, \citenamefont {Singh},
  \citenamefont {Mok}, \citenamefont {Totz}, \citenamefont {Vidakovic},
  \citenamefont {Eckhardt}, \citenamefont {Dunkel},\ and\ \citenamefont
  {Drescher}}]{Jeckel:2019}%
  \BibitemOpen
  \bibfield  {author} {\bibinfo {author} {\bibfnamefont {H.}~\bibnamefont
  {Jeckel}}, \bibinfo {author} {\bibfnamefont {E.}~\bibnamefont {Jelli}},
  \bibinfo {author} {\bibfnamefont {R.}~\bibnamefont {Hartmann}}, \bibinfo
  {author} {\bibfnamefont {P.~K.}\ \bibnamefont {Singh}}, \bibinfo {author}
  {\bibfnamefont {R.}~\bibnamefont {Mok}}, \bibinfo {author} {\bibfnamefont
  {J.~F.}\ \bibnamefont {Totz}}, \bibinfo {author} {\bibfnamefont
  {L.}~\bibnamefont {Vidakovic}}, \bibinfo {author} {\bibfnamefont
  {B.}~\bibnamefont {Eckhardt}}, \bibinfo {author} {\bibfnamefont
  {J.}~\bibnamefont {Dunkel}},\ and\ \bibinfo {author} {\bibfnamefont
  {K.}~\bibnamefont {Drescher}},\ }\bibfield  {title} {\bibinfo {title}
  {{Learning the space-time phase diagram of bacterial swarm expansion.}},\
  }\href {https://doi.org/10.1073/pnas.1811722116} {\bibfield  {journal}
  {\bibinfo  {journal} {Proc. Nat. Acad. Sci. U.S.A.}\ }\textbf {\bibinfo
  {volume} {116}},\ \bibinfo {pages} {1489} (\bibinfo {year}
  {2019})}\BibitemShut {NoStop}%
\bibitem [{\citenamefont {Liu}\ \emph {et~al.}(2019)\citenamefont {Liu},
  \citenamefont {Cremer}, \citenamefont {Li}, \citenamefont {Hwa},\ and\
  \citenamefont {Liu}}]{Liu:51NDZHe4}%
  \BibitemOpen
  \bibfield  {author} {\bibinfo {author} {\bibfnamefont {W.}~\bibnamefont
  {Liu}}, \bibinfo {author} {\bibfnamefont {J.}~\bibnamefont {Cremer}},
  \bibinfo {author} {\bibfnamefont {D.}~\bibnamefont {Li}}, \bibinfo {author}
  {\bibfnamefont {T.}~\bibnamefont {Hwa}},\ and\ \bibinfo {author}
  {\bibfnamefont {C.}~\bibnamefont {Liu}},\ }\bibfield  {title} {\bibinfo
  {title} {{An evolutionarily stable strategy to colonize spatially extended
  habitats}},\ }\href {https://doi.org/10.1038/s41586-019-1734-x} {\bibfield
  {journal} {\bibinfo  {journal} {Nature}\ }\textbf {\bibinfo {volume} {575}},\
  \bibinfo {pages} {664} (\bibinfo {year} {2019})}\BibitemShut {NoStop}%
\bibitem [{\citenamefont {Spormann}(1999)}]{Spormann.1999}%
  \BibitemOpen
  \bibfield  {author} {\bibinfo {author} {\bibfnamefont {A.~M.}\ \bibnamefont
  {Spormann}},\ }\bibfield  {title} {\bibinfo {title} {Gliding motility in
  bacteria: Insights from studies of {Myxococcus} xanthus},\ }\href@noop {}
  {\bibfield  {journal} {\bibinfo  {journal} {Microbiology and Molecular
  Biology Reviews}\ }\textbf {\bibinfo {volume} {63}},\ \bibinfo {pages} {621}
  (\bibinfo {year} {1999})}\BibitemShut {NoStop}%
\bibitem [{\citenamefont {Weber}\ \emph {et~al.}(2014)\citenamefont {Weber},
  \citenamefont {Poxleitner}, \citenamefont {Hebisch}, \citenamefont {Frey},\
  and\ \citenamefont {Opitz}}]{weber_chemical_2014}%
  \BibitemOpen
  \bibfield  {author} {\bibinfo {author} {\bibfnamefont {M.~F.}\ \bibnamefont
  {Weber}}, \bibinfo {author} {\bibfnamefont {G.}~\bibnamefont {Poxleitner}},
  \bibinfo {author} {\bibfnamefont {E.}~\bibnamefont {Hebisch}}, \bibinfo
  {author} {\bibfnamefont {E.}~\bibnamefont {Frey}},\ and\ \bibinfo {author}
  {\bibfnamefont {M.}~\bibnamefont {Opitz}},\ }\bibfield  {title} {\bibinfo
  {title} {{Chemical warfare and survival strategies in bacterial range
  expansions}},\ }\href {https://doi.org/10.1098/rsif.2014.0172} {\bibfield
  {journal} {\bibinfo  {journal} {Journal of The Royal Society Interface}\
  }\textbf {\bibinfo {volume} {11}},\ \bibinfo {pages} {20140172} (\bibinfo
  {year} {2014})}\BibitemShut {NoStop}%
\bibitem [{\citenamefont {Bertin}\ \emph {et~al.}(2006)\citenamefont {Bertin},
  \citenamefont {Droz},\ and\ \citenamefont {Gregoire}}]{Bertin06}%
  \BibitemOpen
  \bibfield  {author} {\bibinfo {author} {\bibfnamefont {E.}~\bibnamefont
  {Bertin}}, \bibinfo {author} {\bibfnamefont {M.}~\bibnamefont {Droz}},\ and\
  \bibinfo {author} {\bibfnamefont {G.}~\bibnamefont {Gregoire}},\ }\bibfield
  {title} {\bibinfo {title} {Boltzmann and hydrodynamic description for
  self-propelled particles},\ }\href {10.1103/PhysRevE.74.022101} {\bibfield
  {journal} {\bibinfo  {journal} {Phys. Rev. E}\ }\textbf {\bibinfo {volume}
  {74}},\ \bibinfo {pages} {022101} (\bibinfo {year} {2006})}\BibitemShut
  {NoStop}%
\bibitem [{\citenamefont {{Bertin}}\ \emph {et~al.}(2009)\citenamefont
  {{Bertin}}, \citenamefont {{Droz}},\ and\ \citenamefont
  {{Gr{\'e}goire}}}]{Bertin09}%
  \BibitemOpen
  \bibfield  {author} {\bibinfo {author} {\bibfnamefont {E.}~\bibnamefont
  {{Bertin}}}, \bibinfo {author} {\bibfnamefont {M.}~\bibnamefont {{Droz}}},\
  and\ \bibinfo {author} {\bibfnamefont {G.}~\bibnamefont {{Gr{\'e}goire}}},\
  }\bibfield  {title} {\bibinfo {title} {{Hydrodynamic equations for
  self-propelled particles: microscopic derivation and stability analysis}},\
  }\href {https://doi.org/10.1088/1751-8113/42/44/445001} {\bibfield  {journal}
  {\bibinfo  {journal} {Journal of Physics A: Mathematical General.}\ }\textbf
  {\bibinfo {volume} {42}},\ \bibinfo {pages} {445001} (\bibinfo {year}
  {2009})}\BibitemShut {NoStop}%
\bibitem [{\citenamefont {Huber}\ \emph {et~al.}(2018)\citenamefont {Huber},
  \citenamefont {Suzuki}, \citenamefont {Kr{\"u}ger}, \citenamefont {Frey},\
  and\ \citenamefont {Bausch}}]{Huber18}%
  \BibitemOpen
  \bibfield  {author} {\bibinfo {author} {\bibfnamefont {L.}~\bibnamefont
  {Huber}}, \bibinfo {author} {\bibfnamefont {R.}~\bibnamefont {Suzuki}},
  \bibinfo {author} {\bibfnamefont {T.}~\bibnamefont {Kr{\"u}ger}}, \bibinfo
  {author} {\bibfnamefont {E.}~\bibnamefont {Frey}},\ and\ \bibinfo {author}
  {\bibfnamefont {A.~R.}\ \bibnamefont {Bausch}},\ }\bibfield  {title}
  {\bibinfo {title} {Emergence of coexisting ordered states in active matter
  systems},\ }\href {10.1126/science.aao5434} {\bibfield  {journal} {\bibinfo
  {journal} {Science}\ }\textbf {\bibinfo {volume} {361}},\ \bibinfo {pages}
  {255} (\bibinfo {year} {2018})}\BibitemShut {NoStop}%
\bibitem [{\citenamefont {Suzuki}\ \emph {et~al.}(2015)\citenamefont {Suzuki},
  \citenamefont {Weber}, \citenamefont {Frey},\ and\ \citenamefont
  {Bausch}}]{Suzuki:2015}%
  \BibitemOpen
  \bibfield  {author} {\bibinfo {author} {\bibfnamefont {R.}~\bibnamefont
  {Suzuki}}, \bibinfo {author} {\bibfnamefont {C.~A.}\ \bibnamefont {Weber}},
  \bibinfo {author} {\bibfnamefont {E.}~\bibnamefont {Frey}},\ and\ \bibinfo
  {author} {\bibfnamefont {A.~R.}\ \bibnamefont {Bausch}},\ }\bibfield  {title}
  {\bibinfo {title} {{Polar pattern formation in driven filament systems
  requires non-binary particle collisions}},\ }\href
  {https://doi.org/10.1038/nphys3423} {\bibfield  {journal} {\bibinfo
  {journal} {Nature Physics}\ }\textbf {\bibinfo {volume} {11}},\ \bibinfo
  {pages} {839} (\bibinfo {year} {2015})}\BibitemShut {NoStop}%
\bibitem [{\citenamefont {Denk}\ and\ \citenamefont {Frey}(2020)}]{Denk:2020}%
  \BibitemOpen
  \bibfield  {author} {\bibinfo {author} {\bibfnamefont {J.}~\bibnamefont
  {Denk}}\ and\ \bibinfo {author} {\bibfnamefont {E.}~\bibnamefont {Frey}},\
  }\bibfield  {title} {\bibinfo {title} {Pattern-induced local symmetry
  breaking in active-matter systems},\ }\href {10.1073/pnas.2010302117}
  {\bibfield  {journal} {\bibinfo  {journal} {Proceedings of the National
  Academy of Sciences}\ }\textbf {\bibinfo {volume} {117}},\ \bibinfo {pages}
  {31623} (\bibinfo {year} {2020})}\BibitemShut {NoStop}%
\bibitem [{\citenamefont {Frey}\ and\ \citenamefont
  {Reichenbach}(2011)}]{BacterialGames}%
  \BibitemOpen
  \bibfield  {author} {\bibinfo {author} {\bibfnamefont {E.}~\bibnamefont
  {Frey}}\ and\ \bibinfo {author} {\bibfnamefont {T.}~\bibnamefont
  {Reichenbach}},\ }\bibfield  {title} {\bibinfo {title} {Bacterial games},\
  }\href {https://doi.org/10.1007/978-3-642-18137-5_13} {\bibfield  {journal}
  {\bibinfo  {journal} {Principles of Evolution: From the Planck Epoch to
  Complex Multicellular Life, The Frontiers Collection}\ ,\ \bibinfo {pages}
  {297}} (\bibinfo {year} {2011})}\BibitemShut {NoStop}%
\bibitem [{\citenamefont {Th\"uroff}\ \emph {et~al.}(2014)\citenamefont
  {Th\"uroff}, \citenamefont {Weber},\ and\ \citenamefont {Frey}}]{Thueroff14}%
  \BibitemOpen
  \bibfield  {author} {\bibinfo {author} {\bibfnamefont {F.}~\bibnamefont
  {Th\"uroff}}, \bibinfo {author} {\bibfnamefont {C.~A.}\ \bibnamefont
  {Weber}},\ and\ \bibinfo {author} {\bibfnamefont {E.}~\bibnamefont {Frey}},\
  }\bibfield  {title} {\bibinfo {title} {Numerical treatment of the boltzmann
  equation for self-propelled particle systems},\ }\href
  {10.1103/PhysRevX.4.041030} {\bibfield  {journal} {\bibinfo  {journal} {Phys.
  Rev. X}\ }\textbf {\bibinfo {volume} {4}},\ \bibinfo {pages} {041030}
  (\bibinfo {year} {2014})}\BibitemShut {NoStop}%
\bibitem [{\citenamefont {Denk}\ \emph {et~al.}(2016)\citenamefont {Denk},
  \citenamefont {Huber}, \citenamefont {Reithmann},\ and\ \citenamefont
  {Frey}}]{DenkHuber16}%
  \BibitemOpen
  \bibfield  {author} {\bibinfo {author} {\bibfnamefont {J.}~\bibnamefont
  {Denk}}, \bibinfo {author} {\bibfnamefont {L.}~\bibnamefont {Huber}},
  \bibinfo {author} {\bibfnamefont {E.}~\bibnamefont {Reithmann}},\ and\
  \bibinfo {author} {\bibfnamefont {E.}~\bibnamefont {Frey}},\ }\bibfield
  {title} {\bibinfo {title} {Active curved polymers form vortex patterns on
  membranes},\ }\href {10.1103/PhysRevLett.116.178301} {\bibfield  {journal}
  {\bibinfo  {journal} {Phys. Rev. Lett.}\ }\textbf {\bibinfo {volume} {116}},\
  \bibinfo {pages} {178301} (\bibinfo {year} {2016})}\BibitemShut {NoStop}%
\bibitem [{\citenamefont {Peshkov}\ \emph {et~al.}(2014)\citenamefont
  {Peshkov}, \citenamefont {Bertin}, \citenamefont {Ginelli},\ and\
  \citenamefont {Chat{\'e}}}]{Peshkov14}%
  \BibitemOpen
  \bibfield  {author} {\bibinfo {author} {\bibfnamefont {A.}~\bibnamefont
  {Peshkov}}, \bibinfo {author} {\bibfnamefont {E.}~\bibnamefont {Bertin}},
  \bibinfo {author} {\bibfnamefont {F.}~\bibnamefont {Ginelli}},\ and\ \bibinfo
  {author} {\bibfnamefont {H.}~\bibnamefont {Chat{\'e}}},\ }\bibfield  {title}
  {\bibinfo {title} {Boltzmann-{G}inzburg-{L}andau approach for continuous
  descriptions of generic {V}icsek-like models},\ }\href
  {https://doi.org/10.1140/epjst/e2014-02193-y} {\bibfield  {journal} {\bibinfo
   {journal} {Eur. Phys. J. Spec. Top.}\ }\textbf {\bibinfo {volume} {223}},\
  \bibinfo {pages} {1315} (\bibinfo {year} {2014})}\BibitemShut {NoStop}%
\bibitem [{\citenamefont {Peshkov}\ \emph
  {et~al.}(2012{\natexlab{a}})\citenamefont {Peshkov}, \citenamefont {Aranson},
  \citenamefont {Bertin}, \citenamefont {Chat{\'e}},\ and\ \citenamefont
  {Ginelli}}]{PeshkovAranson12}%
  \BibitemOpen
  \bibfield  {author} {\bibinfo {author} {\bibfnamefont {A.}~\bibnamefont
  {Peshkov}}, \bibinfo {author} {\bibfnamefont {I.~S.}\ \bibnamefont
  {Aranson}}, \bibinfo {author} {\bibfnamefont {E.}~\bibnamefont {Bertin}},
  \bibinfo {author} {\bibfnamefont {H.}~\bibnamefont {Chat{\'e}}},\ and\ \bibinfo
  {author} {\bibfnamefont {F.}~\bibnamefont {Ginelli}},\ }\bibfield  {title}
  {\bibinfo {title} {Nonlinear field equations for aligning self-propelled
  rods},\ }\href {10.1103/PhysRevLett.109.268701} {\bibfield  {journal}
  {\bibinfo  {journal} {Phys. Rev. Lett.}\ }\textbf {\bibinfo {volume} {109}},\
  \bibinfo {pages} {268701} (\bibinfo {year} {2012}{\natexlab{a}})}\BibitemShut
  {NoStop}%
\bibitem [{\citenamefont {Peshkov}\ \emph
  {et~al.}(2012{\natexlab{b}})\citenamefont {Peshkov}, \citenamefont {Ngo},
  \citenamefont {Bertin}, \citenamefont {Chat{\'e}},\ and\ \citenamefont
  {Ginelli}}]{PeshkovNgo12}%
  \BibitemOpen
  \bibfield  {author} {\bibinfo {author} {\bibfnamefont {A.}~\bibnamefont
  {Peshkov}}, \bibinfo {author} {\bibfnamefont {S.}~\bibnamefont {Ngo}},
  \bibinfo {author} {\bibfnamefont {E.}~\bibnamefont {Bertin}}, \bibinfo
  {author} {\bibfnamefont {H.}~\bibnamefont {Chat{\'e}}},\ and\ \bibinfo {author}
  {\bibfnamefont {F.}~\bibnamefont {Ginelli}},\ }\bibfield  {title} {\bibinfo
  {title} {Continuous theory of active matter systems with metric-free
  interactions},\ }\href {10.1103/PhysRevLett.109.098101} {\bibfield  {journal}
  {\bibinfo  {journal} {Phys. Rev. Lett.}\ }\textbf {\bibinfo {volume} {109}},\
  \bibinfo {pages} {098101} (\bibinfo {year} {2012}{\natexlab{b}})}\BibitemShut
  {NoStop}%
\bibitem [{\citenamefont {Mayer}(2019)}]{JohannaMasterthesis}%
  \BibitemOpen
  \bibfield  {author} {\bibinfo {author} {\bibfnamefont {J.}~\bibnamefont
  {Mayer}},\ }\bibfield  {title} {\bibinfo {title} {Two species of
  self-propelled particles interacting in a snowdrift game scenario - kinetic
  approach},\ }\href
  {https://www.theorie.physik.uni-muenchen.de/lsfrey/master_thesis_folder/master_thesis_mayer.pdf}
  {\bibfield  {journal} {\bibinfo  {journal} {Master thesis at the Department
  of Statistical and Biological Physics at LMU Munich}\ }
  (\bibinfo {year} {2019})}\BibitemShut {NoStop}%
\bibitem [{\citenamefont {Gr\'egoire}\ and\ \citenamefont
  {Chat{\'e}}(2004)}]{Gregoire04}%
  \BibitemOpen
  \bibfield  {author} {\bibinfo {author} {\bibfnamefont {G.}~\bibnamefont
  {Gr{\'e}goire}}\ and\ \bibinfo {author} {\bibfnamefont {H.}~\bibnamefont
  {Chat{\'e}}},\ }\bibfield  {title} {\bibinfo {title} {Onset of collective and
  cohesive motion},\ }\href {10.1103/PhysRevLett.92.025702} {\bibfield
  {journal} {\bibinfo  {journal} {Phys. Rev. Lett.}\ }\textbf {\bibinfo
  {volume} {92}},\ \bibinfo {pages} {025702} (\bibinfo {year}
  {2004})}\BibitemShut {NoStop}%
\bibitem [{\citenamefont {Caussin}\ \emph {et~al.}(2014)\citenamefont
  {Caussin}, \citenamefont {Solon}, \citenamefont {Peshkov}, \citenamefont
  {Chat{\'e}}, \citenamefont {Dauxois}, \citenamefont {Tailleur}, \citenamefont
  {Vitelli},\ and\ \citenamefont {Bartolo}}]{Caussin14}%
  \BibitemOpen
  \bibfield  {author} {\bibinfo {author} {\bibfnamefont {J.-B.}\ \bibnamefont
  {Caussin}}, \bibinfo {author} {\bibfnamefont {A.}~\bibnamefont {Solon}},
  \bibinfo {author} {\bibfnamefont {A.}~\bibnamefont {Peshkov}}, \bibinfo
  {author} {\bibfnamefont {H.}~\bibnamefont {Chat{\'e}}}, \bibinfo {author}
  {\bibfnamefont {T.}~\bibnamefont {Dauxois}}, \bibinfo {author} {\bibfnamefont
  {J.}~\bibnamefont {Tailleur}}, \bibinfo {author} {\bibfnamefont
  {V.}~\bibnamefont {Vitelli}},\ and\ \bibinfo {author} {\bibfnamefont
  {D.}~\bibnamefont {Bartolo}},\ }\bibfield  {title} {\bibinfo {title}
  {Emergent spatial structures in flocking models: A dynamical system
  insight},\ }\href {10.1103/PhysRevLett.112.148102} {\bibfield  {journal}
  {\bibinfo  {journal} {Phys. Rev. Lett.}\ }\textbf {\bibinfo {volume} {112}},\
  \bibinfo {pages} {148102} (\bibinfo {year} {2014})}\BibitemShut {NoStop}%
\bibitem [{\citenamefont {Vicsek}\ \emph {et~al.}(1995)\citenamefont {Vicsek},
  \citenamefont {Czir\'ok}, \citenamefont {Ben-Jacob}, \citenamefont {Cohen},\
  and\ \citenamefont {Shochet}}]{Vicsek95}%
  \BibitemOpen
  \bibfield  {author} {\bibinfo {author} {\bibfnamefont {T.}~\bibnamefont
  {Vicsek}}, \bibinfo {author} {\bibfnamefont {A.}~\bibnamefont {Czir{\'o}k}},
  \bibinfo {author} {\bibfnamefont {E.}~\bibnamefont {Ben-Jacob}}, \bibinfo
  {author} {\bibfnamefont {I.}~\bibnamefont {Cohen}},\ and\ \bibinfo {author}
  {\bibfnamefont {O.}~\bibnamefont {Shochet}},\ }\bibfield  {title} {\bibinfo
  {title} {Novel type of phase transition in a system of self-driven
  particles},\ }\href {10.1103/PhysRevLett.75.1226} {\bibfield  {journal}
  {\bibinfo  {journal} {Phys. Rev. Lett.}\ }\textbf {\bibinfo {volume} {75}},\
  \bibinfo {pages} {1226} (\bibinfo {year} {1995})}\BibitemShut {NoStop}%
\bibitem [{\citenamefont {Chat{\'e}}\ \emph {et~al.}(2008)\citenamefont
  {Chat{\'e}}, \citenamefont {Ginelli}, \citenamefont {Gr\'egoire},\ and\
  \citenamefont {Raynaud}}]{Chate08}%
  \BibitemOpen
  \bibfield  {author} {\bibinfo {author} {\bibfnamefont {H.}~\bibnamefont
  {Chat{\'e}}}, \bibinfo {author} {\bibfnamefont {F.}~\bibnamefont {Ginelli}},
  \bibinfo {author} {\bibfnamefont {G.}~\bibnamefont {Gr{\'e}goire}},\ and\
  \bibinfo {author} {\bibfnamefont {F.}~\bibnamefont {Raynaud}},\ }\bibfield
  {title} {\bibinfo {title} {Collective motion of self-propelled particles
  interacting without cohesion},\ }\href
  {https://doi.org/10.1103/physreve.77.046113} {\bibfield  {journal} {\bibinfo
  {journal} {Phys. Rev. E}\ }\textbf {\bibinfo {volume} {77}},\ \bibinfo
  {pages} {046113} (\bibinfo {year} {2008})}\BibitemShut {NoStop}%
\bibitem [{\citenamefont {Oberm{\"u}ller}(2019)}]{MichaelMasterthesis}%
  \BibitemOpen
  \bibfield  {author} {\bibinfo {author} {\bibfnamefont {M.}~\bibnamefont
  {Oberm{\"u}ller}},\ }\bibfield  {title} {\bibinfo {title} {Two species of
  self-propelled particles interacting in a snowdrift game scenario -
  agent-based approach},\ }\href
  {https://www.theorie.physik.uni-muenchen.de/lsfrey/master_thesis_folder/main.pdf}
  {\bibfield  {journal} {\bibinfo  {journal} {Master thesis at the Department
  of Statistical and Biological Physics at LMU Munich}\ }
  (\bibinfo {year} {2019})}\BibitemShut {NoStop}%
\bibitem [{Note1()}]{Note1}%
  \BibitemOpen
  \bibinfo {note} {This is the case as long as the game range is not chosen too
  small, i.e.~that the average number of particles participating in an
  interaction is insufficient to roughly achieve the desired ratio given by
  $\lambda $.}\BibitemShut {Stop}%
\bibitem [{\citenamefont {B{\"a}r}\ \emph {et~al.}(2020)\citenamefont
  {B{\"a}r}, \citenamefont {Gro{\ss}mann}, \citenamefont {Heidenreich},\ and\
  \citenamefont {Peruani}}]{Baer_etal:2020}%
  \BibitemOpen
  \bibfield  {author} {\bibinfo {author} {\bibfnamefont {M.}~\bibnamefont
  {B{\"a}r}}, \bibinfo {author} {\bibfnamefont {R.}~\bibnamefont
  {Gro{\ss}mann}}, \bibinfo {author} {\bibfnamefont {S.}~\bibnamefont
  {Heidenreich}},\ and\ \bibinfo {author} {\bibfnamefont {F.}~\bibnamefont
  {Peruani}},\ }\bibfield  {title} {\bibinfo {title} {Self-propelled rods:
  Insights and perspectives for active matter},\ }\bibfield  {journal}
  {\bibinfo  {journal} {Annual Review of Condensed Matter Physics}\ }\textbf
  {\bibinfo {volume} {11}},\ \href
  {https://doi.org/10.1146/annurev-conmatphys-031119-050611}
  {10.1146/annurev-conmatphys-031119-050611} (\bibinfo {year}
  {2020})\BibitemShut {NoStop}%
\bibitem [{\citenamefont {Gude}\ \emph {et~al.}(2020)\citenamefont {Gude},
  \citenamefont {Pince}, \citenamefont {Taute}, \citenamefont {Seinen},
  \citenamefont {Shimizu},\ and\ \citenamefont {Tans}}]{Gude:2020}%
  \BibitemOpen
  \bibfield  {author} {\bibinfo {author} {\bibfnamefont {S.}~\bibnamefont
  {Gude}}, \bibinfo {author} {\bibfnamefont {E.}~\bibnamefont {Pince}},
  \bibinfo {author} {\bibfnamefont {K.~M.}\ \bibnamefont {Taute}}, \bibinfo
  {author} {\bibfnamefont {A.-B.}\ \bibnamefont {Seinen}}, \bibinfo {author}
  {\bibfnamefont {T.~S.}\ \bibnamefont {Shimizu}},\ and\ \bibinfo {author}
  {\bibfnamefont {S.~J.}\ \bibnamefont {Tans}},\ }\bibfield  {title} {\bibinfo
  {title} {{Bacterial coexistence driven by motility and spatial
  competition}},\ }\href {https://doi.org/10.1038/s41586-020-2033-2} {\bibfield
   {journal} {\bibinfo  {journal} {Nature}\ }\textbf {\bibinfo {volume}
  {578}},\ \bibinfo {pages} {588} (\bibinfo {year} {2020})}\BibitemShut
  {NoStop}%
\bibitem [{\citenamefont {Grubmueller}\ \emph {et~al.}(1991)\citenamefont
  {Grubmueller}, \citenamefont {Heller}, \citenamefont {Windemuth},\ and\
  \citenamefont {Schulten}}]{BoxAlgo}%
  \BibitemOpen
  \bibfield  {author} {\bibinfo {author} {\bibfnamefont {H.}~\bibnamefont
  {Grubmueller}}, \bibinfo {author} {\bibfnamefont {H.}~\bibnamefont {Heller}},
  \bibinfo {author} {\bibfnamefont {A.}~\bibnamefont {Windemuth}},\ and\
  \bibinfo {author} {\bibfnamefont {K.}~\bibnamefont {Schulten}},\ }\bibfield
  {title} {\bibinfo {title} {Generalized verlet algorithm for efficient
  molecular dynamics simulations with long-range interactions},\ }\href
  {https://doi.org/10.1080/08927029108022142} {\bibfield  {journal} {\bibinfo
  {journal} {Mol. Sim.}\ }\textbf {\bibinfo {volume} {6}},\ \bibinfo {pages}
  {121} (\bibinfo {year} {1991})}\BibitemShut {NoStop}%
\end{thebibliography}
\end{document}